\pdfoutput=1
\documentclass[a4paper,11pt]{article}
\usepackage{graphicx}
\usepackage{fullpage}
\usepackage{jcappub} 
\usepackage{bbm}
\usepackage{amsfonts}
\usepackage{slashed}
\usepackage{subfigure}
\usepackage{array}
\usepackage{verbatim}
\usepackage{booktabs}
\usepackage{color, soul}
\usepackage{mathrsfs}
\usepackage{epsfig}
\usepackage{graphicx}
\usepackage{dcolumn}
\usepackage{bm}
\usepackage{amsmath}
\usepackage{amssymb}
\usepackage{multirow}

\setstcolor{red}

\title{\sffamily Heavy quark-philic scalar dark matter with a vector-like fermion portal}

\author[a,b]{Seungwon Baek,}
\author[a,c]{Pyungwon Ko,}
\author[a]{Peiwen Wu}

\affiliation[a]{School of Physics, KIAS, 85 Hoegiro, Seoul 02455, Republic of  Korea}
\affiliation[b]{Department of Physics, Korea University, Seoul 02841, Republic of Korea}
\affiliation[c]{Quantum Universe Center, KIAS, 85 Hoegiro, Seoul 02455, Republic of Korea}

\emailAdd{swbaek@kias.re.kr}
\emailAdd{pko@kias.re.kr}
\emailAdd{pwwu@kias.re.kr}

\abstract{The absence of confirmed signal in dark matter (DM) direct detection (DD) may suggest weak interaction strengths between DM and the abundant constituents inside nucleon, i.e. gluons and valence light quarks. In this work we consider a real scalar dark matter $S$ 
interacting only with $SU(2)_L$ singlet Up-type quarks $U_i=u_R,c_R,t_R$ via a vector-like fermion $\psi$ which has the same quantum number as $U_i$. The DM-nucleon scattering can proceed through both $h$-mediated Higgs portal (HP) and $\psi$-mediated vector-like portal (VLP), in which HP can receive sizable radiative corrections through the new fermions. We first study the separate constraints on the new Yukawa couplings $y_i$
and find that the constraints of XENON1T results are strong on $y_1$ from VLP scattering and on $y_3$ from its radiative contributions to HP scattering. Since both DM-light quark interactions and HP have been well studied in the existing literature, we move forward to focus on DM-heavy quark interactions. Since there is no valence $c,t$ quark inside nucleons at $\mu_{\rm had}\sim 1$ GeV, $y_2,y_3$ interactions are manifested in DM-gluon scattering at loop level.
We find that renormalization group equation (RGE) and heavy quark threshold effects are important if one calculates the DM-nucleon scattering rate $\sigma^{\rm SI}_{p}$ at $\mu_{\rm had}\sim 1\, {\rm GeV}$ while constructing the effective theory at $\mu_{\rm EFT}\sim m_Z$. For the benchmarks $y_3=0.5, y_2=0.5, 1, 3$, combined results from $\Omega_{\rm DM} h^2\simeq 0.12$, XENON1T, Fermi-LAT, 13 TeV LHC data have almost excluded $m_S<m_t/2$ when only DM-$\{c,t\}$ interactions are considered. FCNC of top quark can be generated at both tree level $t\to \psi^{(*)}S \to cSS$ and loop level $t\to c+\gamma/g/Z$, of which the branching fractions are typically below $10^{-9}$ after passing the other constraints, which are still safe from the current top quark width measurements.}

\begin{document}
\maketitle \indent
\newpage

\section{\label{introduction}Introduction}

The existence of Dark Matter (DM) has been strongly suggested in various astrophysical and cosmological observations \cite{Sofue2000,PlanckCollaboration2013,Clowe2006}. Despite the fact that about $80\%$ of the matter content of the Universe is composed of DM and many experiments have been searching for possible DM signals including the Direct Detection (DD), Indirect Detection (ID) and collider searches, no confirmed non-gravitational properties of DM has been established.

Given the fact that the dominant component in proton and neutron are gluons and the light quarks, the null confirmed signal from the DD experiments motivate the possibility that DM may couple weakly to the first generation quarks if DM is a colorless particle in the first place, the origin of which may be attributed to loop nature and/or destructive cancellation. The preference of DM couplings to specific fermion flavors in the Standard Model (SM) has also been suggested in the explanation of the Galactic Center gamma-ray Excess (GCE) \cite{Goodenough_0910.2998,Hooper_1010.2752,Hooper_1110.0006, Abazajian_1207.6047,Gordon_1306.5725,Abazajian_1402.4090,Hooper_1302.6589,Hooper_1402.6703,Calore_1409.0042,Murgia-Fermi}. The gamma-ray spectrum fitting of DM annihilations into different SM final states showed that some fermion species are more suitable than others \cite{Calore2014-1411.4647}.
The sensitivities of collider searches for DM also rely on whether and/or how strongly DM couples to different quark
flavors, which would affect both the DM production cross section via light quark fusions and the visible signal types in the final states associated with the DM.

The above observations have led to the proposal of flavor structure in the interaction between DM and the SM, the so-called quark-philic (or leptophilic)
\cite{Lopez-Honorez2013,Jackson2009,Agrawal2015a,Baek2016,Bai2013,Baek2015,Chao2016-1606.07174,Hamze2014,Kile2014,Kile2013,Agrawal2014b,Agrawal2014,Yu2014,Chang2014,Lee2014,Calibbi2015,Agrawal2015,Cheung2013,DiFranzo2013,Freitas2014,Cohen2009,Schmidt2012,Chao2010,Baltz2002,Cai2014,Isidori2012,Isidori2010,Agrawal2011,Kamenik2011,Batell2011,Kile2011,Abe2016-FDM,Okawa2017-FDM,Bhattacharya:2015xha-FDM,Kumar2013,Kilic2015,Arina2016,Gomez2014,Zhang2012,Batell2013,Baek2016a}. Extensive
studies have been carried out with different assumptions of the quantum numbers of the particles participating in the
interaction including flavor charge assignment, the particle spin configurations as well as the Lorentz structure of the
interaction vertex. Many studies imposed the assumption of Minimal Flavor Violation (MFV) to evade the  strong
constraints from various flavor measurements, which assumes that the origin of flavor mixing comes only from
the SM Yukawa interaction \cite{DAmbrosio2002}. However, implementation of more general flavor structures beyond the MFV
assumption have also been discussed \cite{Chen2015,Agrawal2014a}.

Given the special role played by the top quark in the SM, top-philic DM has been an interesting scenario receiving special attention \cite{Kumar2013,Kilic2015,Arina2016,Gomez2014,Zhang2012,Batell2013,Baek2016a}.
Due to the heavy top mass and the wide mass window below the top threshold, the calculations related to the DM thermal relic are more complicated for DM lighter than the top quark, in which case co-annihilations are usually important and one needs a more complete calculation without neglecting the top quark mass. On the contrary, for light quark- or lepton-philic DM scenario, DM mass is easily above the light fermion threshold if DM is a Weakly Interacting Massive Particle (WIMP). The large top quark mass is also advantageous if one considers the scalar DM scenario, especially the real scalar DM $S$, which receives chiral suppression in both $s-$ and $p-$wave components of $\langle \sigma v\rangle_{SS\to \bar{f}f}\simeq s + p\,v^2$, where $f$ are SM fermions and $s,p\propto m_f^2$. 

Another interesting point worth emphasizing is that the absence of valence top quark in nucleons makes DM-gluon scattering at loop level crucial and the future improvement of the DD sensitivities very important.
In our previous work \cite{Baek2016a} we studied a phenomenologically motivated top-philic DM model in which the DM is a SM singlet real scalar and interacts exclusively with the right-handed (RH) top quark through a vector-like fermion mediator. We found that the projected XENON1T
experiment may test the new Yukawa coupling to $\mathcal{O}(1)$ and the future LZ project may further test to
$\mathcal{O}(0.5)$. The loop nature of DM-nucleon scattering in the top-philic DM scenario manifests itself in the challenges in the direct detection.

Interestingly, in an opposite direction, \cite{Giacchino:2015hvk} has presented a comprehensive study on a scalar DM which interacts only with light quarks $-y S \overline{\psi_L} q_R + h.c.$, where $q=u,d$ and $\psi$ is a vector-like mediator carrying the same quantum number as $q_R$. Considering that the DM self-annihilation $\langle \sigma v\rangle_{SS\to \bar{q}q}\simeq d\, v^4 $ is $d$-wave suppressed due to $m_q\to 0$, \cite{Giacchino:2015hvk} studied two classes of higher-order processes: the internal Bremsstrahlung of a gauge boson $V=\gamma, Z, g$ in $SS\to q\bar{q}V$, and $SS \to VV'$ induced via 1-loop box diagram. Furthermore, non-perturbative Sommerfeld corrections to the co-annihilation processes $\psi\psi, \overline{\psi}\,\overline{\psi}, \overline{\psi}\psi \to {\rm SM}$ are also calculated. Other phenomenology including DM direct/indirect detections and collider signals are also investigated in \cite{Giacchino:2015hvk}.

In this work we extend our previous study \cite{Baek2016a} focusing on DM-top quark interaction and \cite{Giacchino:2015hvk} focusing on DM-light quark interactions, by including the DM-charm quark interaction in the following Lagrangian
\begin{eqnarray}
\mathcal{L} \supset -y_1 S \overline{\psi_L} u_R -y_2 S \overline{\psi_L} c_R  -y_3 S \overline{\psi_L} t_R +  h.c.,
\label{eq-L}
\end{eqnarray}
where $S$ is the SM singlet real scalar DM and $\psi$ is a vector-like mediator carrying the same quantum number as $u_R,c_R,t_R$. We use $y_1,y_2,y_3$ to denote the new Yukawa couplings.

Note that we will not repeat the quite complete analyses in \cite{Giacchino:2015hvk} for $S-u$ coupling $y_1$, especially the complicated higher-order calculations of DM annihilations. Nevertheless we highlight the scale effects in calculating DM-nucleon scattering rate $\sigma^{\rm SI}_{p}$ due to wide mass gaps existing in $u_R,c_R,t_R$, including the renormalization group equation (RGE) and heavy quark threshold matchings. Our observations that $S-u$ coupling $y_1$ receives strong constraints from the current DM direct detection experiments agrees well with \cite{Giacchino:2015hvk}, while we regard our analysis on scale effects as a complementarity to \cite{Giacchino:2015hvk} in which destructive interferences in effective DM-nucleon coupling will be moderately shifted.

On the other hand, since we are considering a scalar DM, apart from the phenomenology-irrelevant DM self interaction $\lambda_S S^4$, the Higgs portal interaction $\lambda_{SH}S^2H^2$ would naturally exist \cite{Silveira:1985rk,McDonald:1993ex,Burgess:2000yq}. The Higgs portal has been extensively studied in the literature and strong constraints on it have been found from the measurement of invisible Higgs decay and DM direct detection experiments (see e.g. \cite{Cline2013,Beniwal2015-HP}). Both \cite{Baek2016a} and \cite{Giacchino:2015hvk} assumed the Higgs portal to be small and focused on the new interactions mediated by the vector-like $\psi$ portal. However, $\lambda_{SH}$ can receive high order corrections starting at 1-loop level from Eq.(\ref{eq-L}). In this work we upgrade our previous analysis by calculating the radiatively generated Higgs portal $\lambda_{SH}^{\rm 1PI}$ at $\mathcal{O}(y_3^2 \,y_t^2)$ with $y_t$ being the SM top Yukawa. This correction $\lambda_{SH}^{\rm 1PI}$ is negligible in \cite{Giacchino:2015hvk} for $S-u$ interaction due to $y_q\to 0$, while it turns out to be significant with $S-t$ interaction and puts strong constraints on $y_3$ if $\lambda_{SH}=0$ at tree-level is assumed. The general case with free parameter $\lambda_{SH}\neq 0$ at tree-level, however, will bring the total Higgs portal contributions back to the well-studied scenario. Therefore, except for using $\lambda_{SH}^{\rm 1PI}$ to illustrate the interesting interplay between Higgs portal and vector-like portal, we will still assume that the total Higgs portal strength up to 1-loop level is small compared to the vector-like portal which can be easily realized by turning the tree-level free parameter $\lambda_{SH}$.

After the general analysis and observing that both the $S-u$ coupling $y_1$ and the Higgs portal have been well discussed in the existing literature and strongly constrained, especially from the DM direct detection results, their existence would make it difficult to observe the effects of DM-heavy quark interactions $y_2,y_3$ through the new vector-like $\psi$ portal. In the latter part of this work we will focus on the DM-heavy quark $c,t$ interactions by imposing
\begin{eqnarray}
\lambda_{SH}^{\rm ren.}=0,
\end{eqnarray} 
which is the total renormalized Higgs portal up to 1-loop level and
\begin{eqnarray}
y_1=0.
\end{eqnarray} 
In this case the new Lagrangian we consider is reduced to
\begin{eqnarray}
\mathcal{L} \supset -y_2 S \overline{\psi_L} c_R  -y_3 S \overline{\psi_L} t_R +  h.c..
\label{eq-L-tc}
\end{eqnarray}


This paper is organized as follows. In Section \ref{section-Modeldescription} we describe our model in more detail. In Section \ref{section-DD-review} we briefly review the general framework of calculating the DM-nucleon scattering rate. In Section \ref{section-DD} we perform a careful discussion of the direct detection signals in this model including both scalar-type and twist-2 operator contributions, which are missed in some papers. 
Starting from Section \ref{section-DM-heavy-quark} we concentrate on DM interactions with heavy quarks $c,t$ with the alleviation of direct detection constraints. In Section \ref{section-omg} we explore the model parameter space to produce a correct DM thermal relic. In Section \ref{section-DD-heavy} we investigate the XENON1T constraints on the model's mass plane.
In Section \ref{section-ID} we consider the constraints from DM indirect detection signals from Fermi gamma-ray observations of dwarf galaxies. In Section \ref{section-FCNC} we discuss the predictions of top quark FCNC observables in this model. In Section \ref{section-collider} we study the collider signals of this model using the $36\, fb^{-1}$ data at LHC.
We combine the various results in Section \ref{section-combined} and present our conclusion in Section \ref{section-conclusion}.

\section{Model description}
\label{section-Modeldescription}

We extend the SM with a real scalar singlet DM $S$ which couples exclusively to the $SU(2)_L$ singlet up-type quarks $u_R,c_R,t_R$ via a vector-like fermion mediator $\psi$. The new interactions beyond the SM are
\begin{eqnarray}
\label{eq-L-full}
{\mathcal L_{\rm new}} &=&  {\mathcal L_{\rm fermion}} + {\mathcal L_{\rm scalar}} + {\mathcal L_{\rm Yukawa}},\\ \nonumber
{\mathcal L_{\rm fermion}} &=& \bar{\psi}(i \slashed D-m_\psi)\psi,\\ \nonumber
{\mathcal L_{\rm scalar}} &=& \frac{1}{2}\partial^\mu S \partial_\mu S - \frac{1}{2}m_S^2 S^2 - \frac{1}{4!} \lambda_S S^4 - \frac{1}{2}\lambda_{SH} S^2 H^2, \\ \nonumber
{\mathcal L_{\rm Yukawa}} &=& -y_1 S \overline{\psi_L} u_R -y_2 S \overline{\psi_L} c_R -y_3 S \overline{\psi_L} t_R +  h.c.,
\end{eqnarray}
where $\psi$ carries the same gauge quantum number as $u_R,c_R,t_R$ and $D_\mu$ is the $SU(3)\times SU(2)\times U(1)$ covariant derivative in the SM. We assume $m_S < m_\psi$ to meet the DM scenario and impose an odd $Z_2$ parity to $S,\psi$ to stabilize the DM. Because of this $Z_2$ parity and the chosen gauge quantum numbers for $S,\psi$, only $u_R,c_R,t_R$ sector in the SM are involved in the new Yukawa interaction indicated by ${\mathcal L_{\rm Yukawa}}$ in Eq.(\ref{eq-L-full}). This $Z_2$ parity also forbids the mass mixing between DM $S$ and SM Higgs $h$. Similarly, the mixings between $\psi$ and SM quarks $u,c,t$ are also forbidden which implies that the results of LHC searches for exotic heavy quarks $Q'$ do not apply, as the targeted decay channels $Q' \to Q+h/Z$ in those searches are based on the mass mixing (see e.g. \cite{CMSCollaboration2017-VLQuark-1,CMSCollaboration2017a-VLQuark-2,TheATLASCollaboration2017-VLQuark-1,ATLASCollaboration2017-VLQuark-2}).

Compared to our previous study \cite{Baek2016a} focusing on a phenomenologically motivated top-philic DM, Eq.(\ref{eq-L-full}) including all $u_R,c_R,t_R$ introduces several interesting changes:
\begin{itemize}
\item  DM-nucleon scattering in the direct detection is a low energy non-relativistic process occurring at $\mu_{\rm had}\sim 1\, {\rm GeV}$, while other DM phenomenology can happen at much higher energy scales such as the thermal freeze-out in the early Universe at $\mathcal{O}(T_f)$ with $T_f\simeq m_{\rm DM}/20$ and collider detections at $\mathcal{O}(1)$ TeV. Although one has the freedom of choosing at which scale to match the ultraviolet (UV)-complete Lagrangian Eq.(\ref{eq-L-full}) to an effective field theory (EFT) between DM and the SM, this EFT at $\mu_{\rm EFT}$ still needs to be matched to another one at $\mu_{\rm had}\sim 1\, {\rm GeV}$ with only a few active degree of freedom ({\it d.o.f}) including DM, light quarks and gluons. If there is a sizable gap between $\mu_{\rm EFT}$ and $\mu_{\rm had}$, e.g. $\mu_{\rm EFT}\sim m_Z$, the renormalization group equation (RGE) effects and heavy quark threshold matching with respect to (w.r.t) $c_R,t_R$ may be important.
\item Since gluon is the most abundant constituents inside nucleon, DM-gluon interactions generated through loops may also be important despite the loop factor suppression. In this case, loops connecting DM and gluon involving $u_R,c_R,t_R$ should be distinguished in the sense of short-distance and long-distance parts characterized by different energy scales \cite{Hisano2010-Gluon}.
\item More annihilation channels $\overline{U_i}U_j+\overline{U_j}U_i$ with $U_{i,j}=u,c,t$ and $i,j=1,2,3$ are available to produce the observed DM relic abundance $\Omega_{\rm DM}h^2\sim0.12$, which are helpful when DM mass is below top quark threshold and $SS\to\bar{t}t$ is kinematically difficult to open.
\item Flavor Changing Neutral Current (FCNC) processes of top quark can be generated at both tree level $t\to \psi^{(*)}S \to u/c+SS$ and loop level $t\to u/c+\gamma/g/Z$, which can provide additional tests of this model.
\end{itemize}

\section{Brief review of DM-nucleon effective theory}
\label{section-DD-review}
We start the discussion with the DM-nucleon scattering rate in direct detection experiment. We follow the effective theories of DM-nucleon interaction constructed in \cite{Hisano2015-DMEFT,Hisano2015-GluonWino} which carefully distinguishes long-distance (LD) and short-distance (SD) components in the loop-generated DM-gluon coupling \cite{Hisano2010-Gluon}.

To calculate the DM-nucleon scattering rate at $\mu_{\rm had}\sim 1\, {\rm GeV}$, the following steps are included:
\begin{itemize}
\item First, one integrates out the mediator particles and constructs the effective theory for the DM, quarks, and gluons at a certain scale $\mu_{\rm EFT}$. The effective interactions usually consist of higher-dimensional operators.
\item Second, the Wilson coefficients (WC) of the effective operators are evolved according to the RGE down to $\mu_{\rm had}\sim 1\, {\rm GeV}$ where the nucleon matrix elements of the operators are evaluated. In this process one usually needs to pass several heavy quark thresholds where the corresponding heavy quarks are also integrated out.
\item Finally, one expresses the DM-nucleon effective coupling in terms of the Wilson coefficients and the nucleon matrix elements. Then the scattering cross sections are obtained from this DM-nucleon effective coupling.
\end{itemize}
In the first step, the effective theory for the DM, quarks, and gluons can be expressed as
\begin{equation}
\label{eq-Leff-master}
\mathcal{L}_{\rm eff}
=\sum_{p=q,g}C^p_S\mathcal{O}^p_S
+\sum_{p=q,g}C^p_{T_2}\mathcal{O}^p_{T_2}~,
\end{equation} 
with
\begin{align}
\label{eq-twist-0}
\mathcal{O}^q_S&\equiv   S^2 m_q\bar{q}q~, \quad\quad\quad\quad\quad \,\,  \mathcal{O}^g_S\equiv 
 S^2  \frac{\alpha_s}{\pi}G^{A\mu\nu}G^A_{\mu\nu}~,\nonumber\\
\mathcal{O}^q_{T_2}&\equiv\frac{1}{m_S^2} S i \partial^\mu i
 \partial^\nu S \mathcal{O}^q_{\mu\nu}~, \quad \mathcal{O}^g_{T_2}\equiv\frac{1}{m_S^2} S i\partial^\mu i\partial^\nu S \mathcal{O}^g_{\mu\nu}~.
\end{align}
In the above, $\mathcal{O}^q_S,\mathcal{O}^g_S$ are scalar-type operators while ${\cal O}^q_{\mu\nu}$ and ${\cal O}^g_{\mu\nu}$ are the twist-2 operators of quarks and gluon defined by
\begin{align}
\label{eq-twist-2}
 {\cal O}^q_{\mu\nu}&\equiv \frac{1}{2}\overline{q}i\biggl(
D_\mu^{}\gamma_\nu^{} +D_\nu^{}\gamma_\mu^{}-\frac{1}{2}g_{\mu\nu}^{}
\slashed{D}\biggr)q~,\nonumber \\
{\cal O}^g_{\mu\nu}&\equiv 
G^{A\rho}_{\mu} G^{A}_{\nu\rho}-\frac{1}{4}g_{\mu\nu}^{}
G^A_{\rho\sigma}G^{A\rho\sigma}~.
\end{align}
Then one performs the evolutions of Wilson coefficients according to RGE. For scalar-type interactions we have
\begin{align}
C^q_S(\mu)&= C^q_S(\mu_0)-4C^g_S(\mu_0)
\frac{\alpha^2_s(\mu_0)}{\beta(\alpha_s(\mu_0))}
(\gamma_m(\mu)-\gamma_m(\mu_0)) ~,\label{eq:cqsrge} \\
C^g_S(\mu)&=\frac{\beta(\alpha_s(\mu))}{\alpha_s^2(\mu)}
\frac{\alpha_s^2(\mu_0)}
 {\beta (\alpha_s(\mu_0))}C^g_S(\mu_0)
~,
\end{align}
where the beta-function of $\alpha_s$ and the anomalous dimension of quark mass operator are given by
\begin{align}
 \beta(\alpha_s)=(2b_1)\frac{\alpha_s^{2}}{4\pi}+(2b_2)\frac{\alpha_s^3}{(4\pi)^2} ~, \quad\quad
  \gamma_m&=-6C_F \frac{\alpha_s^{}}{4\pi}  ~,
\end{align} 
with 
\begin{equation}
 b_1=-\frac{11}3N_c+\frac23 N_f~, \quad\quad
  b_2=-\frac{34}{3}N_c^2 +\frac{10}{3}N_cN_f+2C_F N_f~.
\label{b1b2}
\end{equation}
In the above expressions, $N_c=3$ is the number of $SU(3)$ colors, $N_f$ denotes the number of active quark
flavors in an effective theory and $C_F$ is the quadratic Casimir invariant  $C_F\equiv \frac{N_c^2-1}{2N_c}$.
    
For the twist-2 type interactions the RGEs are
\begin{equation}
 \mu \frac{d}{d\mu}(C^q_{T_i}, C^g_{T_i})=
(C^q_{T_i}, C^g_{T_i})~ \Gamma_{\rm T}~,
\end{equation}
with $i=1,2$ in Eq.(\ref{eq-twist-0}) and $\Gamma_{\rm T}$ being a $(N_f+1)\times (N_f+1)$ matrix:
\begin{equation}
 \Gamma_{\rm T}=
\begin{pmatrix}
 \gamma_{qq}&0&\cdots&0&\gamma_{qg}\\
 0 &\gamma_{qq}&&\vdots&\vdots\\
\vdots&&\ddots&0&\vdots\\
0&\cdots&0&\gamma_{qq}&\gamma_{qg}\\
\gamma_{gq}&\cdots&\cdots&\gamma_{gq}&\gamma_{gg}
\end{pmatrix}
~,
\end{equation}
where
\begin{align}
 \gamma_{qq}&=\frac{16}{3}C_F \, \frac{\alpha_s}{4\pi}, \quad  \gamma_{qg}=\frac{4}{3} \, \frac{\alpha_s}{4\pi}~,\nonumber \\
 \gamma_{gq}&=\frac{16}{3}C_F \, \frac{\alpha_s}{4\pi}, \quad  \gamma_{gg}=\frac{4}{3}N_f \, \frac{\alpha_s}{4\pi}~.
\end{align}
The solution of twist-2 operator RGEs can be found in \cite{Hill2014-I,Hill2014a-II}.

Before reaching $\mu_{\rm had}\sim 1\, {\rm GeV}$, we need to pass the heavy quark thresholds such as $\mu_b\sim m_b$ and $\mu_c\sim m_c$. Taking bottom quark as an example, the matching condition we took are
\begin{equation}
 \frac{1}{\alpha_s(\mu_b)|_{N_f=4}}
= \frac{1}{\alpha_s(\mu_b)|_{N_f=5}}
+\frac{1}{3\pi}\ln\biggl(\frac{\mu_b}{m_b}\biggr) ~,
\end{equation}
and
\begin{align}
 C^q_S(\mu_b)|_{N_f=4}&= C^q_S(\mu_b)|_{N_f=5}~,\nonumber \\
[\alpha_s C^g_S](\mu_b)|_{N_f=4}&=
-\frac{\alpha_s(\mu_b)}{12}
\left[
1+\frac{\alpha_s(\mu_b)}{4\pi}
\left(11+\frac{2}{3}\ln\frac{m_b^2}{\mu_b^2}\right)
\right]
C^b_S(\mu_b)|_{N_f=5}
\nonumber \\ &
+
\left[
1+\frac{\alpha_s(\mu_b)}{4\pi}
\frac{2}{3}
\ln\frac{m_b^2}{\mu_b^2}
\right]
[\alpha_s C^g_S](\mu_b)|_{N_f=5}~,\nonumber \\
C^q_{T_i}(\mu_b)|_{N_f=4}&= C^q_{T_i}
(\mu_b)|_{N_f=5}~,\nonumber \\
 C^g_{T_i}(\mu_b)|_{N_f=4}&= 
\left[
1+\frac{\alpha_s(\mu_b)}{4\pi}
\frac{2}{3}
\ln\frac{m_b^2}{\mu_b^2}
\right]
C^g_{T_i}(\mu_b)|_{N_f=5}
+\frac{\alpha_s(\mu_b)}{4\pi}\frac{2}{3}\ln \frac{m_b^2}{\mu_b^2}
C^b_{T_i}(\mu_b)|_{N_f=5}
~,
\label{eq:quarkthmatch}
\end{align}
where we have included the next-to-leading order $\alpha_s$ contribution to $C_S^g$ since the effect is known to be large \cite{Djouadi2000-GluonDM}, especially for charm quark.
In our calculation, we take $\mu_b=m_b$ and $\mu_c=m_c$ for simplification.

Finally, the effective interaction between scalar DM and a nucleon is defined by 
\begin{eqnarray}
\mathcal{L}^{(N)}_{{SI}}=f_N S^2 \overline{N}N~,
\end{eqnarray}
where
\begin{align}
\label{eq-fN}
 f_N/m_N&= \sum_{q=u,d,s}C^q_S(\mu_{\text{had}}) f^{(N)}_{T_q} 
-\frac{8}{9}C^g_S(\mu_{\text{had}}) f^{(N)}_{T_G}
\nonumber \\
&+\frac{3}{4}\sum_{q}^{N_f}
C^q_{T_2}(\mu)[q(2;\mu)+\overline{q}(2;\mu)]
-\frac{3}{4}C^g_{T_2}(\mu)g(2;\mu)~,
\end{align}
and 
\begin{align}
\label{eq-form-factor}
f_{T_q}^{(N)} &\equiv \langle N|m_q \bar{q}q|N\rangle/m_N ~, \nonumber \\
q(2;\mu) = \int^1_0 dx~ x\ q(x,\mu)~, ~~ \bar{q}(2;\mu) &= \int^1_0 dx~ x\ \bar{q}(x,\mu)~, ~~ g(2;\mu) = \int^1_0 dx~ x\ g(x,\mu)~,
\end{align}
with $q(x,\mu)$, $\bar{q}(x,\mu)$ and $g(x,\mu)$ being the PDFs of quarks, anti-quarks and gluon at factorization scale $\mu$, respectively.

Note that the light quark mass fractions $f_{T_q}^{(N)}$ belong to the non-perturbative QCD region and are obtained from the lattice QCD simulations. Accordingly, one needs to use the Wilson coefficients $C^q_S(\mu_{\text{had}})$ at $\mu_{\rm had}\sim 1\, {\rm GeV}$. On the contrary, twist-2 matrix elements $q(2;\mu), \bar{q}(2;\mu), g(2;\mu)$ can be calculated perturbatively using parton PDFs to other scales \cite{Hisano2015-DMEFT,Hisano2015-GluonWino,Hisano2010-Gluon,Hill2014-I,Hill2014a-II,Drees1993-DM}, thus one can choose a convenient scale applied to both the twist-2 Wilson coefficients and matrix elements, e.g. simply the scale $\mu_{\rm EFT}$ of the first-step EFT is defined. In this work we use the same the nucleon matrix elements as those given in \cite{Hisano2010-Gluon,Hill2014a-II}.

Since the DM particle is a scalar in our model, only spin-independent (SI) DM-nucleus interaction is generated. The cross section can be expressed by
\begin{equation}
 \sigma =\frac{1}{\pi}\biggl(\frac{m_T}{m_S+m_T}\biggr)^2
|n_pf_p+n_nf_n|^2~,
\end{equation}
where $m_T$ is the mass of the target nucleus and $n_p,n_n$ are the numbers of protons and neutrons in the nucleus, respectively.

\section{Direct detection in our model}
\label{section-DD}

In this section we choose $\mu_{\rm EFT}=m_Z$ and extract the Wilson coefficients in Eq.(\ref{eq-Leff-master}) from the UV-complete Lagrangian in Eq.(\ref{eq-L-full}) up to $\mathcal{O}(y_i^2)$. The active SM {\it d.o.f}s at this scale include $u,c,g$, while top quark and vector-like fermion $\psi$ are integrated out. Then we include RGE effects and heavy quark threshold matching to calculate DM-nucleon scattering rate at $\mu_{\rm had}\sim 1\, {\rm GeV}$.

\begin{figure}[h]
  \centering
  \includegraphics[width=13cm]{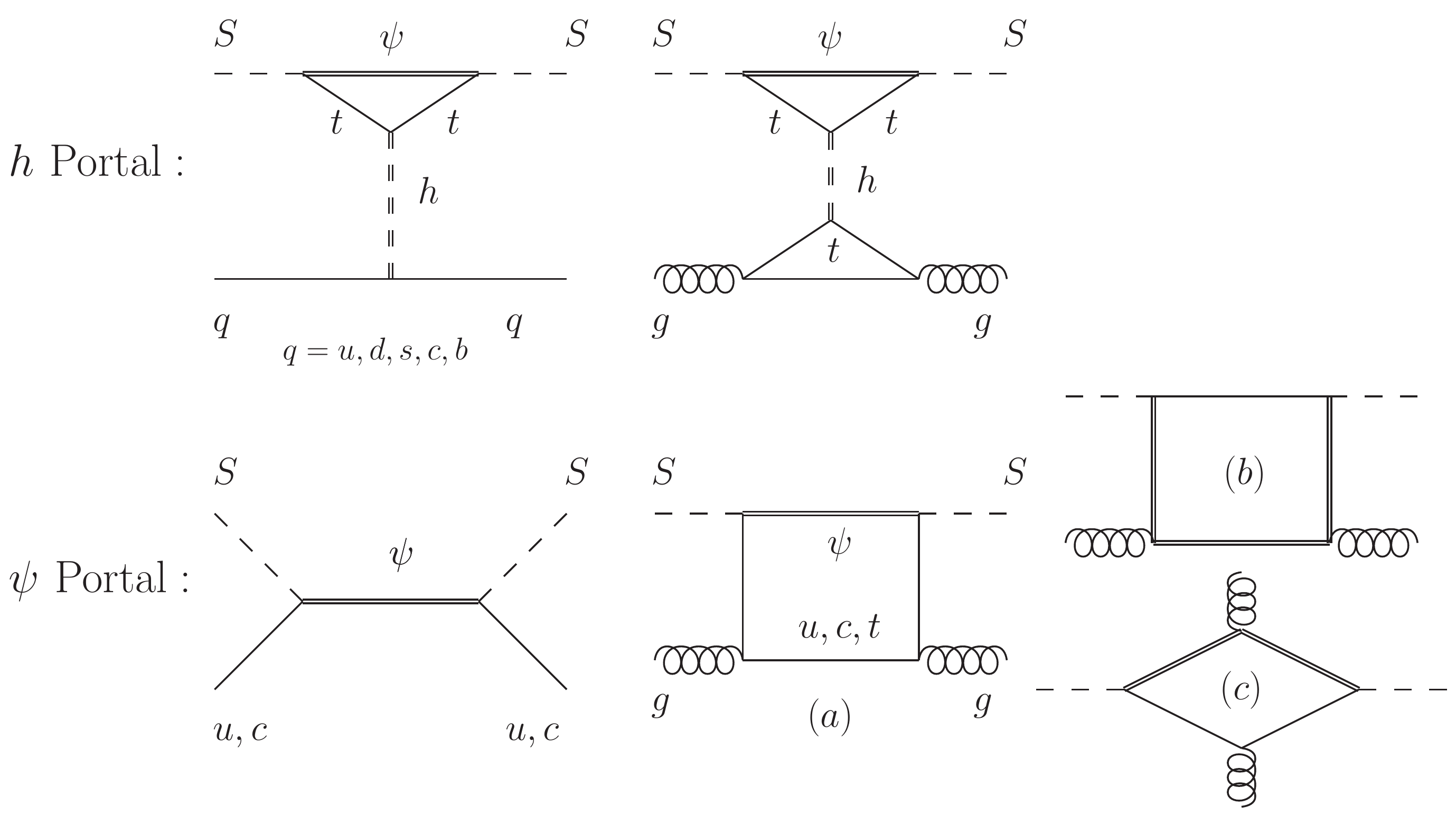} 
  \caption{Feynman diagrams used for calculating the Wilson coefficients, at the order of $\mathcal{O}(y_i^2)$, of the effective operators in Eq.(\ref{eq-Leff-master}) when choosing $\mu_{\rm EFT}=m_Z$. We refer to diagrams mediated by the SM Higgs $h$ as Higgs portal, while denoting others as vector-like $\psi$ portal.}
  \label{fig-DD-diagram}
\end{figure}

Apart from the conventional $t$-channel Higgs-mediated DM-nucleon scattering at tree-level for scalar DM, the new Feynman diagrams we consider in our model are presented in Fig.\ref{fig-DD-diagram}, all of which generate Wilson coefficients in Eq.(\ref{eq-Leff-master}) at the order of $\mathcal{O}(y_i^2)$ in terms of the new Yukawa couplings in Eq.(\ref{eq-L-full}).  We will still identify the $t$-channel Higgs-mediated diagrams to be the Higgs portal (HP), while calling the others the vector-like $\psi$ portal (VLP). Note that the HP diagrams only generate scalar-type interactions in Eq.(\ref{eq-Leff-master}) while VLP generates both scalar-type and twist-2 operators \cite{Hisano2015-GluonWino}. 

\subsection{Higgs portal at $\mathcal{O}(y_i^2)$}
\label{WC-HP}

Since we are considering a scalar DM, apart from the phenomenologically irrelevant DM self interaction $\lambda_S S^4$, the Higgs portal interaction $\lambda_{SH}S^2H^2$ would naturally exist \cite{Silveira:1985rk,McDonald:1993ex,Burgess:2000yq}. The Higgs portal has been extensively studied in the literature and strong constraints on it have been found from the measurements of invisible Higgs decay and DM direct detection experiments (see e.g. \cite{Cline2013,Beniwal2015-HP}). Even if one sets $\lambda_{SH}=0$ in the Lagrangian Eq.(\ref{eq-L-full}), it will receive loop corrections starting from 1-loop level in our model, as shown in the 1PI triangle vertex in Fig.\ref{fig-DD-diagram}. 
\begin{eqnarray}
\label{eq-lamSH-1PI}
\lambda_{SH}^{\rm1PI} &=& \sum_{k=1,2,3} (-8) y_3^2 \, N_c \, (\frac{m_{U_k}}{v})^2  \frac{1}{16\pi^2} \Big( 4 C_{00} + m_S^2(C_{11}+C_{22}) - 2 m_S^2 C_{12}  \Big), \\
C_{ij} &\equiv& C_{ij}(m_S^2, 0, m_S^2; m_\psi^2, m_{U_k}^2, m_{U_k}^2),
\end{eqnarray}
where $N_c=3, U_{k=1,2,3}=\{u,c,t\}$ and $v\simeq 246$ GeV is the SM vacuum expectation value (vev).
In the more general case with $\lambda_{SH}\neq 0$ in the Lagrangian Eq.(\ref{eq-L-full}), one would need a more complete renormalization for $\lambda_{SH}$ at 1-loop level
\begin{eqnarray}
\label{eq-lamSH-FullRen}
\lambda_{SH}^{\rm ren.} &=& \lambda_{SH} + \delta \lambda_{SH} + \lambda_{SH}^{\rm1PI}  + \lambda_{SH} (\frac{1}{2} \delta Z_h + \delta Z_S + \delta v),
\end{eqnarray}
where $\delta Z_h, \delta Z_S$ are field renormalization strength constants in on-shell renormalization scheme for Higgs and scalar DM, while $\delta v$ is the counterterm for the SM vev. The full calculations in usual Higgs portal for $\delta Z_h, \delta Z_S, \delta v$ can be found in, e.g. \cite{Kanemura2004-REN1,Kanemura2016-REN2}, while $\delta Z_S$ in our model receives the following additional contributions
\begin{eqnarray}
\label{eq-lamSH-ZS}
\delta Z_S^{\rm new} &=& \sum_{k=1,2,3} 2\, y_k^2\, N_c\, \frac{1}{16 \pi^2} \Big(  4 B'_{00} + (B_{11}+B_1) + m_S^2 (B'_{11}+B'_1)  \Big),\\
B^{(\prime)}_{i,ij} &\equiv& B^{(\prime)}_{i,ij} (m_S^2; m_\psi^2, m_{U_k}^2).
\end{eqnarray}

In the numerical calculations we impose modified minimal subtraction ($\overline{{\textrm{MS}}}$) scheme by choosing $\delta \lambda_{SH} = (...) \Delta$ to eliminate all $\Delta = 1/\epsilon -\gamma_E + \ln 4 \pi$ in $\lambda_{SH}^{\rm ren.}$. Note that the minimal subtraction scheme is also a common choice, e.g. as utilized in \cite{Kanemura2004-REN1,Kanemura2016-REN2}. We also set the artificial scale originating from dimensional regularization (DR) to be $\mu_{\rm DR}=m_\psi$ throughout this work. By denoting the finite parts in $\lambda_{SH}^{\rm1PI}, \delta Z_h, \delta Z_S$ using the notation $Fin.[...]$, one can rewrite $\lambda_{SH}^{\rm ren.}$ as
\begin{eqnarray}
\label{eq-lamSH-FullRen-Finite}
\lambda_{SH}^{\rm ren.} &=& \lambda_{SH} \Big(1+ Fin.[\frac{1}{2} \delta Z_h + \delta Z_S + \delta v] \Big) + Fin.[\lambda_{SH}^{\rm1PI}].
\end{eqnarray}
When combined with the Higgs-quark and Higgs-gluon interactions in Fig.\ref{fig-DD-diagram}, the Wilson coefficients in Eq.(\ref{eq-Leff-master}) from Higgs portal are:
\begin{eqnarray}
\label{eq-WC-HP}
C_{S,{\rm HP}}^{q} = \frac{\lambda_{SH}^{\rm ren.}}{m_h^2}, \quad \quad C_{S,{\rm HP}}^{g} = -\frac{\lambda_{SH}^{\rm ren.}}{12 m_h^2},
\end{eqnarray}
where $q=u,d,s,c,b$ and we use $m_h=125$ GeV. Note that there is no $\frac{\alpha_s}{\pi}$ factor in $C_{S,{\rm HP}}^{g}$ which has been absorbed in the definition of scalar-type gluon operator $\frac{\alpha_s}{\pi} G^{A\mu\nu}G^A_{\mu\nu}$ in Eq.(\ref{eq-twist-0}), as clarified in \cite{Hisano2015-GluonWino}.

In this work we will consider two special choices for $\lambda_{SH}$ in Eq.(\ref{eq-lamSH-FullRen-Finite}).
\begin{itemize}
\item Case-I:
\begin{eqnarray}
\label{eq-WC-lamSH-Case-I}
\lambda_{SH} &=& 0,\\
\lambda_{SH}^{\rm ren.} &=& Fin.[\lambda_{SH}^{\rm1PI}].
\end{eqnarray}
This choice can provide a direct comparison to the vector-like $\psi$ portal contribution to DM-nucleon scattering at the same order $\mathcal{O}(y_i^2)$,
\begin{eqnarray}
\label{eq-WC-HP-Case-I}
C_{S,{\rm HP}}^{q} = \frac{Fin.[\lambda_{SH}^{\rm 1PI}]}{m_h^2} \propto y_3^2, \quad \quad C_{S,{\rm HP}}^{g} = -\frac{Fin.[\lambda_{SH}^{\rm 1PI}]}{12 m_h^2}  \propto y_3^2,
\end{eqnarray}
where we ignored the $U_1=u, U_2=c$ contributions in Eq.(\ref{eq-lamSH-1PI}) for $\lambda_{SH}^{\rm 1PI}$ due to the small $m_u, m_c$.
\item Case-II:
\begin{eqnarray}
\label{eq-WC-lamSH-Case-II}
\lambda_{SH} &=& - \frac{Fin.[\lambda_{SH}^{\rm1PI}]}{1+ Fin.[\frac{1}{2} \delta Z_h + \delta Z_S + \delta v]}, \\
\lambda_{SH}^{\rm ren.} &=& 0.
\end{eqnarray}
In this case, the Higgs portal at 1-loop level vanishes in the effective theory defined at $\mu_{\rm EFT}= m_Z$
\begin{eqnarray}
\label{eq-WC-HP-Case-II}
C_{S,{\rm HP}}^{q} =  C_{S,{\rm HP}}^{g} = 0.
\end{eqnarray}
\end{itemize}



\subsection{Vector-like $\psi$ portal at $\mathcal{O}(y_i^2)$}
\label{WC-VLP}

In Fig.\ref{fig-DD-diagram}, the DM-quark Wilson coefficients in Eq.(\ref{eq-Leff-master}) from vector-like $\psi$ portal at $\mathcal{O}(y_i^2)$ have been given in \cite{Hisano2015-DMEFT} as follows:
\begin{eqnarray}
\label{eq-WC-VLP-tree-matching}
C_{S,{\rm VLP}}^{u,c} &=& \frac{y_{1,2}^2}{4} \frac{2 m_\psi^2 - m_S^2}{\Big(m_\psi^2 - (m_S+m_{u,c})^2 \Big)^2},\\
C_{T_2,{\rm VLP}}^{u,c} &=& y_{1,2}^2 \frac{m_S^2}{\Big(m_\psi^2 - (m_S+m_{u,c})^2 \Big)^2},
\end{eqnarray}
where we have explicitly kept the small quark masses $m_{u,c}$ in the denominators for the convenience of the following discussion.

To calculate DM-gluon Wilson coefficients $C_{S,{\rm VLP}}^{g}$ in Fig.\ref{fig-DD-diagram}, one needs to distinguish diagram (a) from (b,c), in which (a) belongs to long distance (LD) effect characterized by $q^{\rm loop}\sim m_{u,c,t}$ while diagram (b,c) belong to short distance (SD) effects characterized by $q^{\rm loop}\sim m_\psi$ \cite{Hisano2010-Gluon}. Only those effects coming from energy scales above $\mu_{\rm EFT}= m_Z$ should be integrated into $C_S^g$, which include all SD effects as well as top quark LD\footnote{Although the case $m_\psi<m_Z$ will slightly change the argument, a colored fermion particle lighter than $m_Z$ will be strongly constrained given the current LHC data. We do not consider such light colored mediator in this work.}.
\begin{eqnarray}
\label{eq-CSg-mZ}
C_{S,{\rm VLP}}^{g} &=& C_{S,{\rm VLP}}^{g}|_t^{\rm LD+SD} + C_{S,{\rm VLP}}^{g}|_{u,c}^{\rm SD}~,\\ \nonumber
C_{S,{\rm VLP}}^{g}|_t^{\rm LD+SD} &=& \frac{1}{4}\frac{y_3^2}{2} \Big(f^{(a)}_+ + f^{(b)}_+ + f^{(c)}_+\Big)(m_S;m_t,m_{\psi})~,\\ \nonumber
C_{S,{\rm VLP}}^{g}|_{u,c}^{\rm SD} &=&  \frac{1}{4}\frac{y_{1,2}^2}{2} \Big(f^{(b)}_+ + f^{(c)}_+\Big)(m_S;m_{u,c},m_{\psi})~,
\end{eqnarray}
where the expressions of $f^{(a,b,c)}_{+}(m_S;m_{u,c,t},m_\psi)$ can be found in \cite{Hisano2015-DMEFT}.

Furthermore, one needs to distinguish diagram (a) containing $u$ and $c$ quarks characterized by $q^{\rm loop}\sim m_{u,c}$. Tiny up quark mass $m_u<\Lambda_{\rm QCD}$ is in the non-perturbative QCD energy region, therefore diagram (a) for up quark should be included in the matrix element $\langle N |m_u \bar{u}u| N\rangle$ obtained from lattice QCD simulations, but not in the perturbative evaluation of $C_{S,{\rm VLP}}^{g}$.
On the contrary, $m_c>\Lambda_{\rm QCD}$ can be included in perturbative calculations and the LD diagram (a) for charm quark comes into place when one performs charm quark threshold matching near $m_c$, i.e. by integrating out charm quark into DM-gluon operator. More interestingly, an explicit calculation of $C_{S,{\rm VLP}}^{g}|_c^{\rm LD}$ from LD diagram (a) for charm quark at $\mu_{\rm EFT}= m_Z$ can be related to DM-quark Wilson coefficient $C_{S,{\rm VLP}}^{c}$ using the following relation \cite{Shifman1978-QQGG}
\begin{eqnarray}
m_Q \overline{Q}Q=-\frac{\alpha_s}{12\pi} G^{A\mu\nu}G^A_{\mu\nu},
\end{eqnarray}
where $Q$ denotes heavy quarks $c,b,t$. This implies
\begin{eqnarray}
\label{eq-WC-VLP-loop-matching}
C_{S,{\rm VLP}}^{c} =\Big(-\frac{1}{12} \Big)^{-1} \,C_{S,{\rm VLP}}^{g}|_c^{\rm LD},
\end{eqnarray}
which can be regarded as loop-diagram matching for $C_{S,{\rm VLP}}^{c}$. It has been shown that \cite{Gondolo2013-pole}, contrary to the pole at $m_\psi=m_S+m_c$ in Eq.(\ref{eq-WC-VLP-tree-matching}) obtained from tree-diagram matching using left-bottom diagram in Fig.\ref{fig-DD-diagram}, Eq.(\ref{eq-WC-VLP-loop-matching}) is regular at $m_\psi=m_S+m_c$. However, our numerical calculations do not cover this small mass parameter region $\sim m_c$. Therefore both matching results are applicable.

As for the Wilson coefficient $C_{T_2,{\rm VLP}}^{g}$ for the twist-2 gluon operator, it turns out to be of order $\mathcal{O}(y_i^2 \alpha_s^1)$ and thus suppressed by $\frac{\alpha_s}{\pi}$. As clarified in \cite{Hisano2010-Gluon,Hisano2015-DMEFT,Hisano2015-GluonWino} and also exhibited in \cite{Hill2014-I,Hill2014a-II}, for a chosen loop diagram generating DM-gluon interaction (which is up to 1-loop considered in this work), the $\frac{\alpha_s}{\pi}$ difference in the definitions of scalar-type gluon operator $\frac{\alpha_s}{\pi} G^{A\mu\nu}G^A_{\mu\nu}$ in Eq.(\ref{eq-twist-0}) and twist-2 gluon operator $G^{A\rho}_{\mu} G^{A}_{\nu\rho}-\frac{1}{4}g_{\mu\nu}G^A_{\rho\sigma}G^{A\rho\sigma}$ in Eq.(\ref{eq-twist-2}) results in an $\frac{\alpha_s}{\pi}$ suppression in $C_{T_2,{\rm VLP}}^{g}$ compared to $C_{S,{\rm VLP}}^{g}$. Considering that the matrix element of scalar and twist-2 gluon operators inside nucleon are at the same order $\mathcal{O}(1)$ \cite{Hisano2010-Gluon,Hill2014a-II}, the twist-2 gluon operator contribution is negligible and we ignore it in Eq.(\ref{eq-fN}) for the DM-nucleon effective coupling $f_N$.

\subsection{Differences between $\mu_{\rm EFT}=m_Z$ and $\mu_{\rm EFT}=\mu_{\rm had}$}
\label{section-difference-EFT-scale}

Before presenting the numerical results, we point out the main differences if one directly chooses $\mu_{\rm EFT}=\mu_{\rm had}\sim 1$ GeV instead of $\mu_{\rm EFT}=m_Z$ for the DM direct detection calculation.
\begin{itemize}
\item When choosing $\mu_{\rm EFT}=\mu_{\rm had}\sim 1$ GeV, calculation of DM-nucleon scattering rate is simplified and no RGE running and heavy quark threshold effects is applied to the Wilson coefficients.
\item Eq.(\ref{eq-fN}) for the DM-nucleon effective coupling $f_N$ only includes $C_{S,T_2}^{u,g}$ at $\mu_{\rm had}\sim 1$ GeV for this model, without the additional $d,s$ quark operators induced by RGE running and mixing in the $\mu_{\rm EFT}=m_Z$ case.
\item $u$-quark twist-2 operator matrix element $u(2;\mu)+\overline{u}(2;\mu)$ should take the value at $\mu=\mu_{\rm had}\sim 1$ GeV if one uses the Wilson coefficient $C_{T_2}^u(\mu_{\rm had})$ matched at $\mu_{\rm EFT}=\mu_{\rm had}$.
\item Both LD and SD effects of $c,t$ quark loops in Fig.\ref{fig-DD-diagram} are integrated into $C_{S,{\rm VLP}}^{g}$ as follows compared to Eq.(\ref{eq-CSg-mZ})
\begin{eqnarray}
\label{eq-CSg-1GeV}
C_{S,{\rm VLP}}^{g} &=& C_{S,{\rm VLP}}^{g}|_{c,t}^{\rm LD+SD} + C_{S,{\rm VLP}}^{g}|_{u}^{\rm SD}~,\\ \nonumber
C_{S,{\rm VLP}}^{g}|_{c,t}^{\rm LD+SD} &=& \frac{1}{4}\frac{y_{2,3}^2}{2} \Big(f^{(a)}_+ + f^{(b)}_+ + f^{(c)}_+\Big)(m_S;m_{c,t},m_{\psi})~,\\ \nonumber
C_{S,{\rm VLP}}^{g}|_{u}^{\rm SD} &=&  \frac{1}{4}\frac{y_{1}^2}{2} \Big(f^{(b)}_+ + f^{(c)}_+\Big)(m_S;m_{u},m_{\psi})~.
\end{eqnarray}
\end{itemize}

\subsection{Numerical results}
\label{section-DD-numerical}

\begin{figure}[h]
  \centering
  \includegraphics[width=8.5cm]{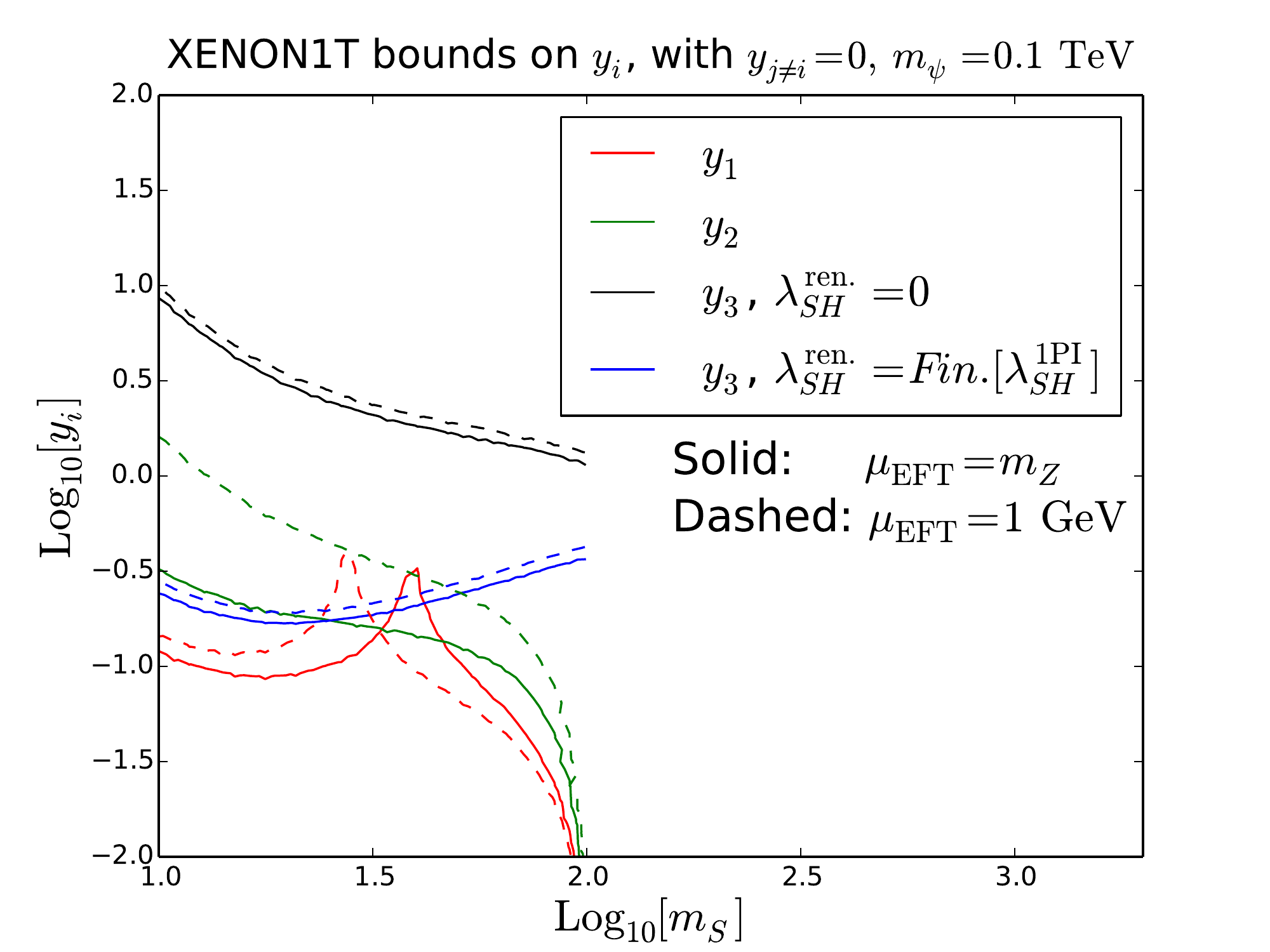} 
  \includegraphics[width=8.5cm]{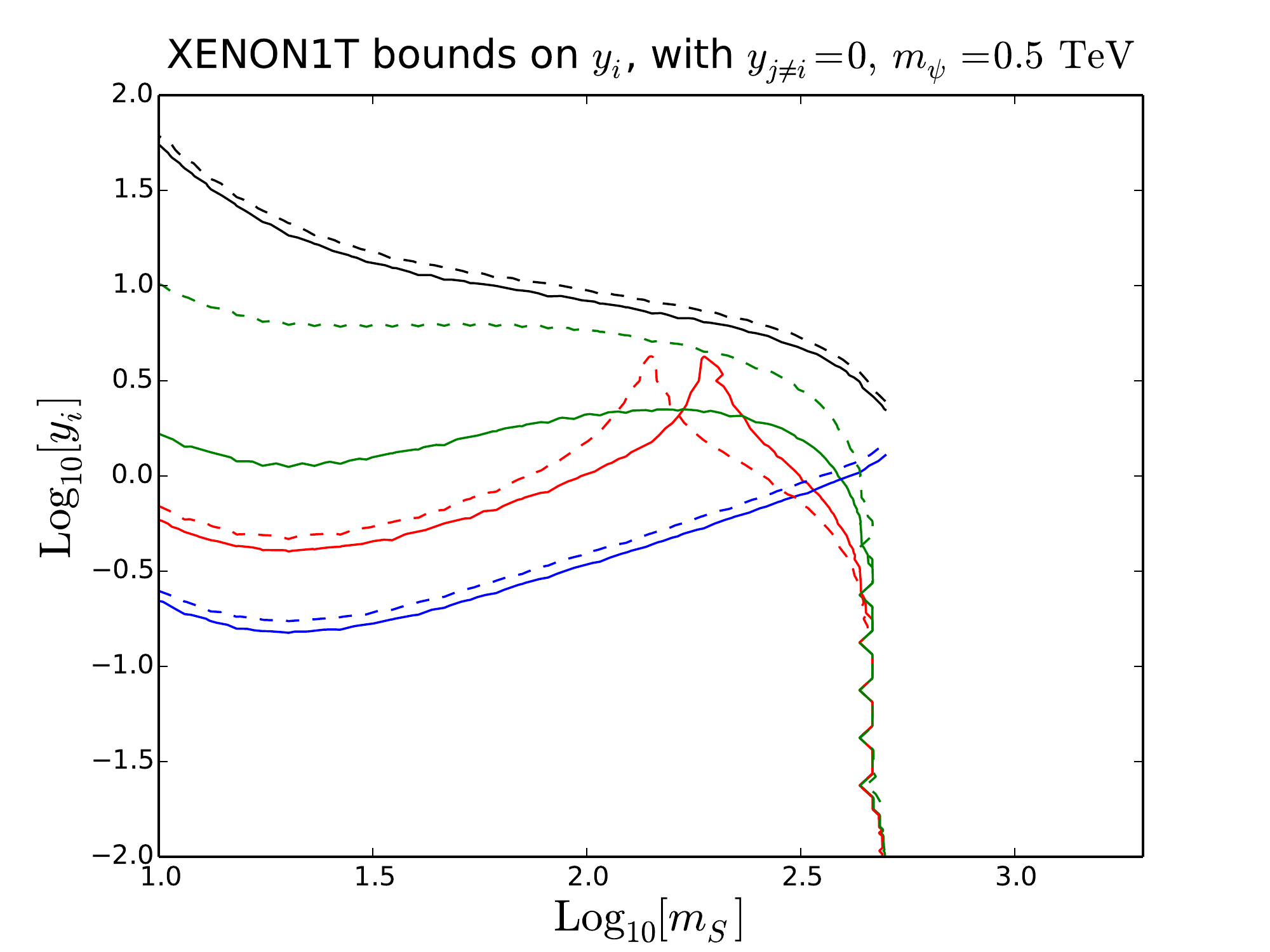} 
  \includegraphics[width=8.5cm]{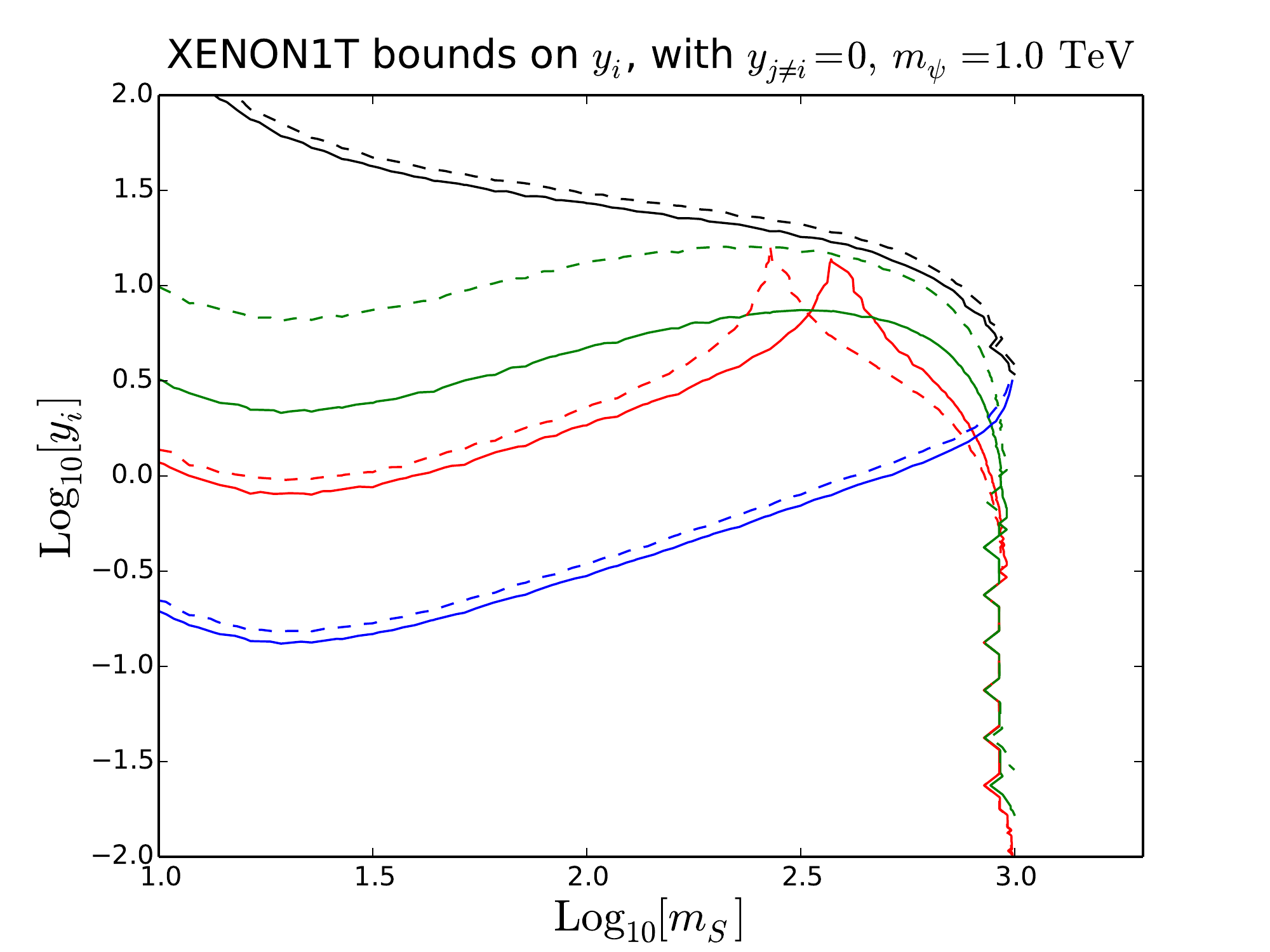}
  \caption{The constraints on a single $y_1,y_2,y_3$ when the other two are set to be zero on the plane of $m_S$ versus $y_i$, from the latest XENON1T results on spin independent DM-nucleon scattering rate.}
  \label{fig_DDySQ-Psi}
\end{figure}

After combining the Higgs portal in Section \ref{WC-HP} and vector-like $\psi$ portal in Section \ref{WC-VLP}, in Fig.\ref{fig_DDySQ-Psi} we present the constraints on a single $y_1,y_2,y_3$ when the other two are set to be zero on the plane of $m_S$ versus $y_i, (i=1,2,3)$, from the latest XENON1T results on spin independent DM-nucleon scattering rate. We choose three benchmark mediator masses $m_\psi = 100, 500, 1000$ GeV indicated by red, green and blue color, and use solid (dashed) lines to denote the choice of $\mu_{\rm EFT}=m_Z$ ( $\mu_{\rm had}$) with (without) RGE and heavy quark threshold effects. Since Eq.(\ref{eq-WC-HP-Case-I}) indicates that only top quark contributes sizably to $Fin.[\lambda_{SH}^{\rm1PI}]$ at $\mathcal{O}(y_3^2)$, we use blue and black lines to distinguish bounds on $y_3$ for the two benchmark choices of $\lambda_{SH}$ in Eq.(\ref{eq-WC-lamSH-Case-I}) and Eq.(\ref{eq-WC-lamSH-Case-II}).

We have the following interesting observations:
\begin{itemize}
\item RGE and heavy quark threshold effects are important if one chooses $\mu_{\rm EFT}=m_Z$ while calculating the DM-nucleon scattering rate at $\mu_{\rm had}\sim 1$ GeV.
We regard the bounds on $y_1$ embedding the scale effects in Fig.\ref{fig-DD-diagram} as a complementarity to \cite{Giacchino:2015hvk} which performed a comprehensive study for scalar DM interactions with light quarks through a vector-like mediator, including a similar part $-y_1 S \overline{\psi_L} u_R$ in this work.
\item When focusing only on the vector-like $\psi$ portal (red, green, black), DM-up quark coupling $y_1$ generally receives the strongest constraints, followed by the weaker bounds on $y_2$ and $y_3$, except for a peak structure coming from the destructive cancellations between DM-gluon and DM-up quark interactions (observed also in \cite{Giacchino:2015hvk,Hill2014a-II}). Choices of $\mu_{\rm EFT}=m_Z$ and $\mu_{\rm EFT}=\mu_{\rm had}$ result in different peak locations.
\item Constraints on $y_3$ with $\lambda_{SH}^{\rm ren.}=Fin.[\lambda_{SH}^{\rm1PI}]$ (blue) is about $\mathcal{O}(10^1)$ stronger than the case of $\lambda_{SH}^{\rm ren.}=0$ (black), and can be even stronger than the bounds on $y_1$ with $m_\psi \gtrsim 200$ GeV. This implies that $y_3$ can contribute significantly through the Higgs portal $Fin.[\lambda_{SH}^{\rm1PI}]/m_h^2$ in Eq.(\ref{eq-lamSH-1PI}), but very little through the vector-like $\psi$ portal $C_{S,{\rm VLP}}^{g}|_{t}$ in Eq.(\ref{eq-CSg-mZ}). Note that strong constraints from DM direct detection experiments on the Higgs portal have already been well studied in, e.g. \cite{Cline2013,Beniwal2015-HP} and references therein.
\end{itemize}

\section{DM-heavy quark $c,t$ interactions}
\label{section-DM-heavy-quark}

Since both the $S-u$ coupling $y_1$ and the Higgs portal have been well discussed in the existing literature and strongly constrained, especially from the DM direct detection results, their existence would make it difficult to observe the effects of DM-heavy quark interactions $y_2,y_3$ through the vector-like $\psi$ portal. In the latter part of this work we will focus on the DM-heavy quark $c,t$ interactions by imposing
\begin{eqnarray}
\lambda_{SH}^{\rm ren.}=0,
\end{eqnarray} 
and
\begin{eqnarray}
y_1=0.
\end{eqnarray} 
In this case the new Lagrangian we consider is reduced to
\begin{eqnarray}
\mathcal{L} \supset -y_2 S \overline{\psi_L} c_R  -y_3 S \overline{\psi_L} t_R +  h.c..
\label{eq-L-tc}
\end{eqnarray}

\subsection{Relic Abundance}
\label{section-omg}
\begin{figure}[h]
  \centering
  \includegraphics[height=3cm,width=12cm]{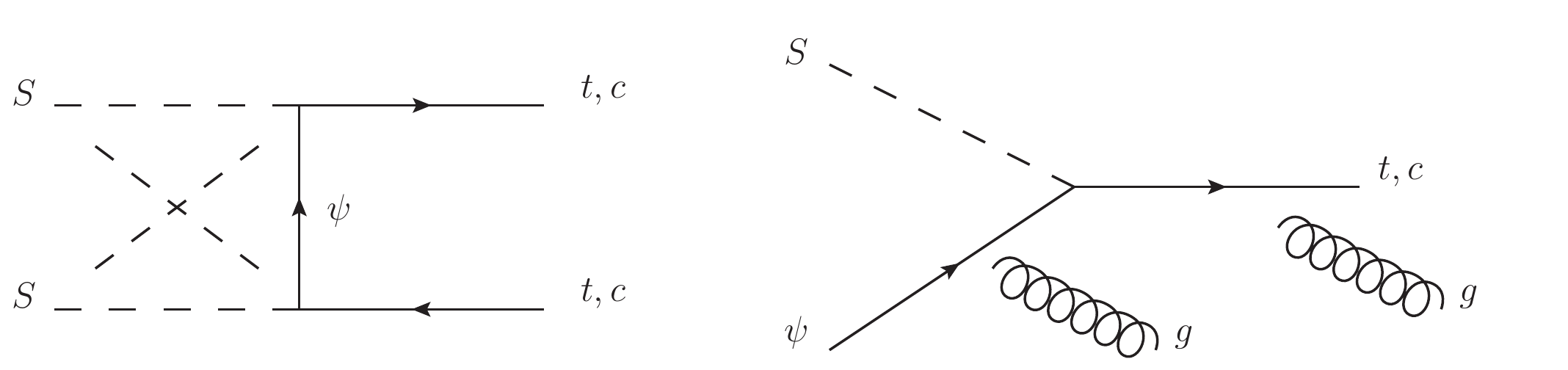} 
  \caption{Most relevant DM annihilation channels in this work.}
  \label{fig-omgPlot}
\end{figure}

The most relevant DM annihilation processes for DM-$\{c,t\}$ interactions are given in Fig.\ref{fig-omgPlot}. The thermally averaged DM annihilation cross sections can be decomposed into partial waves as $\langle \sigma v\rangle\simeq s + p\,v^2$. We implement our model with {\bf FeynRules}
\cite{Alloul2014} and use {\bf micrOMEGAs} \cite{Belanger2015,Belanger2014} to calculate the DM thermal relic density which systematically takes into account co-annihilations, threshold and resonant effects \cite{Griest1991-omg}. Before presenting numerical results, here we give the analytic results obtained using {\bf FeynCalc} \cite{Mertig1991-FeynCalc,Shtabovenko2016-FeynCalc}. We found good agreement with the numerical calculations.

\begin{eqnarray}
\sigma v(SS\to t\bar{t})_s &=& \Big(y_3^4\Big) \frac{3}{4 \pi   m_t^2 }  \frac{(r_S^2-1)^{3/2}}{r_S^3 (r_S^2 (r_\psi^2+1)-1)^2	}~,\\ \nonumber
\sigma v(SS\to t\bar{t})_p &=& \Big(y_3^4\Big)  \frac{1}{32 \pi   m_t^2 }  \frac{(r_S^2-1)^{1/2}}{r_S^3 (r_S^2 (r_\psi^2+1)-1)^4}\\ \nonumber
&\quad&\times (-16 r_S^6 (2 r_\psi^2+1)+r_S^4 (r_\psi^2+1) (9 r_\psi^2+41)-2 r_S^2 (9 r_\psi^2+17)+9)~,
\end{eqnarray}
\begin{eqnarray}
\sigma v(SS\to t\bar{c}+c\bar{t})_s &=& \Big(y_2^2 y_3^2\Big) \frac{3}{16 \pi   m_t^2}  \frac{(1-4 r_S^2)^2 }{ r_S^4 (1-2 r_S^2 (r_\psi^2+1))^2}~,\\ \nonumber
\sigma v(SS\to t\bar{c}+c\bar{t})_p &=& \Big(y_2^2 y_3^2\Big)  \frac{1}{32 \pi   m_t^2 }  \frac{1-4 r_S^2}{r_S^4 (1-2 r_S^2 (r_\psi^2+1))^4}\\ \nonumber
&\quad&\times (64 r_S^6 (2 r_\psi^2+1)-4 r_S^4 (3 r_\psi^4+16 r_\psi^2+17)+2 r_S^2 (7   r_\psi^2+11)-3)~,
\end{eqnarray}
and
\begin{eqnarray}
\sigma v(S\psi\to gt)_s &=& \Big(g_s^2 y_3^2\Big) \frac{1}{24 \pi   m_t^2 } \frac{ r_S^2 (r_\psi+1)^2-1}{r_S^4 r_\psi^2 (r_\psi+1)^3}~,\\ \nonumber
\sigma v(S\psi\to gt)_p &=& \Big(g_s^2 y_3^2\Big) \frac{1}{144 \pi  m_t^2 }\frac{1}{r_S^4 r_\psi^2 (r_\psi+1)^5(r_S^2 (r_\psi+1)^2-1)}\\ \nonumber
&\quad&\times (r_S^4 \Big(13 r_\psi^2-7 r_\psi-4) (r_\psi+1)^4-2 r_S^2 (r_\psi^2-6 r_\psi-4) (r_\psi+1)^2~, \\ \nonumber
&\quad& \quad + 5 r_\psi^2-5 r_\psi-4\Big)
\end{eqnarray}
where $g_s$ is the SM $SU(3)$ gauge coupling and the mass ratios are defined by
\begin{eqnarray}
r_S = \frac{m_S}{m_{t}}, \quad r_\psi = \frac{m_\psi}{m_S}~.
\end{eqnarray}
Results for $SS\to c\bar{c}$ and $S\psi \to gc$ can be obtained by replacing $t\to c$ and $y_3 \to y_2$ in those of $SS\to t\bar{t}$ and $S\psi\to gt$.
Apart from the tree level annihilations, we also have $SS\to gg$ induced by the loop coupling $C_S^g$ and
\begin{eqnarray}
\sigma v(SS\to gg)_s &=& \frac{64}{\pi} |\frac{\alpha_s}{\pi}C_S^g|^2 m_S^2~,\\ \nonumber
\sigma v(SS\to gg)_p &=& \frac{16}{\pi} |\frac{\alpha_s}{\pi}C_S^g|^2 m_S^2~.
\end{eqnarray}

\begin{figure}[h]
  \centering
  \includegraphics[height=6.5cm,width=7.5cm]{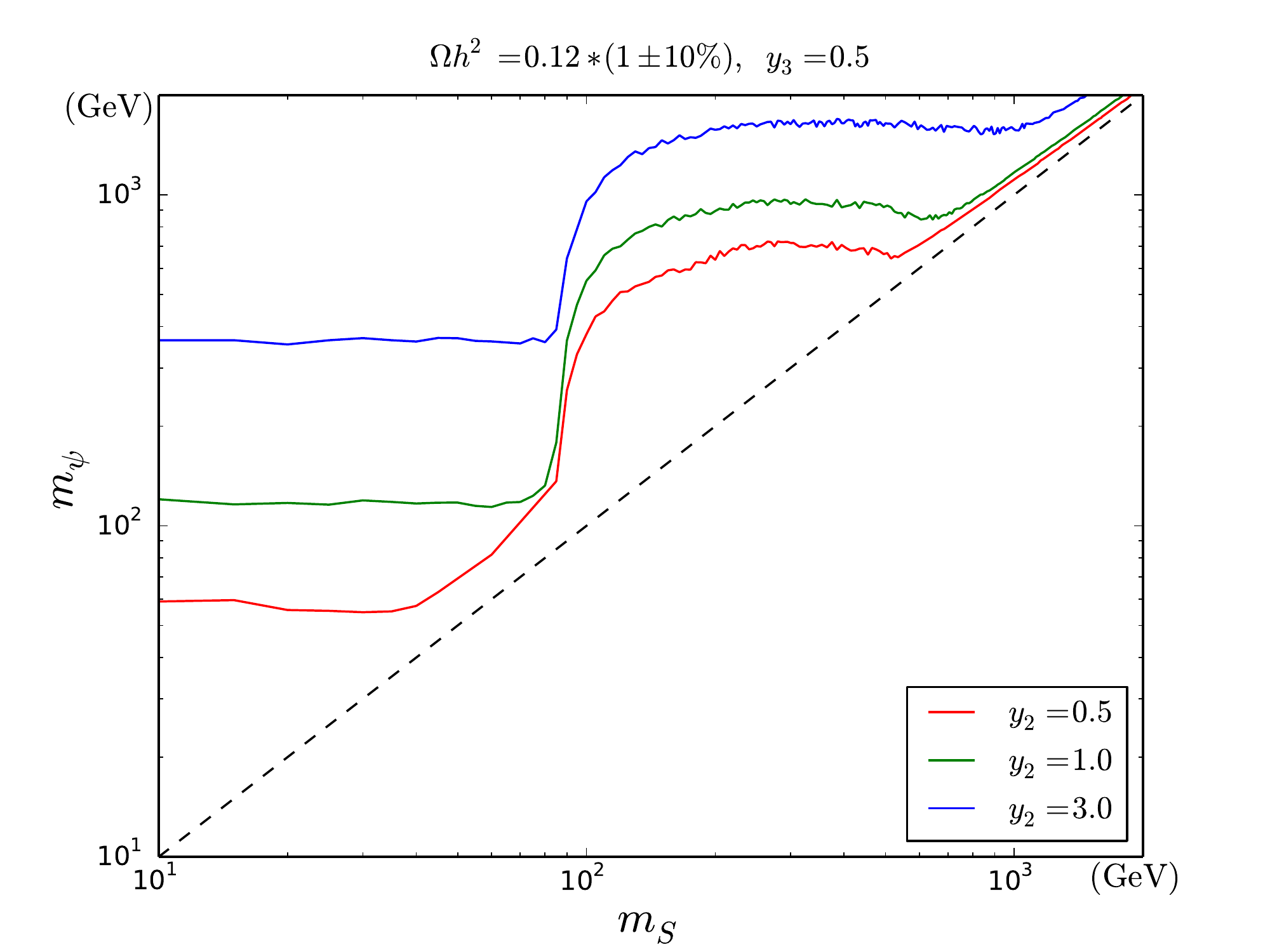}
  \includegraphics[height=6.5cm,width=7.5cm]{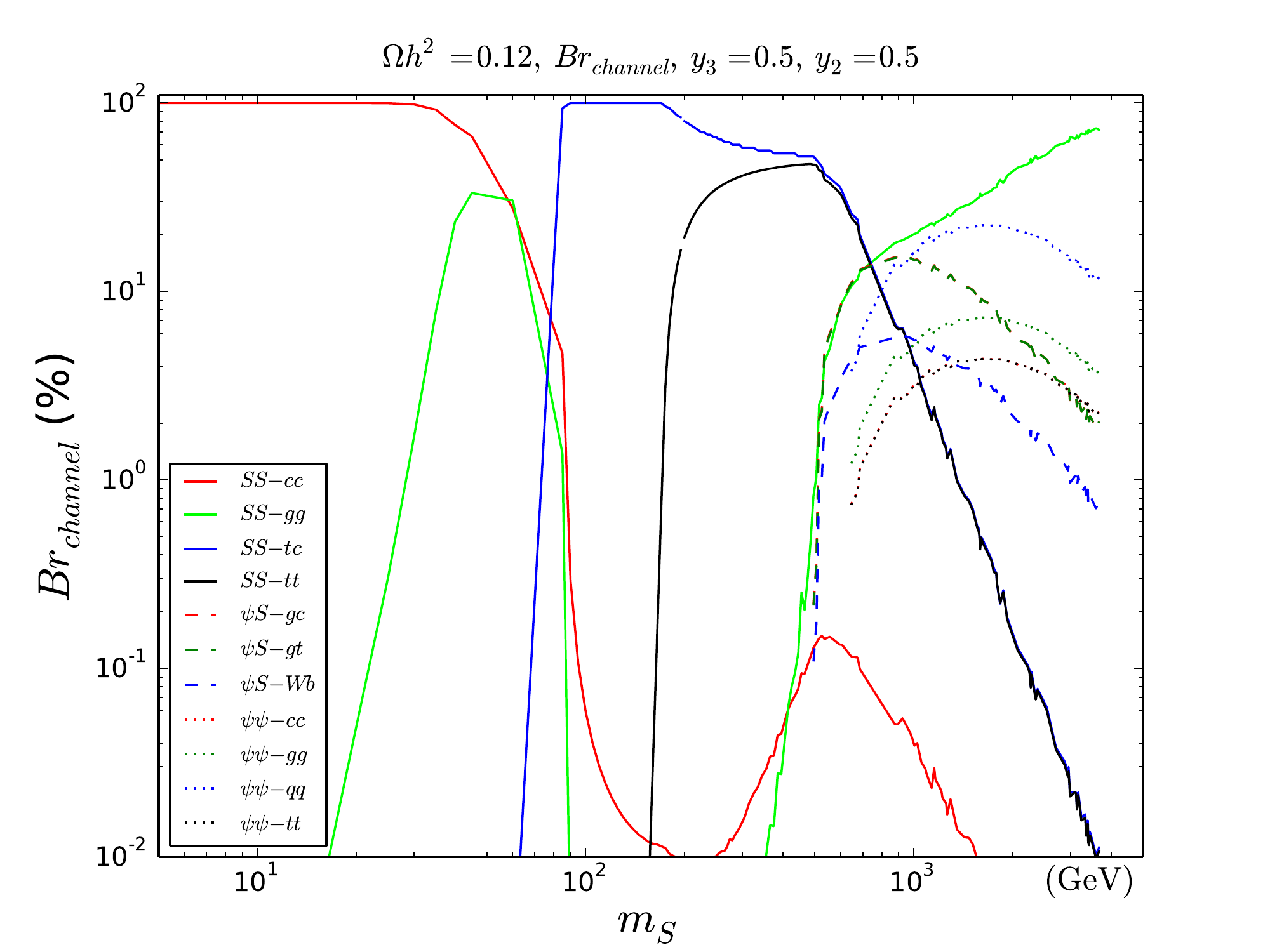} 
  \caption{{\bf Left}: Mass spectrum producing correct relic abundance ($10\%$ uncertainty included); {\bf Right}: Contributions from different annihilation channels for $y_2=y_3=0.5$.}
  \label{fig-Omg}
\end{figure}

In the left panel of Fig.\ref{fig-Omg} we show the proper masses $m_S,m_\psi$ to produce $\Omega_{\rm DM} h^2\simeq 0.12$ \cite{PlanckCollaboration2015} (with an uncertainty of $10\%$ in numerical calculation) for $y_3=0.5$ which can provide a moderate contribution from top quark sector, and $y_2=0.5, 1, 3$ to cover the perturbative but sizable range of DM-charm interaction considering the chiral suppression from $m_c$. We also show the contributions from different annihilation channels for $y_2=y_3=0.5$ in the right panel. Since co-annihilation is only relevant when $(m_\psi/m_S-1)\lesssim 1/x_f$ with $x_f=m_S/T_f\sim 25$ \cite{Griest1991-omg}, the channels with dominant contribution are $SS\to  f\bar{f'}$ with $f,f'=t,c$, except when $m_S \gtrsim 1$ TeV in which case the co-annihilations $\overline{T}T\to {\rm SM}$ and $SS\to gg$ become important. Also note that for $m_S \gtrsim 2m_t$ when both the top and charm quark appear to be massless with respect to $m_S$, the contribution from $SS\to t\bar{c}+c\bar{t}$ approaches that of $SS\to t\bar{t}$ because of $y_2=y_3=0.5$ we choose for the illustration. Also, we remind ourselves that $C_S^g$ obtained in \cite{Hisano2010-Gluon,Hisano2015-DMEFT} assumes small gluon momenta expansion and thus may not sufficiently describe the annihilation process where  full loop calculations may be more suitable. However, the right panel of Fig.\ref{fig-Omg} indicates that tree level annihilations into $c,t$ quarks with $y_2,y_3\sim \mathcal{O}(1)$ can already produce the observed DM relic abundance. Therefore we leave the more complete exploration for the loop-induced DM annihilations in future works (see \cite{Colucci:2018vxz-1804} for a recent discussion and references therein).

\subsection{Direct detection}
\label{section-DD-heavy}

\begin{figure}[h]
  \centering
  \includegraphics[width=10cm]{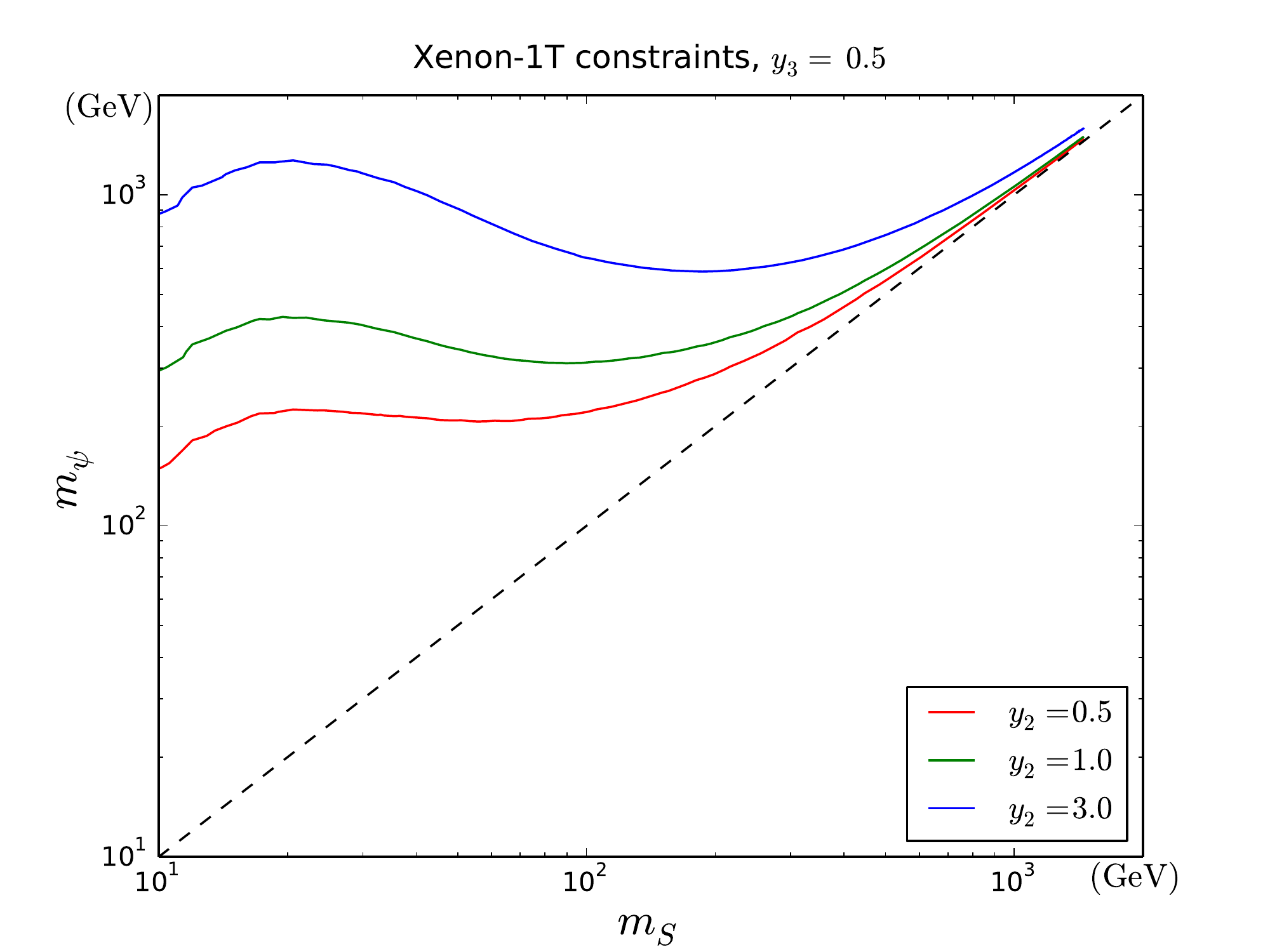} 
  \caption{Constraints from XENON1T results on the mass plane $m_S,m_\psi$ for $y_3=0.5$ and $y_2=0.5, 1, 3$ with $\mu_{\rm EFT}=m_Z$ to be consistent with the phenomenology scales discussed in this section. Regions below solid lines are excluded.}
  \label{fig-DD}
\end{figure}

From the discussions in Section \ref{section-DD}, it is straightforward to obtain the constraints from XENON1T on the mass plane $m_S,m_\psi$ for the benchmark couplings $y_3=0.5$ and $y_2=0.5, 1, 3$ chosen for the thermal relic discussion. We show the contour bounds in Fig.\ref{fig-DD} for $\mu_{\rm EFT}=m_Z$ to be consistent with the phenomenology scales discussed in this section.

\subsection{Indirect detection}
\label{section-ID}
Analogous to the method introduced in our previous work \cite{Baek2016a} (also in \cite{Bringmann:2012vr,Giacchino:2015hvk}), we impose the bounds set by Fermi observation of dwarf galaxy on DM annihilation into certain SM final states to further constrain our model \cite{Ackermann:2015zua}. In today's Universe with DM velocity $v\sim 10^{-3}c$, the relevant DM annihilations are the $s$-wave components of $SS\to  f\bar{f'}$ with $f,f'=t,c$ and $SS\to gg$. We construct the following function of the number of photons produced in a single DM pair annihilation event:
\begin{eqnarray}
N_\gamma^{th} &=&2N_{\gamma,t}(m_S) \times \sigma v(SS\to t\bar{t})_s + 2N_{\gamma,c}(m_S) \times \sigma v(SS\to c\bar{c})_s \\ \nonumber
&\quad& + \Big(N_{\gamma,t}(E_t)+N_{\gamma,c}(E_c)\Big) \times \sigma v(SS\to t\bar{c}+c\bar{t})_s + 2N_{\gamma,g}(m_S) \times \sigma v(SS\to gg)_s~, \\ \nonumber
N_\gamma^{exp} &=& 2N_{\gamma,b}(m_S) \times \sigma v_{b\bar{b}}(m_S),
\end{eqnarray}
where $N_{\gamma,f}(E_f)$ denotes the number of photons produced by a single SM fermion $f$ with energy $E_f$
\begin{eqnarray}
N_{\gamma,f}(E_f)\,= \, \int^{E_{max}}_{E_{min}}  dE_\gamma \frac{dN_{\gamma,f}}{dE_\gamma} (E_f)~,
\end{eqnarray}
and
\begin{eqnarray}
E_t =  \frac{4m_S^2+m_t^2-m_c^2}{4m_S}~, \quad E_c =  \frac{4m_S^2+m_c^2-m_t^2}{4m_S} ~.
\end{eqnarray}
We choose the following photon energy range according to Fermi's report
\begin{eqnarray}
E_{max} \, = \, min\{E_f, 300 \, {\rm GeV}\}~, \quad \quad E_{min} \, = \, 0.3 \, {\rm GeV}.
\end{eqnarray}
We use the photon number spectra $\frac{dN_{\gamma,f}}{dE_\gamma} (E_f)$ calculated by {\bf PPPC4DM} \cite{Buch2015-PPPC4DM,Cirelli2010-PPPC4DM} which contain the electroweak corrections. $\sigma v_{b\bar{b}}(m_S)$ is also taken from Fermi's report. We set the following requirement
\begin{eqnarray}
N_\gamma^{th} \lesssim N_\gamma^{exp}.
\end{eqnarray}

\begin{figure}[h]
  \centering
  \includegraphics[height=6.5cm,width=7.5cm]{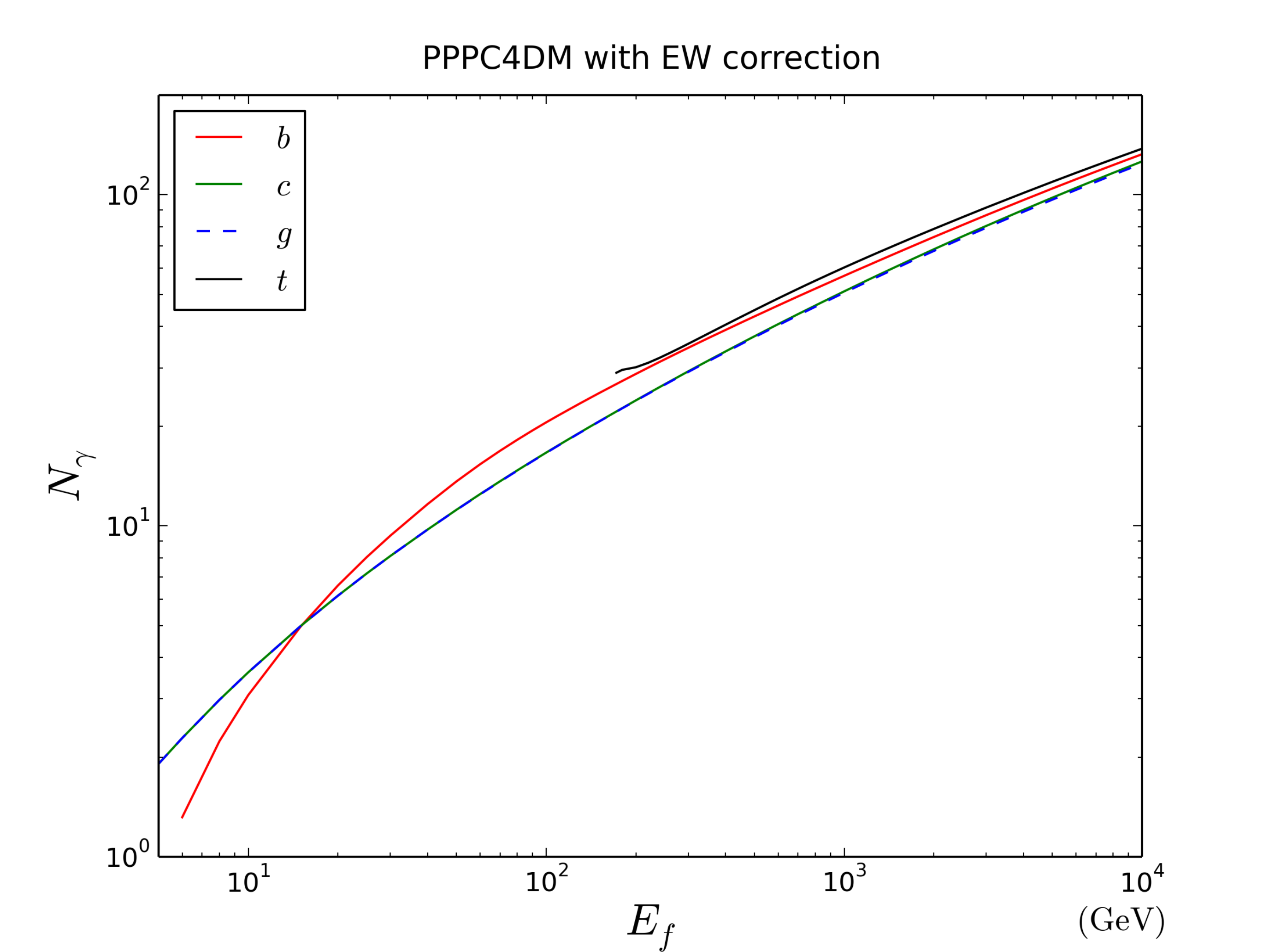} 
  \includegraphics[height=6.5cm,width=7.5cm]{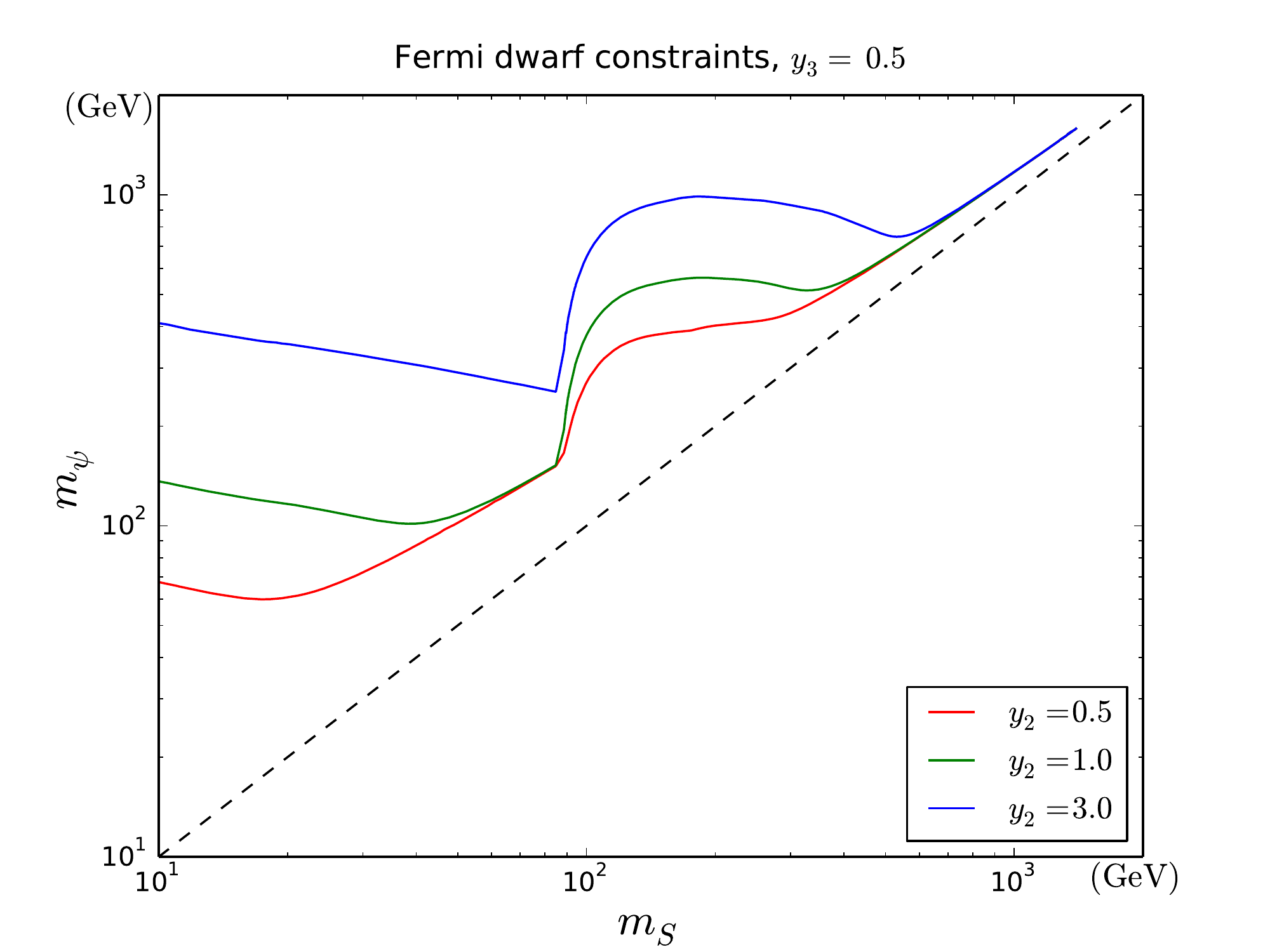} 
  \caption{{\bf Left}: Photon numbers produced by different SM fermions taken from {\bf PPPC4DM} \cite{Buch2015-PPPC4DM,Cirelli2010-PPPC4DM}; {\bf Right}: Constraints from current Fermi dwarf observation \cite{Ackermann:2015zua}, regions below solid lines are excluded.}
  \label{fig-ID}
\end{figure}

In the left panel of Fig.\ref{fig-ID} we show $N_{\gamma,f}(E_f)$ for $f=t,c,b,g$ taken from {\bf PPPC4DM}, from which one can see that the top quark is capable of producing more photons than others, meanwhile photons from charm quark are approximately the same as the those from gluon. This observation suggests that the constraints from Fermi dwarf observation should be stronger when on-shell top quark(s) is produced.
In the right panel we show the bounds on the plane of $m_\psi, m_S$ for $y_3=0.5$ and $y_2=0.5, 1, 3$. As expected, the exclusion becomes stronger when $m_S\gtrsim m_t/2$ in which case a single top quark can be produced. The bounds are even more stringent when $SS\to t\bar{t}$ is kinematically open and the lower bound of the mediator mass can reach $\{400,600,1000\}$ GeV, respectively. We also notice that Fermi dwarf observation is a good complementarity in $m_S>m_t/2$ to the direct detection constraints from XENON1T, which is significant for $m_S<m_t/2$ as shown in Fig.\ref{fig-DD}.

\subsection{FCNC of top quark}
\label{section-FCNC}
\begin{figure}[h]
  \centering
  \includegraphics[height=3cm,width=12cm]{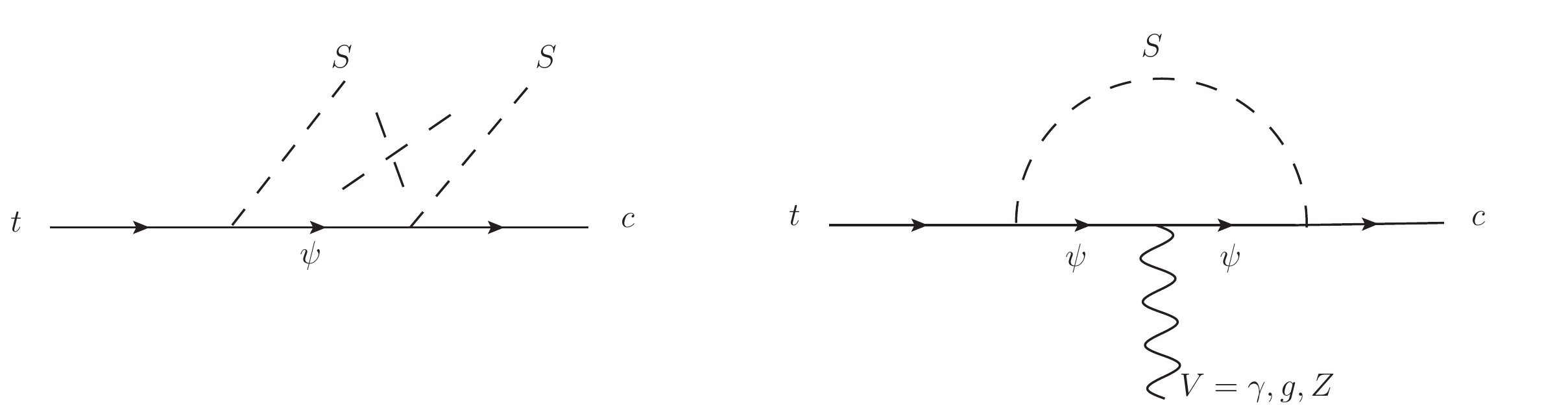} 
  \caption{FCNC processes of top quark in this model.}
  \label{fig-FCNCPlot}
\end{figure}

The FCNC processes for top quark in DM-$\{c,t\}$ interactions can be generated at both tree level $t\to T^{(*)}S \to cSS$ and loop level $t\to c+\gamma/g/Z$, with the diagrams shown in Fig.\ref{fig-FCNCPlot}.

The widths of tree-level processes can be expressed by
\begin{eqnarray}
\Gamma(t\to cSS) &=&  \Big(y_2^2 y_3^2\Big) \frac{1}{1024 \pi^3 m_t} \int^{m^2_{12,max}}_{m^2_{12,min}} dm^2_{12} \int^{m^2_{23,max}}_{m^2_{23,min}} dm^2_{23} \frac{1}{(m_{12}^2-m_\psi^2)^2(m_{23}^2-m_\psi^2)^2}   \\ \nonumber
&\quad& \times \Big( m_{12}^6 (-m_{23}^2)+2 m_{12}^4 m_{23}^4+m_{12}^4 m_{23}^2 m_c^2-m_{12}^2 m_{23}^6+m_{12}^2 m_{23}^4 m_c^2-4 m_{12}^2 m_{23}^2\\ \nonumber
&\quad&\quad  m_c^2 m_\psi^2+m_{12}^2 m_c^2 m_\psi^4+m_t^2 (m_{12}^4 (m_{23}^2+m_c^2-m_S^2)+m_{12}^2 (m_{23}^4+2 m_c^2 (m_{23}^2-2 m_\psi^2) \\ \nonumber
&\quad&\quad  -4 m_{23}^2 m_\psi^2+2 m_S^2 m_\psi^2+m_\psi^4)+m_{23}^4
   m_c^2-4 m_{23}^2 m_c^2 m_\psi^2 -m_S^2
   (m_{23}^2-m_\psi^2)^2\\ \nonumber
&\quad&\quad +m_{23}^2 m_\psi^4+4 m_c^2
   m_\psi^4-m_S^2 m_\psi^4)+m_S^2 (m_{12}^4
   (m_S^2-m_c^2)+m_{12}^2 (2 m_c^2 m_\psi^2-2
   m_{23}^2 m_S^2)\\ \nonumber
&\quad&\quad  +m_{23}^4 m_S^2-m_c^2
   m_\psi^4)  -m_{23}^4 m_c^2 m_S^2+2 m_{23}^2 m_c^2
   m_S^2 m_\psi^2+m_{23}^2 m_c^2 m_\psi^4-m_c^2 m_S^2
   m_\psi^4 \Big)~,
\end{eqnarray}
where we have kept $m_c$ in the calculation and
\begin{align}
m^2_{12,min} &= (m_S+m_c)^2, \quad m^2_{12,max} = (m_t-m_S)^2~, \\ \nonumber
m^2_{23,min} &= (E^*_2+E^*_3)^2-\left(\sqrt{E^{*2}_2-m_c^2} + \sqrt{E^{*2}_3-m^2_S} \right)^2~, \\ \nonumber
m^2_{23,max} &= (E^*_2+E^*_3)^2-\left(\sqrt{E^{*2}_2-m_c^2} - \sqrt{E^{*2}_3-m^2_S} \right)^2~, \\ \nonumber
E^*_2 &=\frac{m_{12}^2-m_S^2+m_c^2}{2m_{12}}, \quad E^*_3 =\frac{m_t^2-m_{12}^2-m_S^2}{2m_{12}}~.
\end{align}

For the loop level processes $t\to c+\gamma/g/Z$, the generic amplitudes can be written in the following form:
\begin{equation}
i {\mathcal M}_{tcV}=\bar u(p_2)\, \Gamma^\mu  u(p_1)\, \epsilon_\mu(k,\lambda)~,
\label{eq:deltaM}
\end{equation}
where $V=\gamma/g/Z$ and $p_1, p_2$, and $k$ are the 4-momenta of the incoming top quark,
outgoing charm quark and outgoing gauge boson, respectively.
$\epsilon_\mu(k,\lambda)$ is the polarization vector of the outgoing gauge
boson. The vertices $\Gamma^\mu$ with on-shell external particles can be decomposed as \cite{Lopez1997-FCNC}
\begin{eqnarray}
\Gamma^\mu_{tcZ}&=&\gamma^\mu (P_L A_{Z1}+ P_R B_{Z1})
+ i\sigma^{\mu\nu}k_\nu (P_L A_{Z2} + P_R B_{Z2})\, ,
\label{eq:VtcZ}\\
\Gamma^\mu_{tc\gamma}&=&i\sigma^{\mu\nu}k_\nu (P_L A_{\gamma2} + P_R B_{\gamma2})\, ,
\label{eq:Vtcgamma}\\
\Gamma^\mu_{tcg}&=& T^a i\sigma^{\mu\nu}k_\nu (P_L A_{g2} + P_R B_{g2})\,,
\label{eq:Vtcg}
\end{eqnarray}
where $P_{R,L}={1\over2}(1\pm\gamma_5)$, $\sigma^{\mu\nu}={i\over 2}[\gamma^\mu,\gamma^\nu]$ and $T^a$ are the generators
of $SU(3)$ color. Then the widths are
\begin{eqnarray}
\Gamma(t\rightarrow cZ)&=&{m_t^3 \over 32 \pi m_Z^2}(1- {m_Z^2\over m_t^2})^2
\Bigl[(1+2{m_Z^2\over m_t^2})(|A_{Z1}|^2+|B_{Z1}|^2) \nonumber\\
&&-6{m_Z^2 \over m_t}\,Re\Big (A_{Z1}B_{Z2}^*+
A_{Z2} B_{Z1}^*\Big) +m_Z^2(2+
{m_Z^2 \over m_t^2})(|A_{Z2}|^2+|B_{Z2}|^2)\Bigr]\, ,\\
\Gamma(t\rightarrow c\gamma)&=&{m_t^3\over 16 \pi } (|A_{\gamma2}|^2+|B_{\gamma2}|^2)\, ,\\
\Gamma(t\rightarrow cg)&=&C_F\,{m_t^3\over 16 \pi } (|A_{g2}|^2+|B_{g2}|^2)\, ,
\end{eqnarray}
where we have used $m_c=0$ for simplification. After calculating the loop diagram in Fig.\ref{fig-FCNCPlot} and extracting the coefficients \cite{Soares1989,Deshpande1982} we can obtain
\begin{align}
A_{Z1}&=-D_Z\, m_c m_t \,\Big( g_{ZL}\frac{B_t-B_c}{m_t^2-m_c^2} + g_{ZR} (C_{0} + 2 C_{1} + C_{11} + 2C_{12} + 2 C_{2} + C_{22}) \Big)~, \nonumber\\
B_{Z1}&=-D_Z\, g_{ZR} \quad \Big( \frac{m_t^2 \,B_t- m_c^2\, B_c}{m_t^2 - m_c^2} + m_\psi^2 C_0 - 2 C_{00} + m_Z^2 C_{12} + \frac{1}{2}\Big)~, \nonumber\\
A_{Z2}&=D_Z\, g_{ZR}\,  m_t \, (C_{12}+C_2+C_{22}) ,\quad\quad B_{Z2}= D_Z\, g_{ZR}\,  m_c \, (C_{1}+C_{11}+C_{12})~, \nonumber\\
A_{\gamma2}&= D_\gamma\, g_{\gamma R}\,  m_t\,  (C_{12}+C_2+C_{22}),\quad\quad \, \,B_{\gamma2}= D_\gamma\, g_{\gamma R}\,  m_c \, (C_{1}+C_{11}+C_{12}) ~, \nonumber\\
A_{g2}&=D_g\, g_{gR}\,  m_t \, (C_{12}+C_2+C_{22}) ,\quad\quad \,\,\, B_{g2}=D_g\, g_{gR}\,  m_c \, (C_{12}+C_2+C_{22}) ~, 
\end{align}
where
\begin{align}
B_t &= (B_0+B_1)\Big(m_t^2;m_S^2,m_\psi^2\Big)~, \nonumber\\
B_c &= (B_0+B_1)\Big(m_c^2;m_S^2,m_\psi^2\Big)~, \nonumber\\
C_{i,ij}\, &=C_{i,ij}\Big(m_c^2, m_V^2, m_t^2; m_S^2, m_\psi^2, m_\psi^2\Big)~, \nonumber\\
D_Z\, &=-y_2 y_3\,  \frac{g_2}{c_w}\, \frac{i}{16\pi^2}, \quad D_\gamma\, =-y_2 y_3\,  e\,  \frac{i}{16\pi^2}, \quad D_g\, =-y_2 y_3\,  g_s\,  \frac{i}{16\pi^2}~, \nonumber\\
g_{ZL} &=\frac{1}{2}-\frac{2}{3}s_w^2, \quad\quad\quad~~g_{ZR} =-\frac{2}{3}s_w^2, \quad\quad\quad~~~ g_{\gamma R} =g_{gR}=\frac{2}{3},
\end{align}
with $s_w\equiv \sin\theta_w,c_w\equiv \cos\theta_w$ and $\theta_w$ is the Weinberg angle. We checked the loop divergence cancellation in the widths and use {\bf LoopTools} \cite{Hahn1999-LoopTools} based on {\bf FF} package \cite{Hahn1999-FF} to perform numerical calculations.

\begin{figure}[h]
  \centering
  \includegraphics[height=6.5cm,width=7.5cm]{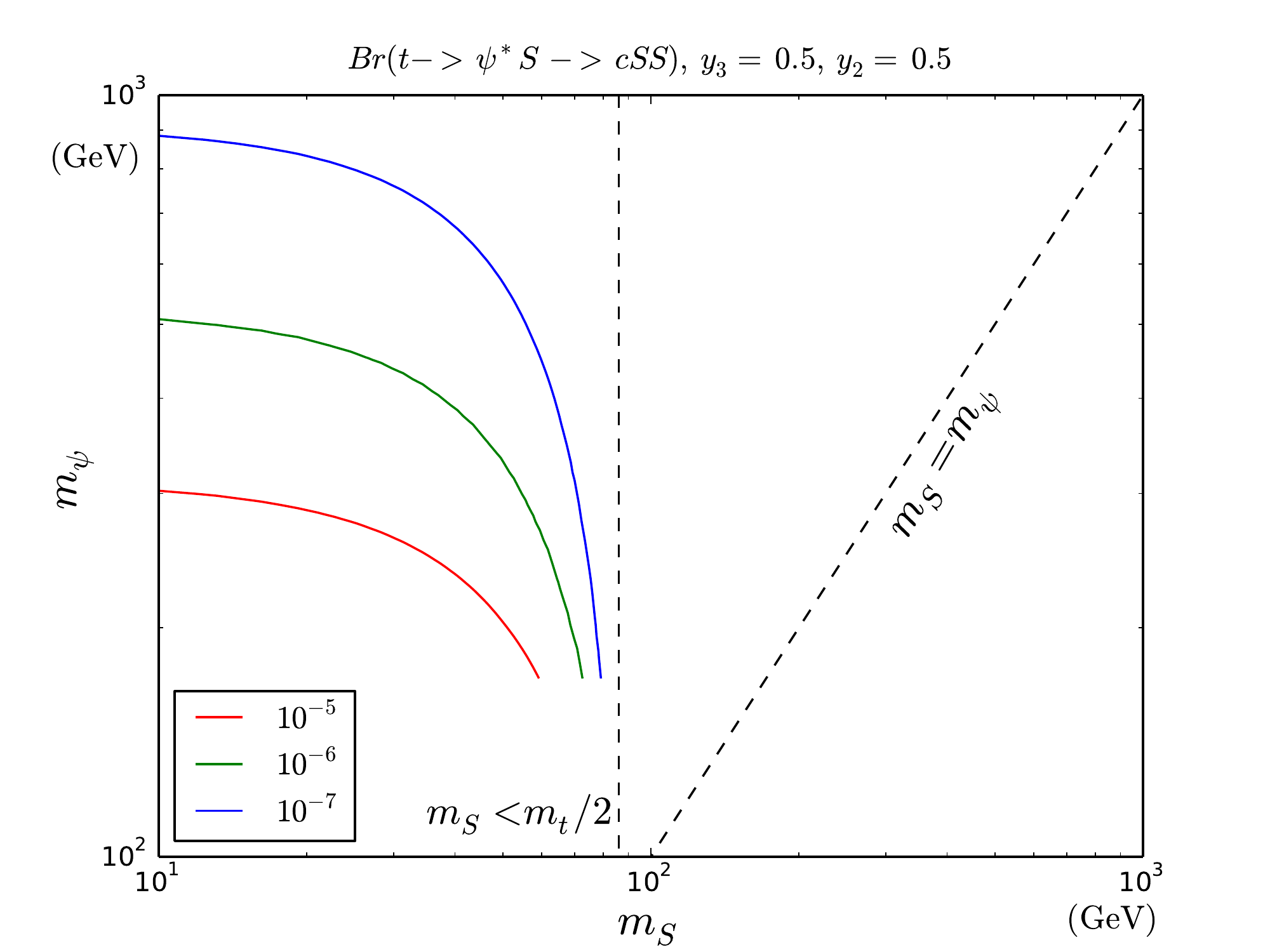} 
  \includegraphics[height=6.5cm,width=7.5cm]{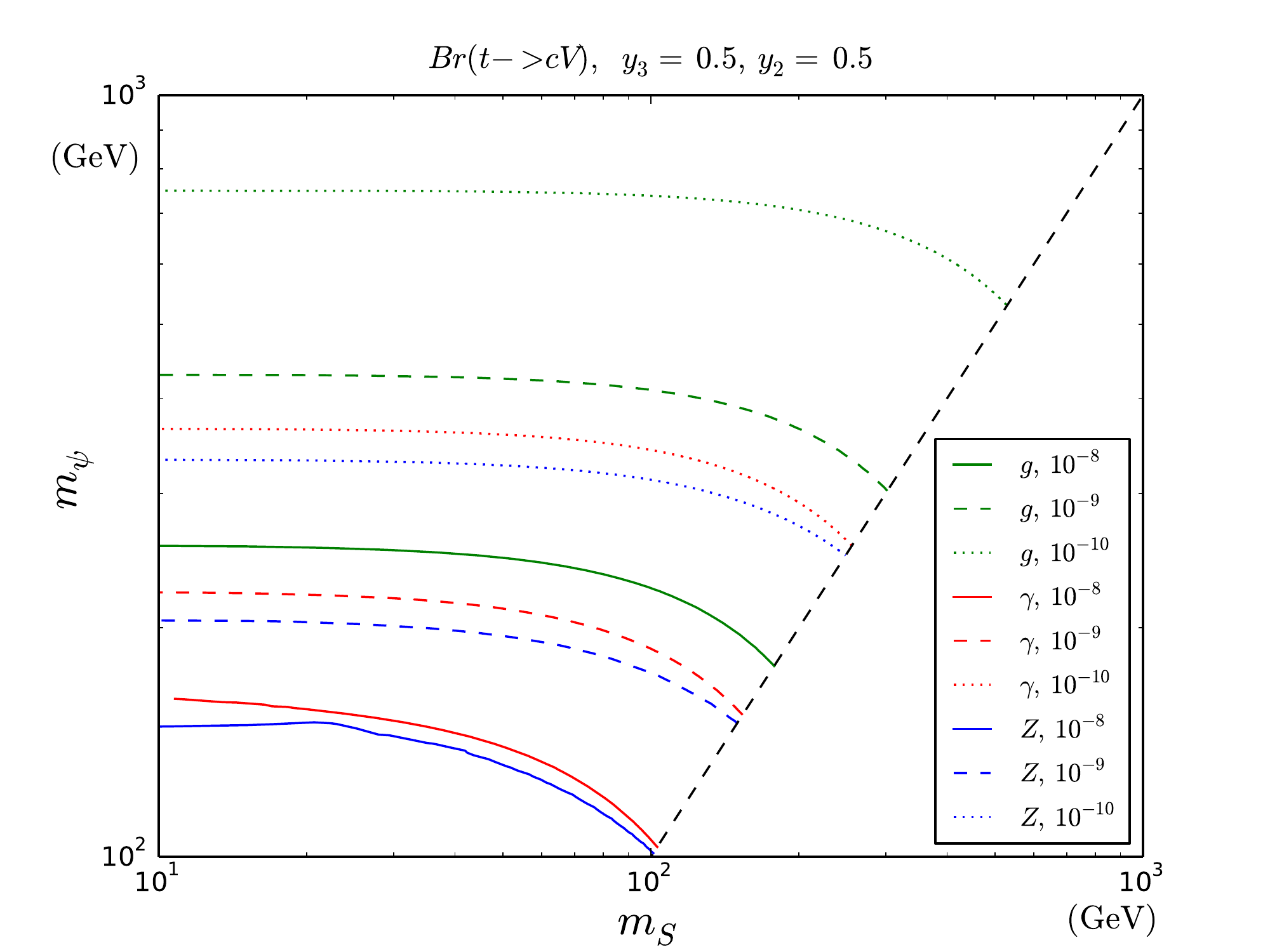} 
  \caption{FCNC of top quark at tree ({\bf left} panel) and loop level ({\bf right} panel).}
  \label{fig-FCNC}
\end{figure}

In Fig.\ref{fig-FCNC} we choose $y_2=y_3=0.5$ as a benchmark point and show the predictions of $Br(t\to cSS)$ and $Br(t\to c+\gamma/g/Z)$ in the left and right panel, respectively. We found that $Br(t\to cSS)$ can exceed $10^{-7}$ with a mediator lighter than 1 TeV. The loop FCNC branching fractions can reach $10^{-8}$ with light mediator $m_\psi\lesssim 300$ GeV but are smaller than $10^{-10}$ with $m_\psi \sim 1$ TeV. Benefiting from the strong coupling strength and color index, $Br(t\to cg)$ is about one order of magnitude larger than the other two.

\subsection{Collider detection}
\label{section-collider}

The typical collider signals for DM-$\{c,t\}$ interactions are $\slashed{E}_T+\{t\bar{t},jj,tj\}$ from the pair production of colored mediator $pp \to \psi\overline{\psi}$ followed by $\psi\to S+t/c$. In this work we consider the $t\bar{t},jj$ signals which are analogous to the supersymmetry (SUSY) searches for stop and first two generation squarks $pp\to \tilde{t}^*\tilde{t}, \tilde{q}^*\tilde{q}$, but with larger $\psi\overline{\psi}$ production cross section compared to $\tilde{t}^*\tilde{t}, \tilde{q}^*\tilde{q}$ with the same mass due to the internal spin degree of freedom. In our previous work \cite{Baek2016a} we performed a detailed study of $\slashed{E}_T+t\bar{t}$ signal for top-philic DM scenario based on ATLAS 13.2 $fb^{-1}$ data at 13 TeV and found that the excluded mediator mass can reach 1150 GeV compared to 800 GeV in stop search \cite{ATLAS-CONF-2016-050}. In the SUSY case, the bounds on first two generation squarks are stronger than stop, which have currently reached 1.6 TeV \cite{ATLAS-CONF-2017-022} for squarks with $36\, fb^{-1}$ data at LHC compared to 950 GeV for stop \cite{ATLAS-CONF-2017-037} using $1\ell + jets + \slashed{E}_T$ signal. As a result, $\slashed{E}_T+jj$ should also be an important signal at collider when we allow DM to couple to both charm and top quark in this work.

The production cross sections of $\slashed{E}_T+t\bar{t},jj$ depend on $Br(\psi\to St,Sc)$. The partial decay width of the mediator is
\begin{eqnarray}
\Gamma(\psi \to SQ)&=& y_Q^2 \frac{m_\psi}{32\pi} (1+x_Q^2-x_S^2)\Big[\Big(1-(x_Q+x_S)^2\Big)\Big(1-(x_Q-x_S)^2\Big)\Big]^{1/2}\\ \nonumber
 &\equiv& y_Q^2 F(m_Q,m_S,m_\psi)~,
\end{eqnarray}
where $Q=c,t$ are on-shell, $y_Q=y_2,y_3$ and $x_Q=m_Q/m_\psi, x_S=m_S/m_\psi$. Since decaying into top quark contains larger phase space suppression compared to the charm quark case, $F(m_t,m_S,m_\psi)$ will be smaller than $F(m_c,m_S,m_\psi)$ except for $\Delta m=m_\psi-m_S\gg m_t$ in which case $F(m_t,m_S,m_\psi)\simeq F(m_c,m_S,m_\psi)$. Then the branching fractions are
\begin{eqnarray}
\label{Eq-BrPsi}
Br(\psi \to St)&=& \frac{1}{1+\frac{y_2}{y_3}\frac{F(m_c,m_S,m_\psi)}{F(m_t,m_S,m_\psi)}}~, \\ \nonumber
Br(\psi \to Sc)&=& \frac{1}{1+\Big(\frac{y_2}{y_3}\frac{F(m_c,m_S,m_\psi)}{F(m_t,m_S,m_\psi)}\Big)^{-1}}~,
\end{eqnarray}
and the signal production cross sections are
\begin{eqnarray}
\label{Xsection}
\sigma\Big(pp \to \psi\overline{\psi} \to \slashed{E}_T+t\bar{t} \Big) = \sigma\Big(pp \to \psi\overline{\psi}\Big) \, Br^2(\psi \to St)~, \\ \nonumber
\sigma\Big(pp \to \psi\overline{\psi} \to \slashed{E}_T+jj \Big) = \sigma\Big(pp \to \psi\overline{\psi}\Big) \, Br^2(\psi \to Sc)~.
\end{eqnarray}

\begin{figure}[h]
  \centering
  \includegraphics[height=6cm,width=7cm]{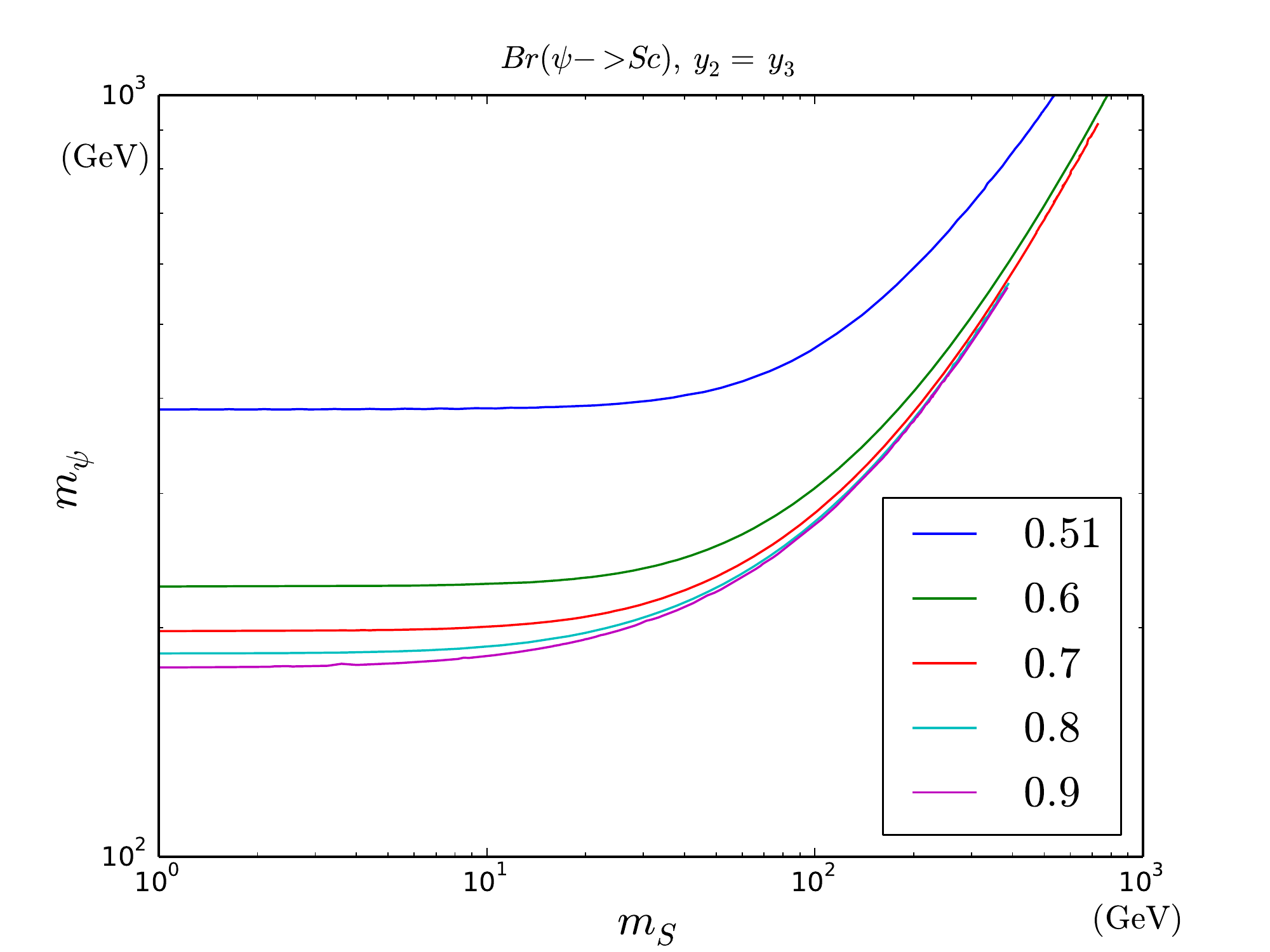} 
  \caption{$Br(\psi\to Sc)$ with $y_2=y_3$ on the plane of $m_S, m_\psi$ with $m_\psi-m_S>m_t$.}
  \label{fig-BrXs}
\end{figure}

In Fig.\ref{fig-BrXs} we set $y_2=y_3$ to show $Br(\psi\to Sc)$ on the plane of $m_S, m_\psi$ with $\Delta m>m_t$. One can easily use $1-Br(\psi\to Sc)$ to obtain the corresponding plot for $Br(\psi\to St)$. We can see that $\Delta m\sim m_t$ makes $Br(\psi\to Sc)$ easily exceed $90\%$, while $\Delta m\gtrsim 400$ GeV would make $F(m_t,m_S,m_\psi)\simeq F(m_c,m_S,m_\psi)$ and result in $Br(\psi\to St)\simeq Br(\psi\to Sc)\simeq 0.5$. $\sigma(pp \to \psi\overline{\psi})$ at $\sqrt{s}=13$ TeV can be found in \cite{ATLASCollaboration2017-VLQuark-2} which uses Top++ v2.0 \cite{Czakon:2011xx} and the MSTW 2008 NNLO PDF set. Next-to-next-to-leading-order (NNLO) QCD corrections and the soft gluon resummation to NNLL accuracy \cite{Cacciari:2011hy,Beneke:2011mq,PhysRevLett.109.132001,Czakon:2012zr,Czakon:2012pz,Czakon:2013goa} are also included.

We use FeynRules \cite{Alloul2014} to implement the model in this work to obtain UFO model file
\cite{Degrande:2011ua-UFO}, which is utilized by MadGraph5 \cite{Alwall2014} to generate the parton level events which
is further linked to PYTHIA8 \cite{Sjostrand:2007gs-PYTHIA81,Sjostrand:2014zea-PYTHIA82} to perform the parton shower
and hadronization. Then the PYTHIA event files are processed by CheckMate2 \cite{Kim2015,Drees2013} which utilizes
Delphes3 \cite{Ovyn2009,DeFavereau2014} to simulate the detector response and calculate the cut efficiency $\epsilon$. The number of signal events after selection cuts are calculated as $N_{sig}=\mathcal{L}*\sigma*\epsilon$ where $\mathcal{L}=36 \,{\rm fb^{-1}}$ is the ATLAS integrated luminosity at 13 TeV in \cite{ATLAS-CONF-2017-022,ATLAS-CONF-2017-037} and $\sigma$ is the $pp\to \psi\overline{\psi}$ production cross section at $\sqrt{s} = 13 \,{\rm TeV}$. We use top++2.0 \cite{Czakon:2011xx} to obtain $\sigma(pp\to \psi\overline{\psi})$ at next-to-next-to-leading order (NNLO) which includes the next-to-next-to-leading logarithmic (NNLL) contributions. We vary the renormalization and factorization scale between $(0.5, 2)m_\psi$ to estimate theoretical uncertainty $\Delta \sigma$ at $1\sigma$ level. CheckMate uses $\Delta \sigma$ and the number of simulated events $N_{MC}$ to calculate $\Delta N_{sig}$, which is the total uncertainty of signal event number. Finally, CheckMATE defines the following quantity:
\begin{eqnarray}
r_{CM} \equiv \frac{N_{sig} - 1.96 \Delta N_{sig}}{N^{95}_{obs}}
\end{eqnarray}
where $N_{obs}^{95}$ is the model independent bounds at $95\%$ Confidence Level (C. L.) on the number of new physics signal events provided in the experimental reports. Then a model can be claimed to be excluded at the $95\%$ C. L. if $r_{CM}>1$. Note that $r_{CM}$-limit is usually weaker than the method based on $S/\sqrt{S + B} < 1.96$ since $r_{CM}$-limit utilizes the total uncertainty on the $N_{sig}$ in a more conservative manner. More detail are provided in \cite{Kim2015,Drees2013}.

\begin{figure}[h]
  \centering
  \includegraphics[height=6cm,width=7cm]{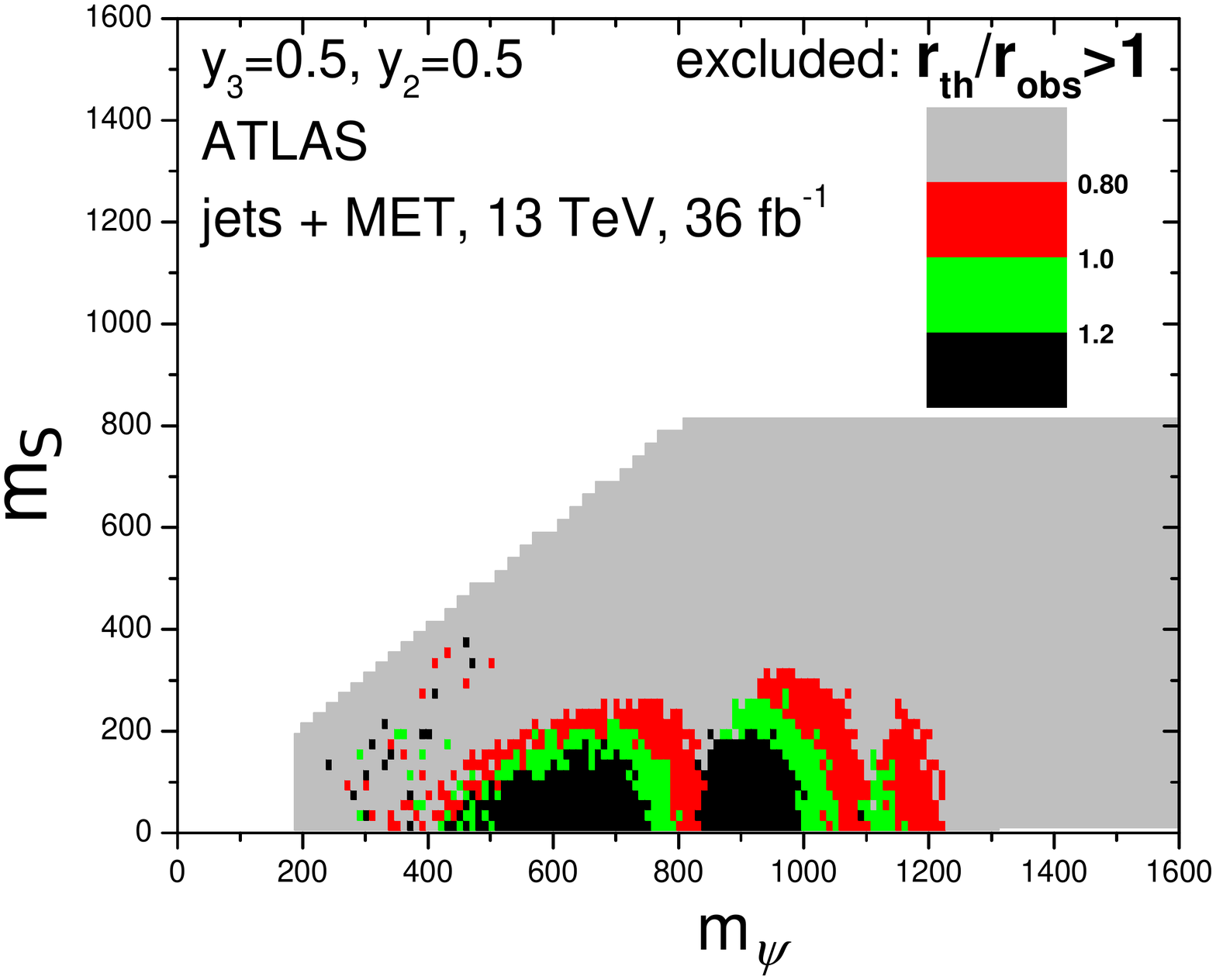}
  \includegraphics[height=6cm,width=7cm]{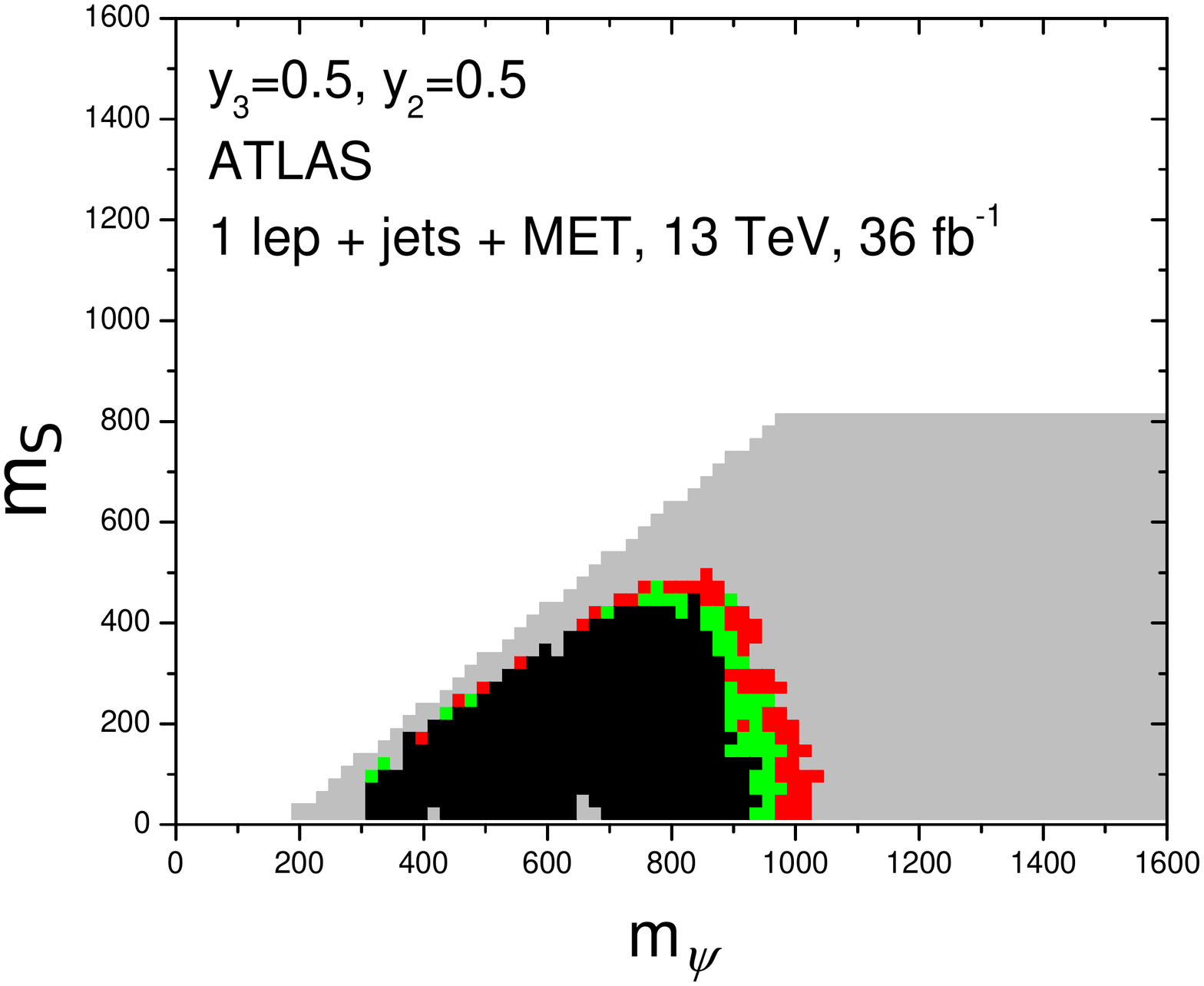}
  \includegraphics[height=6cm,width=7cm]{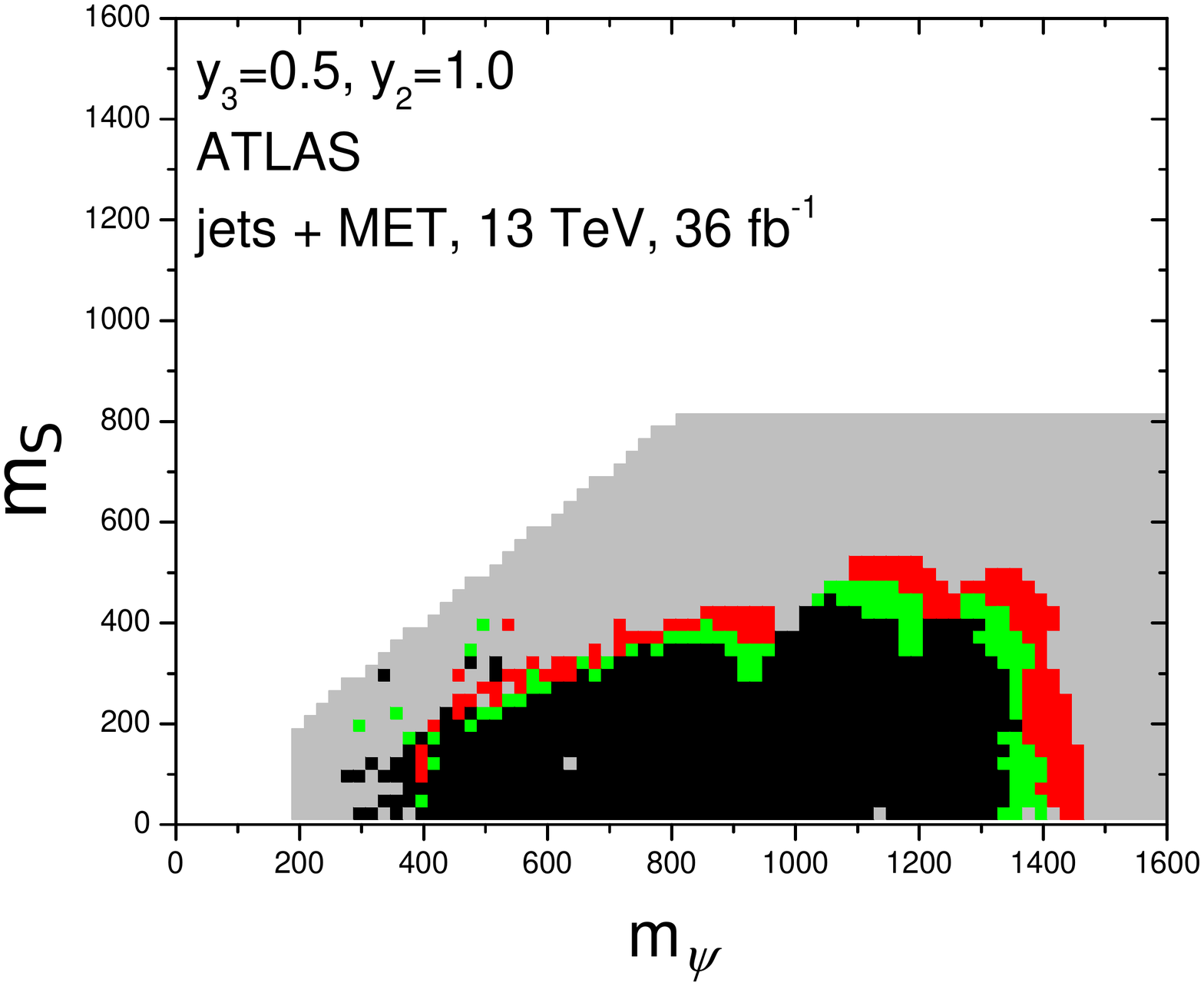}
  \includegraphics[height=6cm,width=7cm]{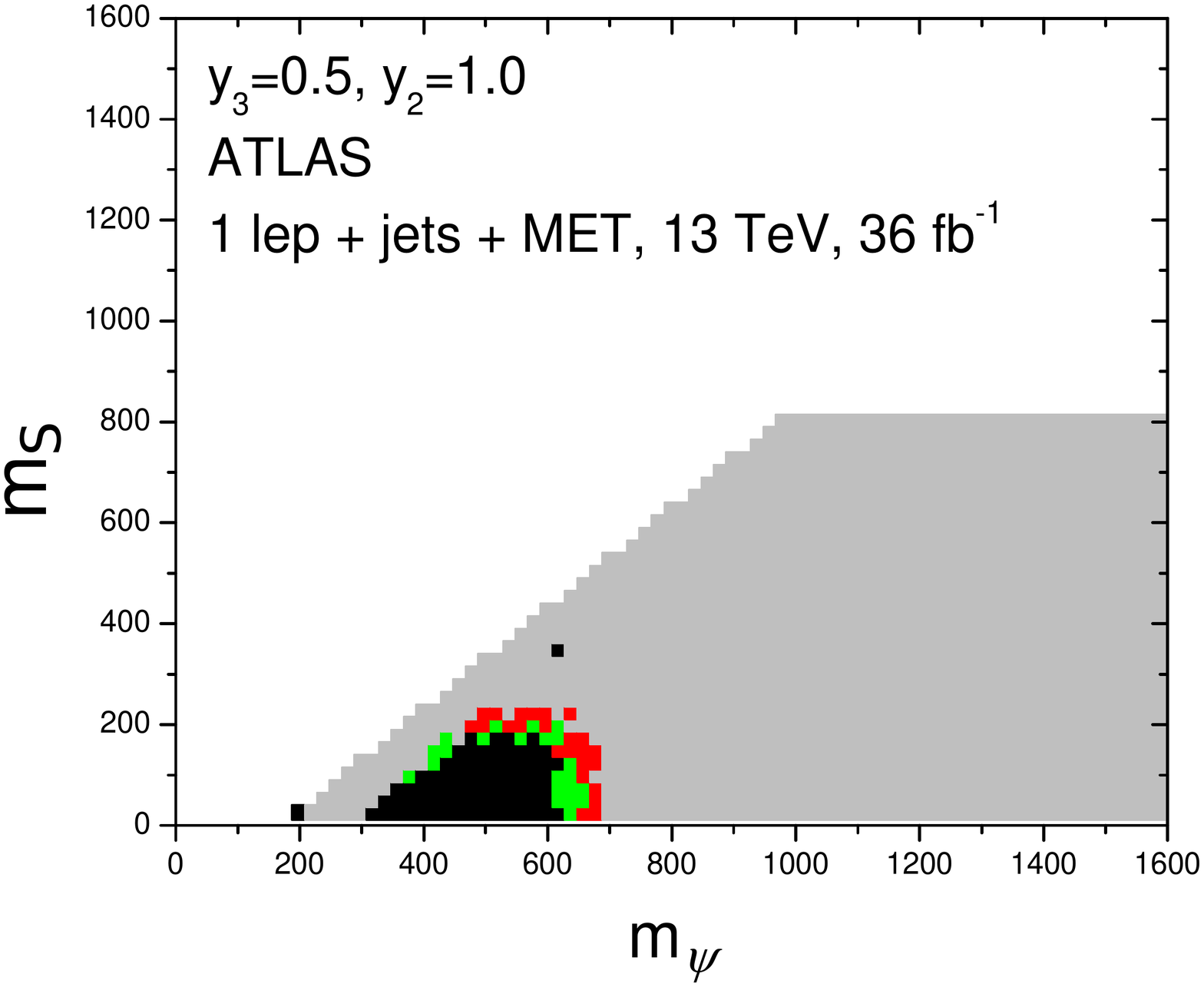}
  \includegraphics[height=6cm,width=7cm]{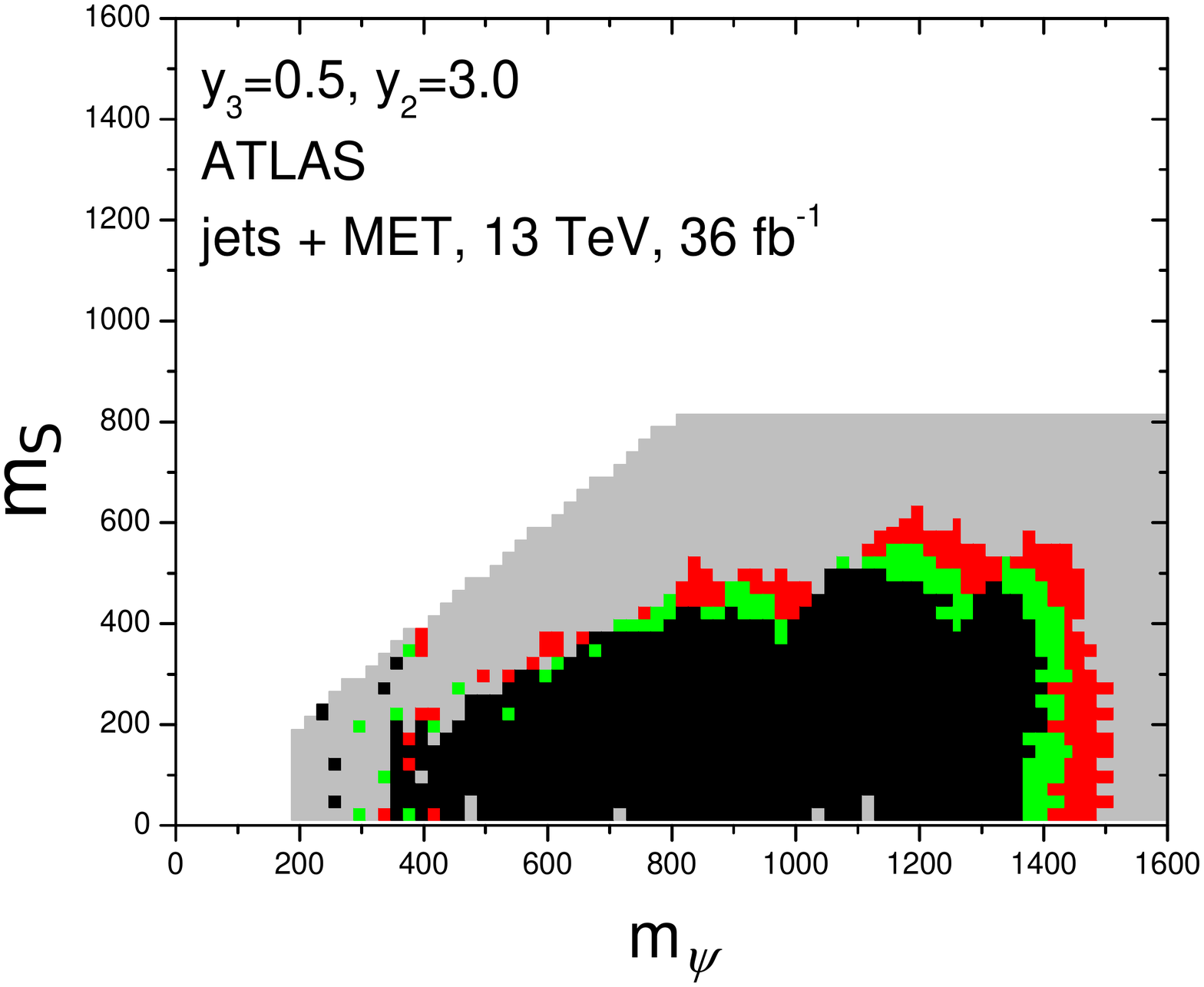}
  \includegraphics[height=6cm,width=7cm]{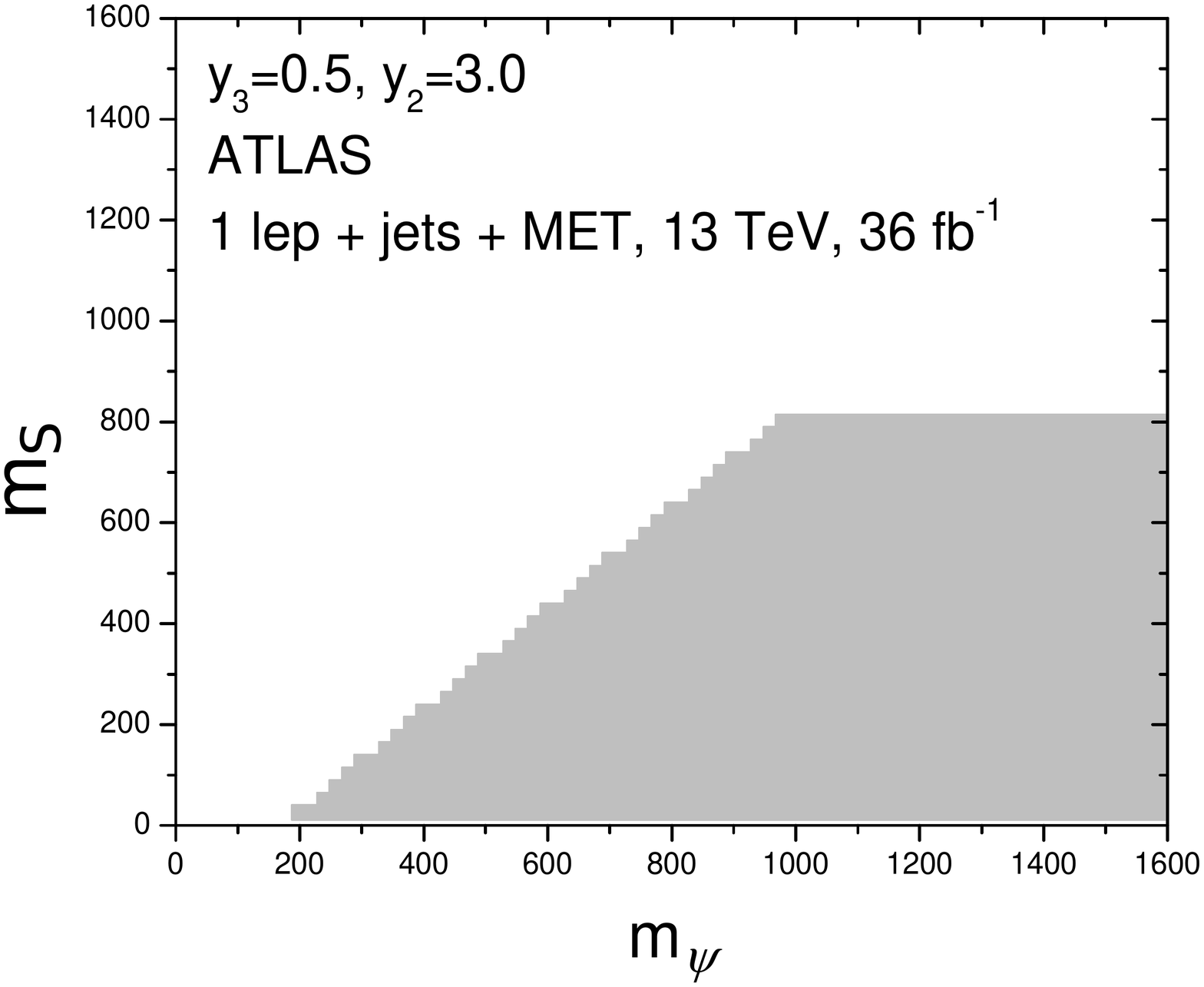}
  \caption{ATLAS bounds on the model of this work using $36\, fb^{-1}$ data at 13 TeV. {\bf Left}: $jets + \slashed{E}_T$ signal; {\bf Right}: $1 \ell + jets + \slashed{E}_T$ signal. Rows from top to bottom correspond to $y_2=0.5,1,3$ with common $y_3=0.5$. All masses are in unit of GeV.}
  \label{fig-Collider}
\end{figure}

In fig.\ref{fig-Collider} we show the collider bounds from $jets + \slashed{E}_T$ searches in \cite{ATLAS-CONF-2017-022}
(left panels) and from $1\ell + jets + \slashed{E}_T$ searches in \cite{ATLAS-CONF-2017-037} (right panels), where
different rows correspond to $y_2=0.5,1,3$ with common $y_3=0.5$, respectively. To account for uncertainties in our
simulation, we highlight $r_{CM}=1\pm 20\%$ samples near the exclusion criteria $r_{CM}=1$. Light grey, red, green,
black samples correspond to $r_{CM}$ values within $r_{CM}<0.8$, $[0.8,1)$, $[1,1.2)$ and $r_{CM}>1.2$,
respectively. The left panels show that with increasing $y_2$ the branching fraction $Br(\psi \to Sc)$ get enhanced and
results in generally larger $jets + \slashed{E}_T$ production cross sections and wider exclusion
region. Correspondingly, the smaller $t\bar{t} + \slashed{E}_T$ production cross sections result in lower exclusion
sensitivity. In the first row with $y_3=y_2=0.5$, we found that $m_\psi$ can be excluded up to around 1100 (950) GeV
using $jets + \slashed{E}_T$ ($1\ell + jets + \slashed{E}_T$) signal, and the sensitivity can reach $m_S$ around 200
(400) GeV. On the contrary, in the lowest row with $y_2=3.0$ and dominating $Br(\psi \to Sc)$, the $jets +
\slashed{E}_T$ sensitivity is greatly amplified and $m_\psi$ can be excluded up to 1400 GeV while the $1\ell + jets +
\slashed{E}_T$ search completely loses the sensitivity.

\subsection{Combined results}
\label{section-combined}

\begin{figure}[h]
  \centering
  \includegraphics[height=6cm,width=7cm]{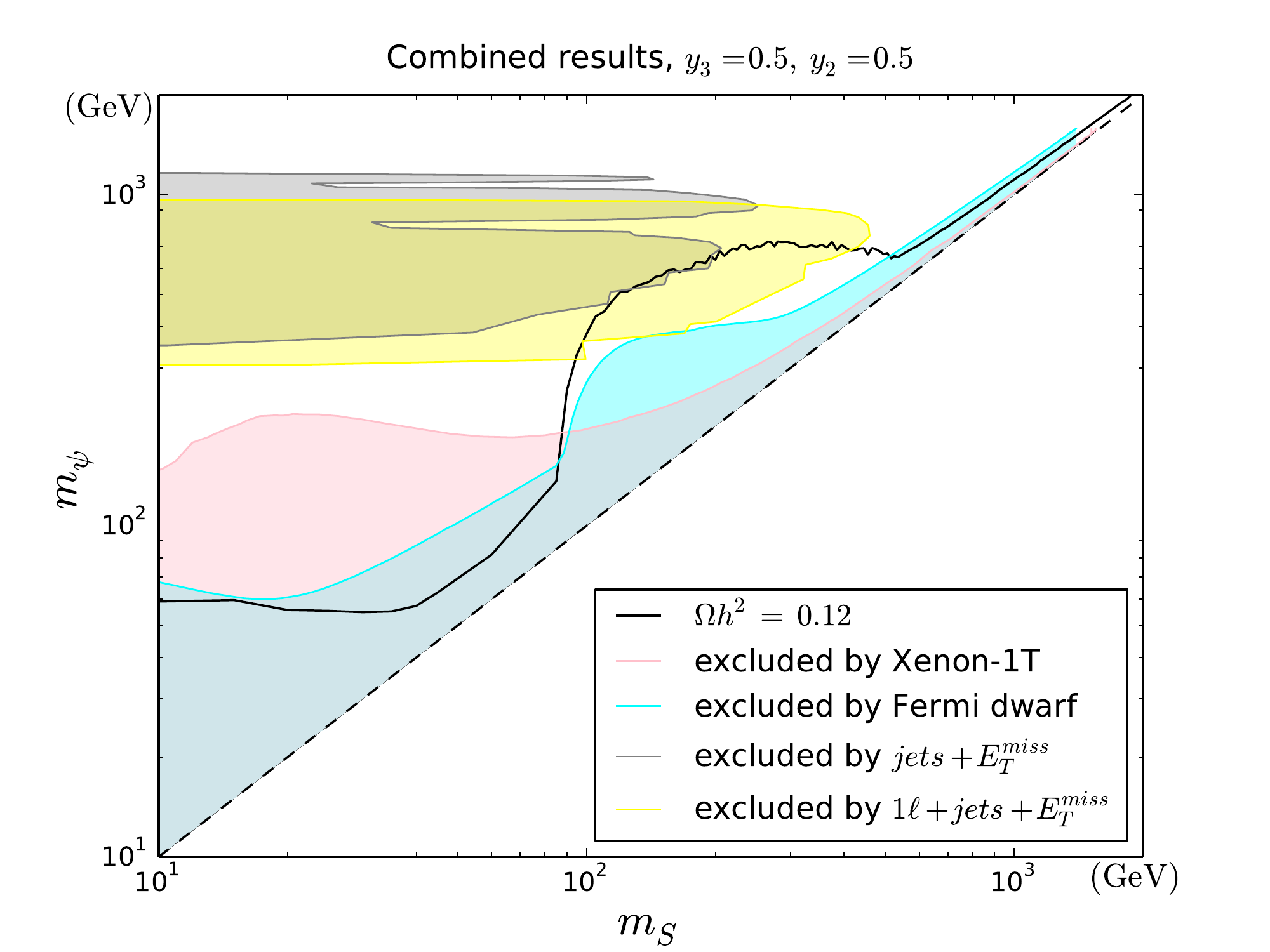}
  \includegraphics[height=6cm,width=7cm]{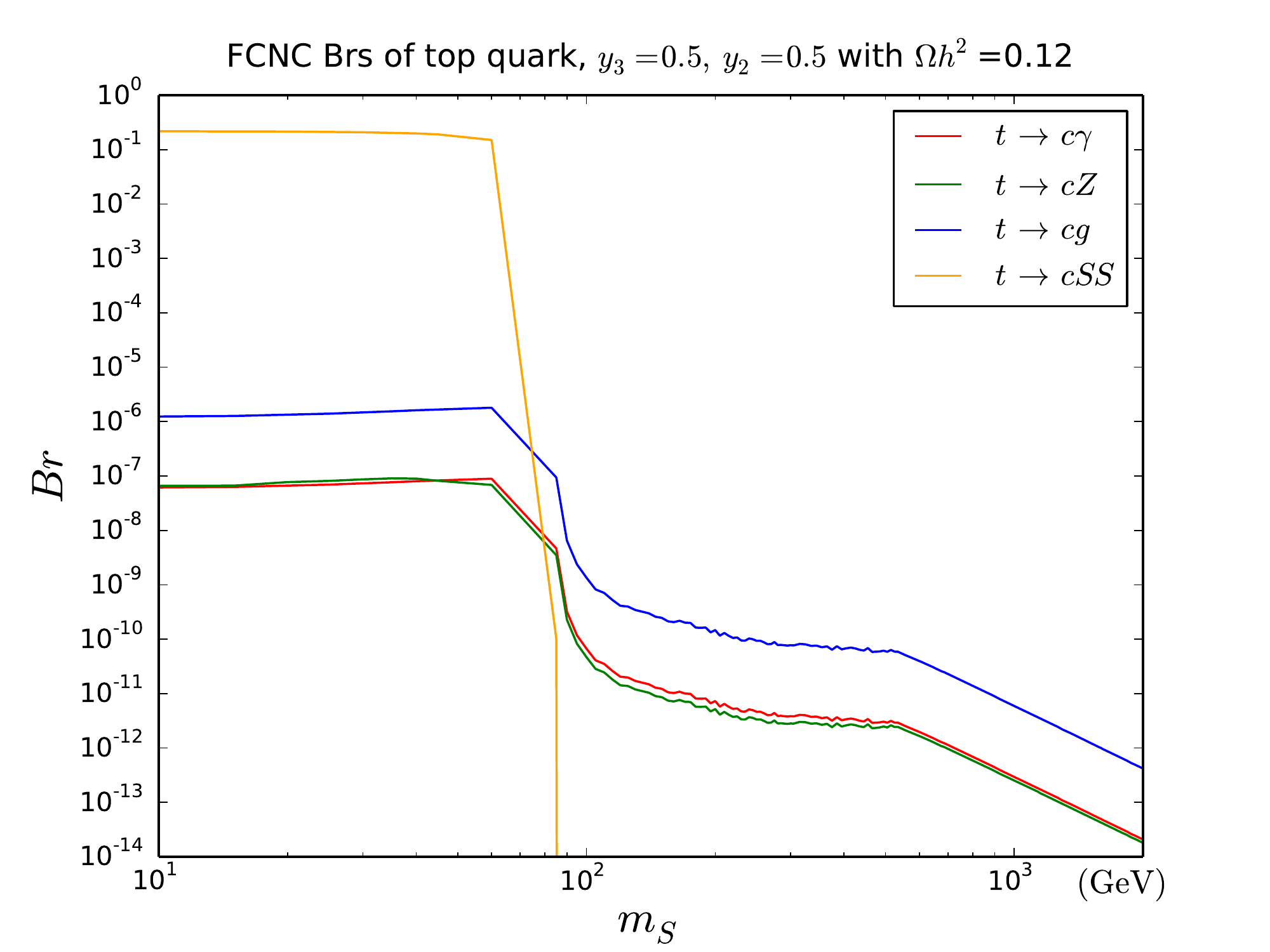}
  \includegraphics[height=6cm,width=7cm]{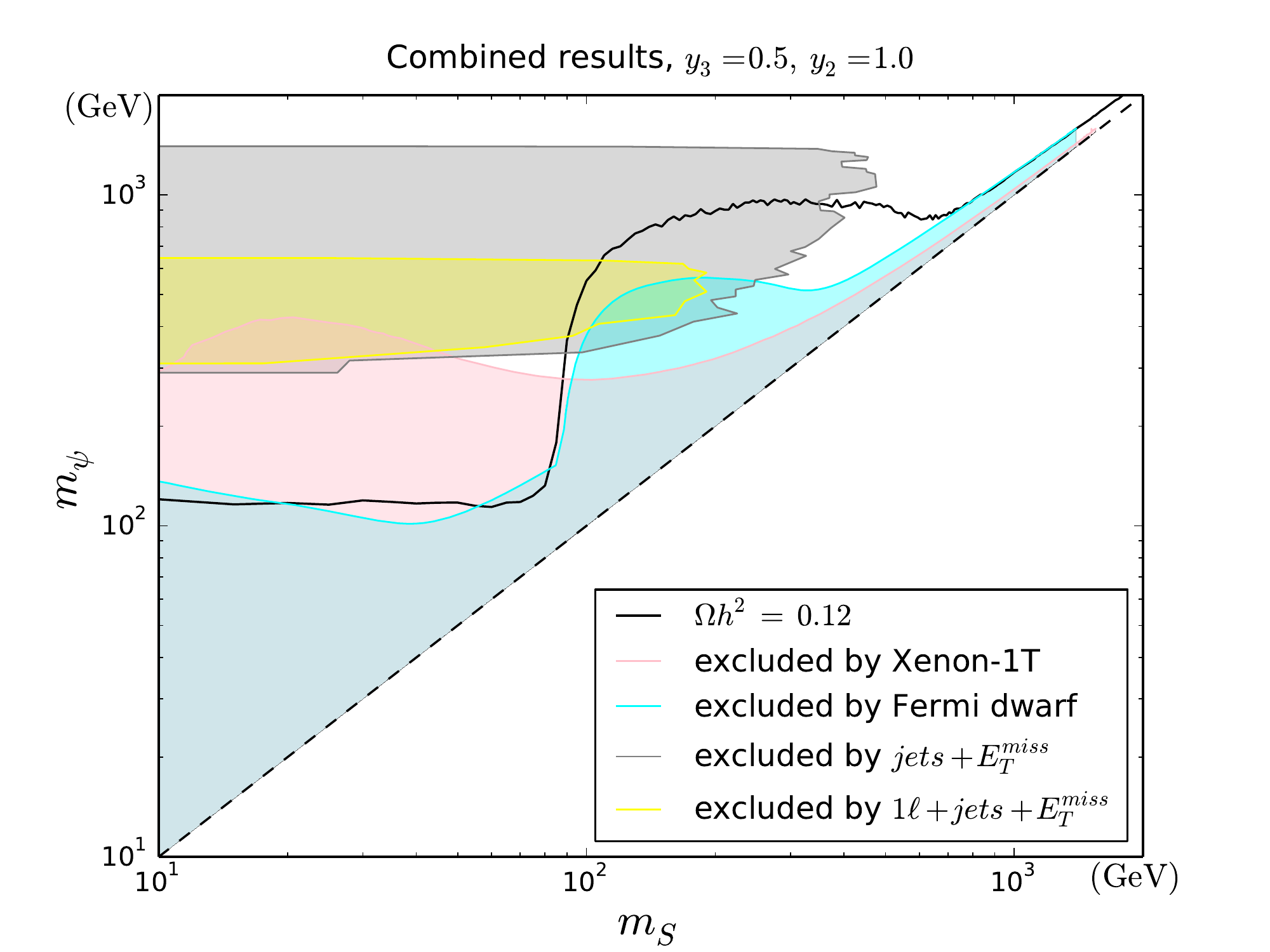}
  \includegraphics[height=6cm,width=7cm]{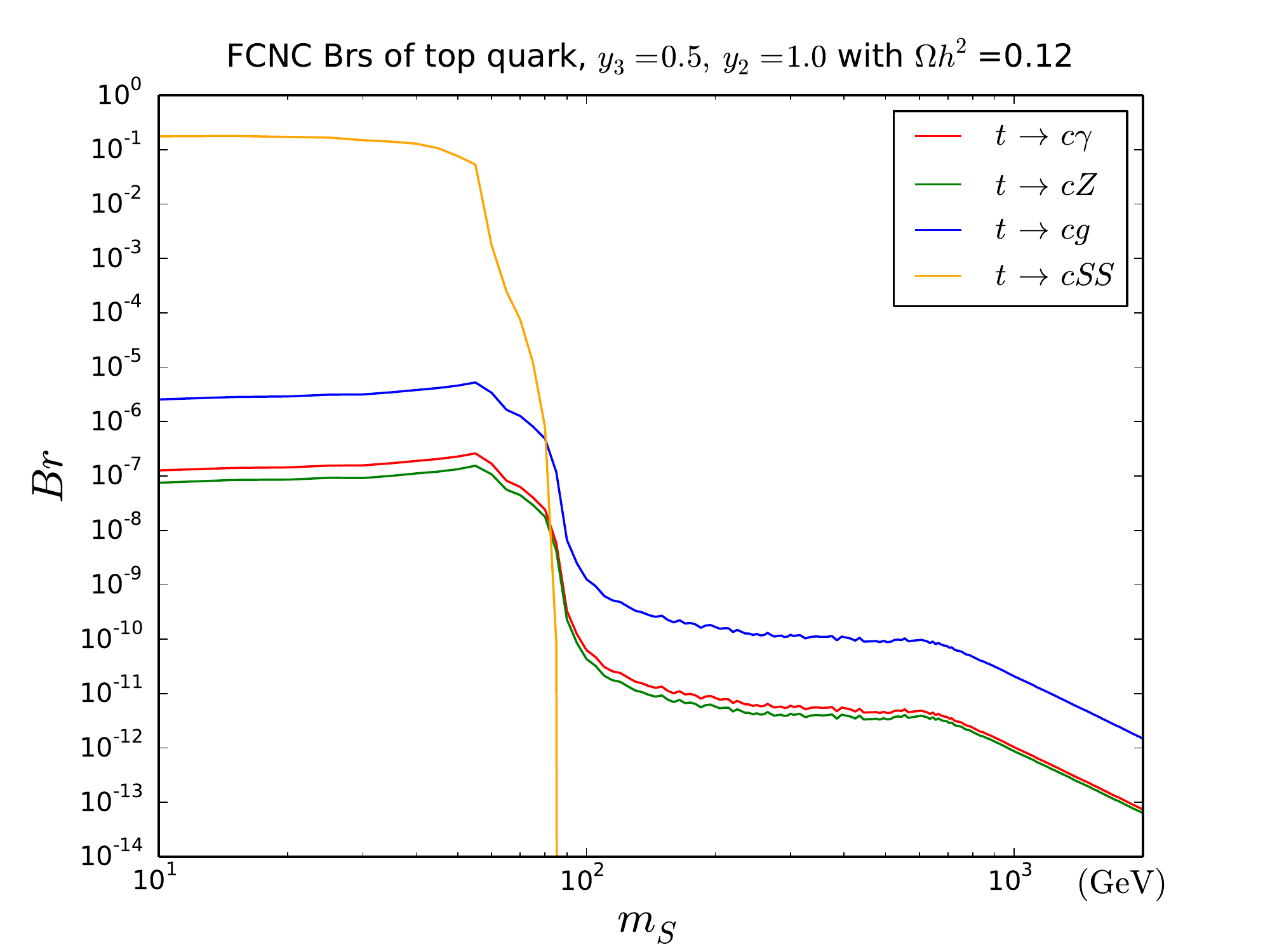}
  \includegraphics[height=6cm,width=7cm]{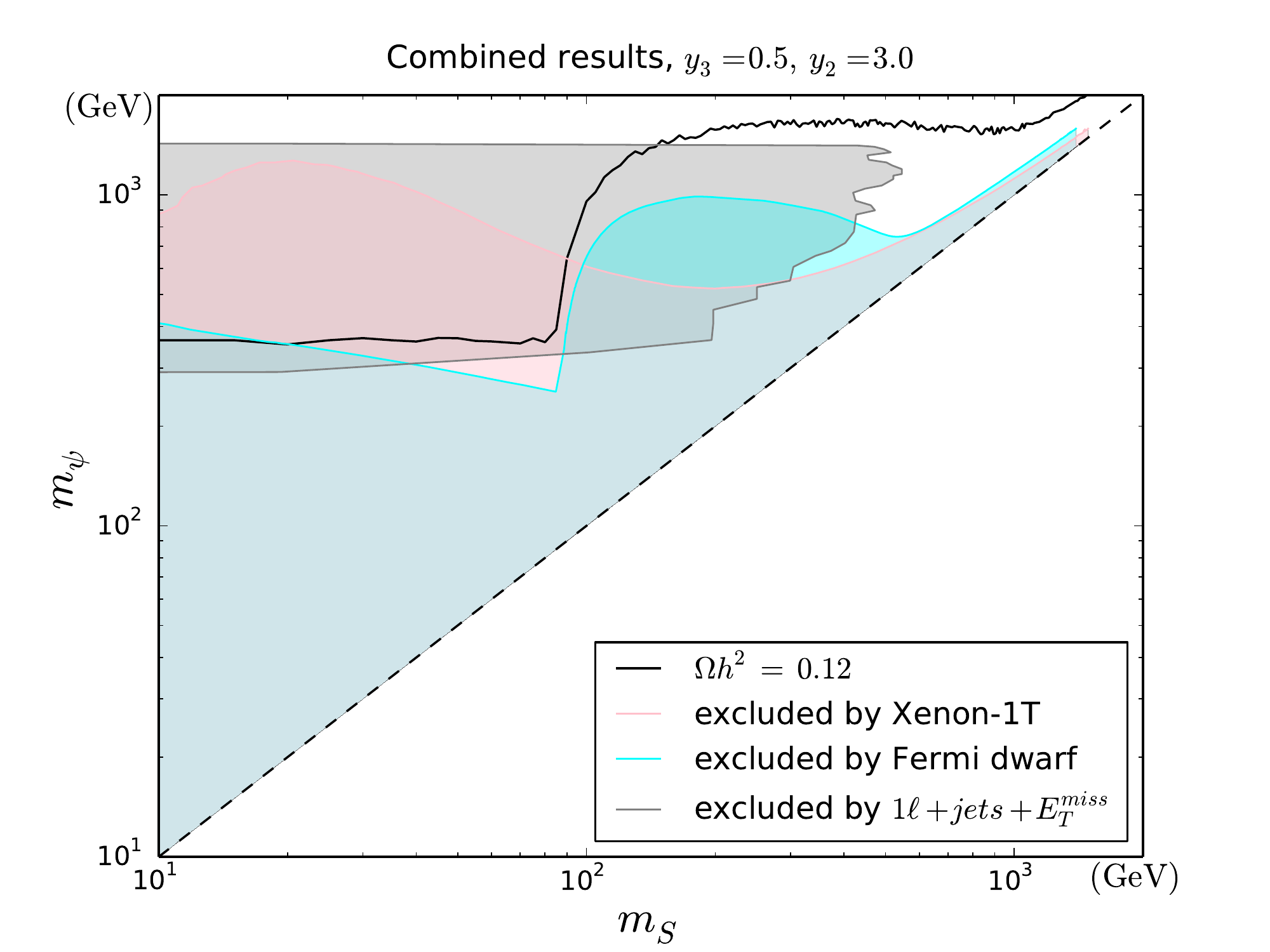}
  \includegraphics[height=6cm,width=7cm]{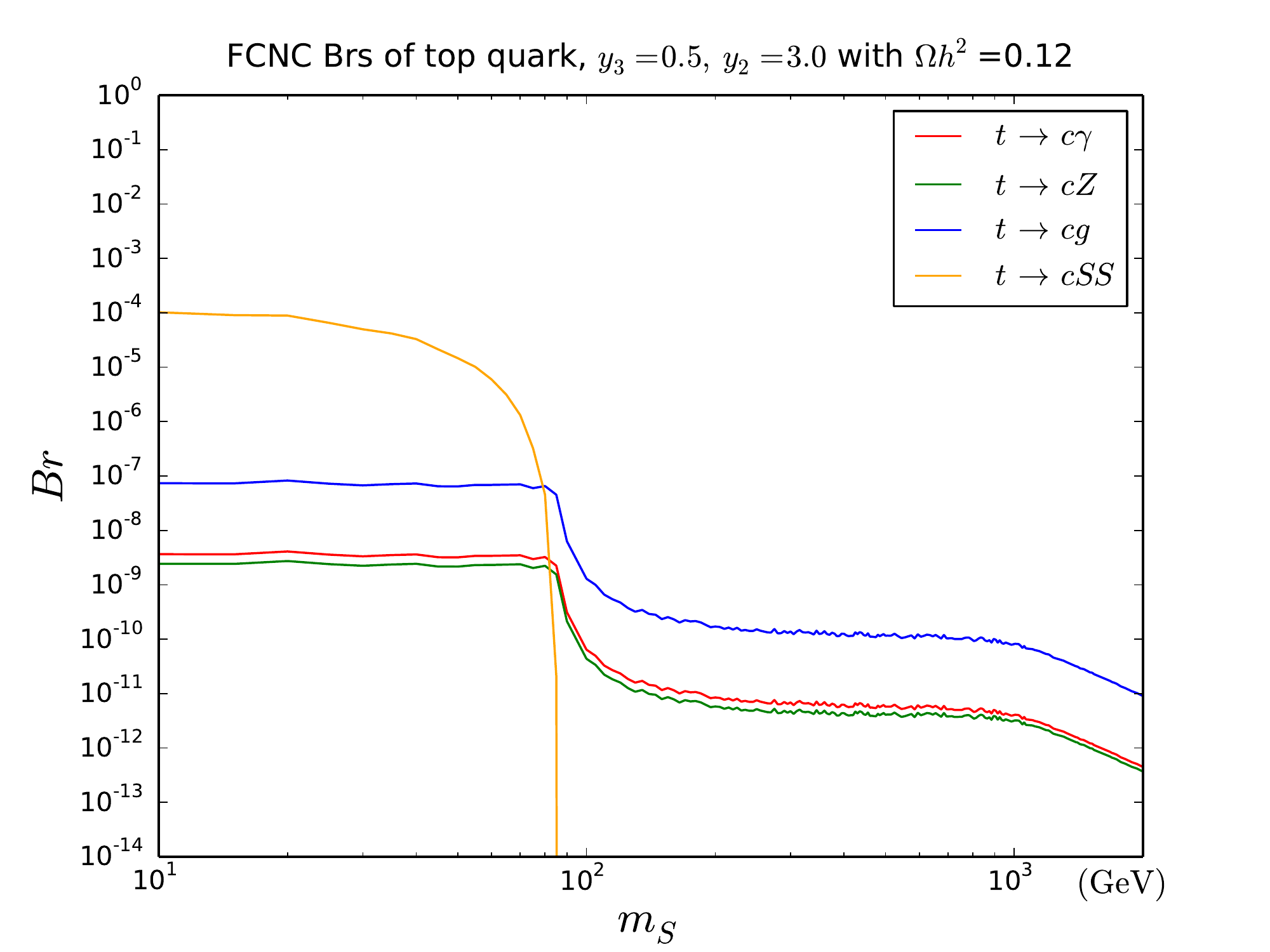}
  \caption{Combined results. {\bf Left: } mass relations required by observed relic abundance confronting the excluded region by direct/indirect detection and 13 TeV LHC data; {\bf Right: } predicted top FCNC branching fractions when satisfying $\Omega_{\rm DM} h^2\simeq 0.12$. Rows from top to bottom correspond to $y_2=0.5,1,3$ with common $y_3=0.5$, respectively.}
  \label{fig-combine}
\end{figure}

In Fig.\ref{fig-combine} we show the combined results from the DM relic abundance requirement $\Omega
h^2\simeq 0.12$, the DM direct/indirect detection constraints, 13 TeV LHC data and the top FCNC predictions for $y_3=0.5$ and $y_2=0.5,1,3$. On the left panels, the pink, cyan, grey and yellow regions are excluded by the current XENON1T,  Fermi dwarf, LHC searches for $jets + \slashed{E}_T$ and $1\ell + jets + \slashed{E}_T$ at 13 TeV, while the black solid lines correspond to the correct DM relic abundance. On the right panels, the solid lines with red, green, blue and orange colors are predictions of $Br(t\to c+\gamma/g/Z/SS)$ when $\Omega_{\rm DM} h^2\simeq 0.12$ is satisfied, respectively. Rows from top to bottom correspond to $y_2=0.5,1,3$ with common $y_3=0.5$.
We found that with increasing $y_2$, the combined results can exclude the whole light DM mass region $m_S\lesssim100$ GeV when scalar $S$ forms the whole DM components, and the top FCNC branching fractions are usually below $10^{-9}$. Note that the current precision of top quark width measurement is around $10\%$ level \cite{CMSCollaboration2014,ATLAS-CONF-2017-056} and the FCNC measurement is around $10^{-4}$ \cite{Bhowmik:2017gzr-FCNCtop}, the benchmark scenarios shown in Fig.\ref{fig-combine} are still safe from top quark width measurements when passing other constraints.
Note that both the experimental measurements ($\sim 10^{-4}$) and the theoretical predictions with $y_2=y_1$ for $Br(t \to u/c+V)$ are close to each other because of $m_u\sim m_c\sim 0$, meanwhile Fig.\ref{fig_DDySQ-Psi} indicates that $y_1$ generally receives stronger limits than $y_2$ from the direct detection (except for a narrow region with destructive cancellation). Therefore, passing the DM direct detection constraints can also make it easy to pass the top FCNC bounds for DM interactions with light quarks through $y_1$ as long as $y_3\sim \mathcal{O}(1)$.

\section{Conclusion}
\label{section-conclusion}

The absence of confirmed signal in dark matter direct detection may suggest weak interaction strengths between DM and the abundant constituents inside nucleon, i.e. gluons and valence light quarks. In this work we consider a real scalar dark matter $S$ interacting only with $SU(2)_L$ singlet Up-type quarks $U_i=u_R,c_R,t_R$ via a vector-like fermion $\psi$ which has the same quantum number as $U_i$. The DM-nucleon scattering can proceed through both $h$-mediated Higgs portal (HP) and $\psi$-mediated vector-like portal (VLP), in which HP can receive sizable radiative corrections through the new fermions.

We first study the separate constraints on the new Yukawa couplings $y_i$ and find that the constraints of XENON1T results are strong on $y_1$ from VLP scattering and on $y_3$ from its radiative contributions to HP scattering. Since both DM-light quark interactions and HP have been well studied in the existing literature, we move forward to focus on DM-heavy quark interactions. Since there is no valence $c,t$ quark inside nucleons at $\mu_{\rm had}\sim 1$ GeV, $y_2,y_3$ interactions are manifested in DM-gluon scattering at loop level.
We find that renormalization group equation and heavy quark threshold effects are important if one calculates the DM-nucleon scattering rate $\sigma^{\rm SI}_{p}$ at $\mu_{\rm had}\sim 1\, {\rm GeV}$ while constructing the effective theory at $\mu_{\rm EFT}\sim m_Z$. For the benchmarks $y_3=0.5, y_2=0.5, 1, 3$, combined results from $\Omega_{\rm DM} h^2\simeq 0.12$, XENON1T, Fermi-LAT, 13 TeV LHC data have almost excluded $m_S<m_t/2$ when only DM-$\{c,t\}$ interactions are considered. FCNC of top quark can be generated at both tree level $t\to \psi^{(*)}S \to cSS$ and loop level $t\to c+\gamma/g/Z$, of which the branching fractions are typically below $10^{-9}$ after passing the other constraints, which are still safe from the current top quark width measurements.

\section*{Acknowledgement}
Peiwen Wu would like to thank Junji Hisano, Natsumi Nagata, Kei Yagyu, Junjie Cao, Yi-Lei Tang, Yangle He, Liangliang
Shang for helpful discussions. We thank the Korea Institute for Advanced Study for providing computing resources (KIAS
Center for Advanced Computation Abacus System) for this work. This research was partly supported by the Munich Institute
for Astro- and Particle Physics (MIAPP) of the DFG cluster of excellence "Origin and Structure of the Universe". 
This work is also supported in part by National Research Foundation of Korea (NRF) Research Grant
NRF-2015R1A2A1A05001869 (SB, PK).

\bibliographystyle{JHEP}
\bibliography{tcDM}

\providecommand{\href}[2]{#2}\begingroup\raggedright\begin{thebibliography}{100}

\bibitem{Sofue2000}
Y.~Sofue and V.~Rubin, \emph{{Rotation Curves of Spiral Galaxies}},
  \href{http://arxiv.org/abs/0010594}{{\tt 0010594}}.

\bibitem{PlanckCollaboration2013}
{Planck Collaboration}, P.~A.~R. Ade, N.~Aghanim, C.~Armitage-Caplan,
  M.~Arnaud, M.~Ashdown et~al., \emph{{Planck 2013 results. XVI. Cosmological
  parameters}},  \href{http://arxiv.org/abs/1303.5076}{{\tt 1303.5076}}.

\bibitem{Clowe2006}
D.~Clowe, M.~Bradac, A.~H. Gonzalez, M.~Markevitch, S.~W. Randall, C.~Jones
  et~al., \emph{{A direct empirical proof of the existence of dark matter}},
  \href{http://arxiv.org/abs/0608407}{{\tt 0608407}}.

\bibitem{Goodenough_0910.2998}
  L.~Goodenough and D.~Hooper,
  arXiv:0910.2998 [hep-ph].

\bibitem{Hooper_1010.2752}
  D.~Hooper and L.~Goodenough,
  Phys.\ Lett.\ B {\bf 697} (2011) 412
  [arXiv:1010.2752 [hep-ph]].

\bibitem{Hooper_1110.0006}
  D.~Hooper and T.~Linden,
  Phys.\ Rev.\ D {\bf 84} (2011) 123005
  [arXiv:1110.0006 [astro-ph.HE]].

\bibitem{Abazajian_1207.6047}
  K.~N.~Abazajian and M.~Kaplinghat,
  Phys.\ Rev.\ D {\bf 86} (2012) 083511
  [arXiv:1207.6047 [astro-ph.HE]].

\bibitem{Gordon_1306.5725}
  C.~Gordon and O.~Macias,
  Phys.\ Rev.\ D {\bf 88} (2013) 083521
   [Erratum-ibid.\ D {\bf 89} (2014) 4,  049901]
  [arXiv:1306.5725 [astro-ph.HE]].

\bibitem{Abazajian_1402.4090}
  K.~N.~Abazajian, N.~Canac, S.~Horiuchi and M.~Kaplinghat,
  Phys.\ Rev.\ D {\bf 90} (2014) 023526
  [arXiv:1402.4090 [astro-ph.HE]].

\bibitem{Hooper_1302.6589}
  D.~Hooper and T.~R.~Slatyer,
  Phys.\ Dark Univ.\  {\bf 2}, 118 (2013).

\bibitem{Hooper_1402.6703}
  T.~Daylan {\it et al.},
  arXiv:1402.6703 [astro-ph.HE].

\bibitem{Calore_1409.0042}
  F.~Calore, I.~Cholis and C.~Weniger,
  arXiv:1409.0042 [astro-ph.CO].

\bibitem{Murgia-Fermi}
 S. Murgia,
 Talk given on Fifth Fermi Symposium, Nagoya, 20-24 October 2014.
 
\bibitem{Calore2014-1411.4647}
F.~Calore, I.~Cholis, C.~McCabe and C.~Weniger, \emph{{A Tale of Tails: Dark
  Matter Interpretations of the Fermi GeV Excess in Light of Background Model
  Systematics}},  \href{http://arxiv.org/abs/1411.4647}{{\tt 1411.4647}}.

\bibitem{Lopez-Honorez2013}
L.~Lopez-Honorez and L.~Merlo, \emph{{Dark matter within the minimal flavour
  violation ansatz}},  \href{http://arxiv.org/abs/1303.1087}{{\tt 1303.1087}}.

\bibitem{Jackson2009}
C.~B. Jackson, G.~Servant, G.~Shaughnessy, T.~M.~P. Tait and M.~Taoso,
  \emph{{Higgs in Space!}},  \href{http://arxiv.org/abs/0912.0004}{{\tt
  0912.0004}}.

\bibitem{Agrawal2015a}
P.~Agrawal, Z.~Chacko, E.~C. F.~S. Fortes and C.~Kilic, \emph{{Skew-Flavored
  Dark Matter}},  \href{http://arxiv.org/abs/1511.06293}{{\tt 1511.06293}}.

\bibitem{Baek2016}
S.~Baek, T.~Nomura and H.~Okada, \emph{{An explanation of one loop induced $h
  \to \mu \tau$ decay}},  \href{http://arxiv.org/abs/1604.03738}{{\tt
  1604.03738}}.

\bibitem{Bai2013}
Y.~Bai and J.~Berger, \emph{{Fermion Portal Dark Matter}},
  \href{http://arxiv.org/abs/1308.0612}{{\tt 1308.0612}}.

\bibitem{Baek2015}
S.~Baek and Z.~Kang, \emph{{Naturally Large Radiative Lepton Flavor Violating
  Higgs Decay Mediated by Lepton-flavored Dark Matter}},
  \href{http://arxiv.org/abs/1510.00100}{{\tt 1510.00100}}.

\bibitem{Chao2016-1606.07174}
W.~Chao, H.-K. Guo and H.-L. Li, \emph{{Tau flavored dark matter and its impact
  on tau Yukawa coupling}},  \href{http://arxiv.org/abs/1606.07174}{{\tt
  1606.07174}}.

\bibitem{Hamze2014}
A.~Hamze, C.~Kilic, J.~Koeller, C.~Trendafilova and J.-H. Yu,
  \emph{{Lepton-Flavored Asymmetric Dark Matter and Interference in Direct
  Detection}},  \href{http://arxiv.org/abs/1410.3030}{{\tt 1410.3030}}.

\bibitem{Kile2014}
J.~Kile, A.~Kobach and A.~Soni, \emph{{Lepton-Flavored Dark Matter}},
  \href{http://arxiv.org/abs/1411.1407}{{\tt 1411.1407}}.

\bibitem{Kile2013}
J.~Kile, \emph{{Flavored Dark Matter: A Review}},
  \href{http://arxiv.org/abs/1308.0584}{{\tt 1308.0584}}.

\bibitem{Agrawal2014b}
P.~Agrawal, Z.~Chacko and C.~B. Verhaaren, \emph{{Leptophilic Dark Matter and
  the Anomalous Magnetic Moment of the Muon}},
  \href{http://arxiv.org/abs/1402.7369}{{\tt 1402.7369}}.

\bibitem{Agrawal2014}
P.~Agrawal, B.~Batell, D.~Hooper and T.~Lin, \emph{{Flavored Dark Matter and
  the Galactic Center Gamma-Ray Excess}},
  \href{http://arxiv.org/abs/1404.1373}{{\tt 1404.1373}}.

\bibitem{Yu2014}
Z.-H. Yu, X.-J. Bi, Q.-S. Yan and P.-F. Yin, \emph{{Tau Portal Dark Matter
  models at the LHC}},  \href{http://arxiv.org/abs/1410.3347}{{\tt 1410.3347}}.

\bibitem{Chang2014}
S.~Chang, R.~Edezhath, J.~Hutchinson and M.~Luty, \emph{{Leptophilic Effective
  WIMPs}},  \href{http://arxiv.org/abs/1402.7358}{{\tt 1402.7358}}.

\bibitem{Lee2014}
C.-J. Lee and J.~Tandean, \emph{{Lepton-Flavored Scalar Dark Matter with
  Minimal Flavor Violation}},  \href{http://arxiv.org/abs/1410.6803}{{\tt
  1410.6803}}.

\bibitem{Calibbi2015}
L.~Calibbi, A.~Crivellin and B.~Zaldivar, \emph{{The Flavour Portal to Dark
  Matter}},  \href{http://arxiv.org/abs/1501.07268}{{\tt 1501.07268}}.

\bibitem{Agrawal2015}
P.~Agrawal, Z.~Chacko, C.~Kilic and C.~B. Verhaaren, \emph{{A Couplet from
  Flavored Dark Matter}},  \href{http://arxiv.org/abs/1503.03057}{{\tt
  1503.03057}}.

\bibitem{Cheung2013}
C.~Cheung and D.~Sanford, \emph{{Simplified Models of Mixed Dark Matter}},
  \href{http://arxiv.org/abs/1311.5896}{{\tt 1311.5896}}.

\bibitem{DiFranzo2013}
A.~DiFranzo, K.~I. Nagao, A.~Rajaraman and T.~M.~P. Tait, \emph{{Simplified
  Models for Dark Matter Interacting with Quarks}},
  \href{http://arxiv.org/abs/1308.2679}{{\tt 1308.2679}}.

\bibitem{Freitas2014}
A.~Freitas and S.~Westhoff, \emph{{Leptophilic Dark Matter in Lepton
  Interactions at LEP and ILC}},  \href{http://arxiv.org/abs/1408.1959}{{\tt
  1408.1959}}.

\bibitem{Cohen2009}
T.~Cohen and K.~M. Zurek, \emph{{Leptophilic Dark Matter from the Lepton
  Asymmetry}},  \href{http://arxiv.org/abs/0909.2035}{{\tt 0909.2035}}.

\bibitem{Schmidt2012}
D.~Schmidt, T.~Schwetz and T.~Toma, \emph{{Direct Detection of Leptophilic Dark
  Matter in a Model with Radiative Neutrino Masses}},
  \href{http://arxiv.org/abs/1201.0906}{{\tt 1201.0906}}.

\bibitem{Chao2010}
W.~Chao, \emph{{Pure Leptonic Gauge Symmetry, Neutrino Masses and Dark
  Matter}},  \href{http://arxiv.org/abs/1005.1024}{{\tt 1005.1024}}.

\bibitem{Baltz2002}
E.~A. Baltz and L.~Bergstrom, \emph{{Detection of Leptonic Dark Matter}},
  \href{http://arxiv.org/abs/0211325}{{\tt 0211325}}.

\bibitem{Cai2014}
Y.~Cai and W.~Chao, \emph{{The Higgs Seesaw Induced Neutrino Masses and Dark
  Matter}},  \href{http://arxiv.org/abs/1408.6064}{{\tt 1408.6064}}.

\bibitem{Isidori2012}
G.~Isidori and D.~M. Straub, \emph{{Minimal Flavour Violation and Beyond}},
  \href{http://arxiv.org/abs/1202.0464}{{\tt 1202.0464}}.

\bibitem{Isidori2010}
G.~Isidori, Y.~Nir and G.~Perez, \emph{{Flavor Physics Constraints for Physics
  Beyond the Standard Model}},  \href{http://arxiv.org/abs/1002.0900}{{\tt
  1002.0900}}.

\bibitem{Agrawal2011}
P.~Agrawal, S.~Blanchet, Z.~Chacko and C.~Kilic, \emph{{Flavored Dark Matter,
  and Its Implications for Direct Detection and Colliders}},
  \href{http://arxiv.org/abs/1109.3516}{{\tt 1109.3516}}.

\bibitem{Kamenik2011}
J.~F. Kamenik and J.~Zupan, \emph{{Discovering Dark Matter Through Flavor
  Violation at the LHC}},  \href{http://arxiv.org/abs/1107.0623}{{\tt
  1107.0623}}.

\bibitem{Batell2011}
B.~Batell, J.~Pradler and M.~Spannowsky, \emph{{Dark Matter from Minimal Flavor
  Violation}},  \href{http://arxiv.org/abs/1105.1781}{{\tt 1105.1781}}.

\bibitem{Kile2011}
J.~Kile and A.~Soni, \emph{{Flavored Dark Matter in Direct Detection
  Experiments and at LHC}},  \href{http://arxiv.org/abs/1104.5239}{{\tt
  1104.5239}}.

\bibitem{Abe2016-FDM}
T.~Abe, J.~Kawamura, S.~Okawa and Y.~Omura, \emph{{Dark matter physics, flavor
  physics and LHC constraints in the dark matter model with a bottom partner}},
   \href{http://arxiv.org/abs/1612.01643}{{\tt 1612.01643}}.

\bibitem{Okawa2017-FDM}
S.~Okawa and Y.~Omura, \emph{{Hidden sector behind the CKM matrix}},
  \href{http://arxiv.org/abs/1703.08789}{{\tt 1703.08789}}.

\bibitem{Bhattacharya:2015xha-FDM}
B.~Bhattacharya, D.~London, J.~M. Cline, A.~Datta and G.~Dupuis,
  \emph{{Quark-flavored scalar dark matter}},
  \href{http://dx.doi.org/10.1103/PhysRevD.92.115012}{\emph{Phys. Rev.} {\bf
  D92} (2015) 115012}, [\href{http://arxiv.org/abs/1509.04271}{{\tt
  1509.04271}}].

\bibitem{Kumar2013}
A.~Kumar and S.~Tulin, \emph{{Top-flavored dark matter and the forward-backward
  asymmetry}},  \href{http://arxiv.org/abs/1303.0332}{{\tt 1303.0332}}.

\bibitem{Kilic2015}
C.~Kilic, M.~D. Klimek and J.-H. Yu, \emph{{Signatures of Top Flavored Dark
  Matter}},  \href{http://arxiv.org/abs/1501.02202}{{\tt 1501.02202}}.

\bibitem{Arina2016}
C.~Arina, M.~Backovi{\'{c}}, E.~Conte, B.~Fuks, J.~Guo, J.~Heisig et~al.,
  \emph{{A comprehensive approach to dark matter studies: exploration of
  simplified top-philic models}},  \href{http://arxiv.org/abs/1605.09242}{{\tt
  1605.09242}}.

\bibitem{Gomez2014}
M.~A. Gomez, C.~B. Jackson and G.~Shaughnessy, \emph{{Dark Matter on Top}},
  \href{http://arxiv.org/abs/1404.1918}{{\tt 1404.1918}}.

\bibitem{Zhang2012}
Y.~Zhang, \emph{{Top Quark Mediated Dark Matter}},
  \href{http://arxiv.org/abs/1212.2730}{{\tt 1212.2730}}.

\bibitem{Batell2013}
B.~Batell, T.~Lin and L.-T. Wang, \emph{{Flavored Dark Matter and R-Parity
  Violation}},  \href{http://arxiv.org/abs/1309.4462}{{\tt 1309.4462}}.

\bibitem{Baek2016a}
S.~Baek, P.~Ko and P.~Wu, \emph{{Top-philic Scalar Dark Matter with a
  Vector-like Fermionic Top Partner}},
  \href{http://arxiv.org/abs/1606.00072}{{\tt 1606.00072}}.

\bibitem{DAmbrosio2002}
G.~D'Ambrosio, G.~F. Giudice, G.~Isidori and A.~Strumia, \emph{{Minimal Flavour
  Violation: an effective field theory approach}},
  \href{http://arxiv.org/abs/0207036}{{\tt 0207036}}.

\bibitem{Chen2015}
M.-C. Chen, J.~Huang and V.~Takhistov, \emph{{Beyond Minimal Lepton Flavored
  Dark Matter}},  \href{http://arxiv.org/abs/1510.04694}{{\tt 1510.04694}}.

\bibitem{Agrawal2014a}
P.~Agrawal, M.~Blanke and K.~Gemmler, \emph{{Flavored dark matter beyond
  Minimal Flavor Violation}},  \href{http://arxiv.org/abs/1405.6709}{{\tt
  1405.6709}}.

\bibitem{Giacchino:2015hvk}
F.~Giacchino, A.~Ibarra, L.~L. Honorez, M.~H.~G. Tytgat and S.~Wild,
  \emph{{Signatures from Scalar Dark Matter with a Vector-like Quark
  Mediator}},
  \href{http://dx.doi.org/10.1088/1475-7516/2016/02/002}{\emph{JCAP} {\bf 1602}
  (2016) 002}, [\href{http://arxiv.org/abs/1511.04452}{{\tt 1511.04452}}].

\bibitem{Silveira:1985rk}
V.~Silveira and A.~Zee, \emph{{SCALAR PHANTOMS}},
  \href{http://dx.doi.org/10.1016/0370-2693(85)90624-0}{\emph{Phys. Lett.} {\bf
  161B} (1985) 136--140}.

\bibitem{McDonald:1993ex}
J.~McDonald, \emph{{Gauge singlet scalars as cold dark matter}},
  \href{http://dx.doi.org/10.1103/PhysRevD.50.3637}{\emph{Phys. Rev.} {\bf D50}
  (1994) 3637--3649}, [\href{http://arxiv.org/abs/hep-ph/0702143}{{\tt
  hep-ph/0702143}}].

\bibitem{Burgess:2000yq}
C.~P. Burgess, M.~Pospelov and T.~ter Veldhuis, \emph{{The Minimal model of
  nonbaryonic dark matter: A Singlet scalar}},
  \href{http://dx.doi.org/10.1016/S0550-3213(01)00513-2}{\emph{Nucl. Phys.}
  {\bf B619} (2001) 709--728}, [\href{http://arxiv.org/abs/hep-ph/0011335}{{\tt
  hep-ph/0011335}}].
 
\bibitem{Cline2013}
J.~M. Cline, K.~Kainulainen, P.~Scott and C.~Weniger, \emph{{Update on scalar
  singlet dark matter}},
  \href{http://dx.doi.org/10.1103/PhysRevD.88.055025}{\emph{Physical Review D}
  {\bf 88} (jun, 2013) 055025}, [\href{http://arxiv.org/abs/1306.4710}{{\tt
  1306.4710}}].

\bibitem{Beniwal2015-HP}
A.~Beniwal, F.~Rajec, C.~Savage, P.~Scott, C.~Weniger, M.~White et~al.,
  \emph{{Combined analysis of effective Higgs portal dark matter models}},
  \href{http://arxiv.org/abs/1512.06458}{{\tt 1512.06458}}.
  
\bibitem{CMSCollaboration2017-VLQuark-1}
{CMS Collaboration}, \emph{{Search for pair production of vector-like T and B
  quarks in single-lepton final states using boosted jet substructure
  techniques at sqrt(s) = 13 TeV}},
  \href{http://arxiv.org/abs/1706.03408}{{\tt 1706.03408}}.

\bibitem{CMSCollaboration2017a-VLQuark-2}
{CMS Collaboration}, \emph{{Search for single production of a vector-like T
  quark decaying to a Z boson and a top quark in proton-proton collisions at
  sqrt(s) = 13 TeV}},  \href{http://arxiv.org/abs/1708.01062}{{\tt
  1708.01062}}.

\bibitem{TheATLASCollaboration2017-VLQuark-1}
{The ATLAS Collaboration}, \emph{{Search for pair production of vector-like top
  quarks in events with one lepton, jets, and missing transverse momentum in 13
  TeV pp collisions with the ATLAS detector}},
  \href{http://arxiv.org/abs/1705.10751}{{\tt 1705.10751}}.

\bibitem{ATLASCollaboration2017-VLQuark-2}
{ATLAS Collaboration}, \emph{{Search for pair production of heavy vector-like
  quarks decaying to high-{\$}p{\_}{\{}mathrm{\{}T{\}}{\}}{\$} {\$}W{\$} bosons
  and {\$}b{\$} quarks in the lepton-plus-jets final state in {\$}pp{\$}
  collisions at 13 TeV with the ATLAS detector}},
  \href{http://arxiv.org/abs/1707.03347}{{\tt 1707.03347}}.

\bibitem{Hisano2010-Gluon}
J.~Hisano, K.~Ishiwata and N.~Nagata, \emph{{Gluon contribution to the dark
  matter direct detection}},  \href{http://arxiv.org/abs/1007.2601}{{\tt
  1007.2601}}.

\bibitem{Hisano2015-DMEFT}
J.~Hisano, R.~Nagai and N.~Nagata, \emph{{Effective Theories for Dark Matter
  Nucleon Scattering}},
  \href{http://dx.doi.org/10.1007/JHEP05(2015)037}{\emph{Journal of High Energy
  Physics} {\bf 2015} (feb, 2015) },
  [\href{http://arxiv.org/abs/1502.02244}{{\tt 1502.02244}}].

\bibitem{Hisano2015-GluonWino}
J.~Hisano, K.~Ishiwata and N.~Nagata, \emph{{QCD Effects on Direct Detection of
  Wino Dark Matter}},  \href{http://arxiv.org/abs/1504.00915}{{\tt
  1504.00915}}.

\bibitem{Hill2014-I}
R.~J. Hill and M.~P. Solon, \emph{{Standard Model anatomy of WIMP dark matter
  direct detection I: weak-scale matching}},
  \href{http://arxiv.org/abs/1401.3339}{{\tt 1401.3339}}.

\bibitem{Hill2014a-II}
R.~J. Hill and M.~P. Solon, \emph{{Standard Model anatomy of WIMP dark matter
  direct detection II: QCD analysis and hadronic matrix elements}},
  \href{http://arxiv.org/abs/1409.8290}{{\tt 1409.8290}}.

\bibitem{Djouadi2000-GluonDM}
A.~Djouadi and M.~Drees, \emph{{QCD corrections to neutralino¨Cnucleon
  scattering}},
  \href{http://dx.doi.org/10.1016/S0370-2693(00)00661-4}{\emph{Physics Letters
  B} {\bf 484} (jul, 2000) 183--191}, [\href{http://arxiv.org/abs/0004205}{{\tt
  0004205}}].

\bibitem{Drees1993-DM}
M.~Drees and M.~M. Nojiri, \emph{{Neutralino-nucleon scattering reexamined}},
  \href{http://dx.doi.org/10.1103/PhysRevD.48.3483}{\emph{Physical Review D}
  {\bf 48} (oct, 1993) 3483--3501}, [\href{http://arxiv.org/abs/9307208}{{\tt
  9307208}}].

\bibitem{Kanemura2004-REN1}
  S.~Kanemura, Y.~Okada, E.~Senaha and C.-P.~Yuan,
  Phys.\ Rev.\ D {\bf 70} (2004) 115002
  doi:10.1103/PhysRevD.70.115002
  [hep-ph/0408364].

\bibitem{Kanemura2016-REN2}
S.~Kanemura, M.~Kikuchi and K.~Yagyu, \emph{{One-loop corrections to the Higgs
  self-couplings in the singlet extension}},
  \href{http://arxiv.org/abs/1608.01582}{{\tt 1608.01582}}.

\bibitem{Shifman1978-QQGG}
M.~Shifman, A.~Vainshtein and V.~Zakharov, \emph{{Remarks on Higgs-boson
  interactions with nucleons}},
  \href{http://dx.doi.org/10.1016/0370-2693(78)90481-1}{\emph{Physics Letters
  B} {\bf 78} (oct, 1978) 443--446}.

\bibitem{Gondolo2013-pole}
P.~Gondolo and S.~Scopel, \emph{{On the sbottom resonance in dark matter
  scattering}},  \href{http://arxiv.org/abs/1307.4481}{{\tt 1307.4481}}.

\bibitem{Alloul2014}
A.~Alloul, N.~D. Christensen, C.~Degrande, C.~Duhr and B.~Fuks,
  \emph{{FeynRules ? 2.0? ??? A complete toolbox for tree-level
  phenomenology}},
  \href{http://dx.doi.org/10.1016/j.cpc.2014.04.012}{\emph{Computer Physics
  Communications} {\bf 185} (aug, 2014) 2250--2300},
  [\href{http://arxiv.org/abs/1310.1921}{{\tt 1310.1921}}].

\bibitem{Belanger2015}
G.~B{\'{e}}langer, F.~Boudjema, A.~Pukhov and A.~Semenov, \emph{{micrOMEGAs4.1:
  Two dark matter candidates}},
  \href{http://dx.doi.org/10.1016/j.cpc.2015.03.003}{\emph{Computer Physics
  Communications} {\bf 192} (jul, 2015) 322--329},
  [\href{http://arxiv.org/abs/1407.6129}{{\tt 1407.6129}}].

\bibitem{Belanger2014}
G.~Belanger, F.~Boudjema and A.~Pukhov, \emph{{micrOMEGAs : a code for the
  calculation of Dark Matter properties in generic models of particle
  interaction}},  \href{http://arxiv.org/abs/1402.0787}{{\tt 1402.0787}}.

\bibitem{Griest1991-omg}
K.~Griest and D.~Seckel, \emph{{Three exceptions in the calculation of relic
  abundances}},
  \href{http://dx.doi.org/10.1103/PhysRevD.43.3191}{\emph{Physical Review D}
  {\bf 43} (may, 1991) 3191--3203}.

\bibitem{Mertig1991-FeynCalc}
R.~Mertig, M.~B{\"{o}}hm and A.~Denner, \emph{{Feyn Calc - Computer-algebraic
  calculation of Feynman amplitudes}},
  \href{http://dx.doi.org/10.1016/0010-4655(91)90130-D}{\emph{Computer Physics
  Communications} {\bf 64} (jun, 1991) 345--359}.

\bibitem{Shtabovenko2016-FeynCalc}
V.~Shtabovenko, R.~Mertig and F.~Orellana, \emph{{New developments in FeynCalc
  9.0}}, \href{http://dx.doi.org/10.1016/j.cpc.2016.06.008}{\emph{Computer
  Physics Communications} {\bf 207} (oct, 2016) 432--444}.

\bibitem{PlanckCollaboration2015}
P.~Collaboration, \emph{{Planck 2015 results. XIII. Cosmological parameters}},
  \href{http://arxiv.org/abs/1502.01589}{{\tt 1502.01589}}.

\bibitem{Colucci:2018vxz-1804}
  S.~Colucci, B.~Fuks, F.~Giacchino, L.~Lopez Honorez, M.~H.~G.~Tytgat and J.~Vandecasteele,
  arXiv:1804.05068 [hep-ph].
  
\bibitem{Bringmann:2012vr}
T.~Bringmann, X.~Huang, A.~Ibarra, S.~Vogl and C.~Weniger, \emph{{Fermi LAT
  Search for Internal Bremsstrahlung Signatures from Dark Matter
  Annihilation}},
  \href{http://dx.doi.org/10.1088/1475-7516/2012/07/054}{\emph{JCAP} {\bf 1207}
  (2012) 054}, [\href{http://arxiv.org/abs/1203.1312}{{\tt 1203.1312}}].

\bibitem{Ackermann:2015zua}
{\scshape Fermi-LAT} collaboration, M.~Ackermann et~al., \emph{{Searching for
  Dark Matter Annihilation from Milky Way Dwarf Spheroidal Galaxies with Six
  Years of Fermi Large Area Telescope Data}},
  \href{http://dx.doi.org/10.1103/PhysRevLett.115.231301}{\emph{Phys. Rev.
  Lett.} {\bf 115} (2015) 231301}, [\href{http://arxiv.org/abs/1503.02641}{{\tt
  1503.02641}}].

\bibitem{Buch2015-PPPC4DM}
J.~Buch, M.~Cirelli, G.~Giesen and M.~Taoso, \emph{{PPPC 4 DM secondary: A Poor
  Particle Physicist Cookbook for secondary radiation from Dark Matter}},
  \href{http://arxiv.org/abs/1505.01049}{{\tt 1505.01049}}.

\bibitem{Cirelli2010-PPPC4DM}
M.~Cirelli, G.~Corcella, A.~Hektor, G.~H{\"{u}}tsi, M.~Kadastik, P.~Panci
  et~al., \emph{{PPPC 4 DM ID: A Poor Particle Physicist Cookbook for Dark
  Matter Indirect Detection}},  \href{http://arxiv.org/abs/1012.4515}{{\tt
  1012.4515}}.

\bibitem{Lopez1997-FCNC}
J.~L. Lopez, D.~V. Nanopoulos and R.~Rangarajan, \emph{{New Supersymmetric
  Contributions to {\$}t-{\textgreater}cV{\$}}},
  \href{http://arxiv.org/abs/9702350}{{\tt 9702350}}.

\bibitem{Soares1989}
J.~M. Soares and A.~Barroso, \emph{{Renormalization of the flavor-changing
  neutral currents}},
  \href{http://dx.doi.org/10.1103/PhysRevD.39.1973}{\emph{Physical Review D}
  {\bf 39} (apr, 1989) 1973--1985}.

\bibitem{Deshpande1982}
N.~G. Deshpande and G.~Eilam, \emph{{Flavor-changing electromagnetic
  transitions}},
  \href{http://dx.doi.org/10.1103/PhysRevD.26.2463}{\emph{Physical Review D}
  {\bf 26} (nov, 1982) 2463--2485}.

\bibitem{Hahn1999-LoopTools}
T.~Hahn and M.~P{\'{e}}rez-Victoria, \emph{{Automated one-loop calculations in
  four and D dimensions}},
  \href{http://dx.doi.org/10.1016/S0010-4655(98)00173-8}{\emph{Computer Physics
  Communications} {\bf 118} (may, 1999) 153--165}.

\bibitem{Hahn1999-FF}
T.~Hahn and M.~P{\'{e}}rez-Victoria, \emph{{Automated one-loop calculations in
  four and D dimensions}},
  \href{http://dx.doi.org/10.1016/S0010-4655(98)00173-8}{\emph{Computer Physics
  Communications} {\bf 118} (may, 1999) 153--165}.

\bibitem{ATLAS-CONF-2016-050}
A.~Collaboration, \emph{{Search for top squarks in final states with one
  isolated lepton, jets, and missing transverse momentum in $\sqrt{s}$ = 13 TeV
  pp collisions with the ATLAS detector}},  Tech. Rep. ATLAS-CONF-2016-050,
  CERN, Geneva, Aug, 2016.

\bibitem{ATLAS-CONF-2017-022}
A.~Collaboration, \emph{{Search for squarks and gluinos in final states with
  jets and missing transverse momentum using 36 fb$^{-1}$ of $\sqrt{s} =13$ TeV
  pp collision data with the ATLAS detector}},  Tech. Rep. ATLAS-CONF-2017-022,
  CERN, Geneva, Apr, 2017.

\bibitem{ATLAS-CONF-2017-037}
A.~Collaboration, \emph{{Search for top squarks in final states with one
  isolated lepton, jets, and missing transverse momentum using 36.1fb$^{-1}$ of
  $\sqrt{13}$ TeV pp collision data with the ATLAS detector}},  Tech. Rep.
  ATLAS-CONF-2017-037, CERN, Geneva, May, 2017.

\bibitem{Czakon:2011xx}
M.~Czakon and A.~Mitov, \emph{{Top++: A Program for the Calculation of the
  Top-Pair Cross-Section at Hadron Colliders}},
  \href{http://dx.doi.org/10.1016/j.cpc.2014.06.021}{\emph{Comput. Phys.
  Commun.} {\bf 185} (2014) 2930}, [\href{http://arxiv.org/abs/1112.5675}{{\tt
  1112.5675}}].

\bibitem{Cacciari:2011hy}
M.~Cacciari, M.~Czakon, M.~Mangano, A.~Mitov and P.~Nason, \emph{{Top-pair
  production at hadron colliders with next-to-next-to-leading logarithmic
  soft-gluon resummation}},
  \href{http://dx.doi.org/10.1016/j.physletb.2012.03.013}{\emph{Phys. Lett.}
  {\bf B710} (2012) 612--622}, [\href{http://arxiv.org/abs/1111.5869}{{\tt
  1111.5869}}].

\bibitem{Beneke:2011mq}
M.~Beneke, P.~Falgari, S.~Klein and C.~Schwinn, \emph{{Hadronic top-quark pair
  production with NNLL threshold resummation}},
  \href{http://dx.doi.org/10.1016/j.nuclphysb.2011.10.021}{\emph{Nucl. Phys.}
  {\bf B855} (2012) 695--741}, [\href{http://arxiv.org/abs/1109.1536}{{\tt
  1109.1536}}].

\bibitem{PhysRevLett.109.132001}
P.~B\"arnreuther, M.~Czakon and A.~Mitov, \emph{Percent-level-precision physics
  at the tevatron: Next-to-next-to-leading order qcd corrections to
  $q\overline{q}\ensuremath{\rightarrow}t\overline{t}\mathbf{+}x$},
  \href{http://dx.doi.org/10.1103/PhysRevLett.109.132001}{\emph{Phys. Rev.
  Lett.} {\bf 109} (Sep, 2012) 132001}.

\bibitem{Czakon:2012zr}
M.~Czakon and A.~Mitov, \emph{{NNLO corrections to top-pair production at
  hadron colliders: the all-fermionic scattering channels}},
  \href{http://dx.doi.org/10.1007/JHEP12(2012)054}{\emph{JHEP} {\bf 12} (2012)
  054}, [\href{http://arxiv.org/abs/1207.0236}{{\tt 1207.0236}}].

\bibitem{Czakon:2012pz}
M.~Czakon and A.~Mitov, \emph{{NNLO corrections to top pair production at
  hadron colliders: the quark-gluon reaction}},
  \href{http://dx.doi.org/10.1007/JHEP01(2013)080}{\emph{JHEP} {\bf 01} (2013)
  080}, [\href{http://arxiv.org/abs/1210.6832}{{\tt 1210.6832}}].

\bibitem{Czakon:2013goa}
M.~Czakon, P.~Fiedler and A.~Mitov, \emph{{Total Top-Quark Pair-Production
  Cross Section at Hadron Colliders Through $O(¦Á\frac{4}{S})$}},
  \href{http://dx.doi.org/10.1103/PhysRevLett.110.252004}{\emph{Phys. Rev.
  Lett.} {\bf 110} (2013) 252004}, [\href{http://arxiv.org/abs/1303.6254}{{\tt
  1303.6254}}].

\bibitem{Degrande:2011ua-UFO}
C.~Degrande, C.~Duhr, B.~Fuks, D.~Grellscheid, O.~Mattelaer and T.~Reiter,
  \emph{{UFO - The Universal FeynRules Output}},
  \href{http://dx.doi.org/10.1016/j.cpc.2012.01.022}{\emph{Comput. Phys.
  Commun.} {\bf 183} (2012) 1201--1214},
  [\href{http://arxiv.org/abs/1108.2040}{{\tt 1108.2040}}].

\bibitem{Alwall2014}
J.~Alwall, R.~Frederix, S.~Frixione, V.~Hirschi, F.~Maltoni, O.~Mattelaer
  et~al., \emph{{The automated computation of tree-level and next-to-leading
  order differential cross sections, and their matching to parton shower
  simulations}}, \href{http://dx.doi.org/10.1007/JHEP07(2014)079}{\emph{Journal
  of High Energy Physics} {\bf 2014} (jul, 2014) 79},
  [\href{http://arxiv.org/abs/1405.0301}{{\tt 1405.0301}}].

\bibitem{Sjostrand:2007gs-PYTHIA81}
T.~Sjostrand, S.~Mrenna and P.~Z. Skands, \emph{{A Brief Introduction to PYTHIA
  8.1}}, \href{http://dx.doi.org/10.1016/j.cpc.2008.01.036}{\emph{Comput. Phys.
  Commun.} {\bf 178} (2008) 852--867},
  [\href{http://arxiv.org/abs/0710.3820}{{\tt 0710.3820}}].

\bibitem{Sjostrand:2014zea-PYTHIA82}
T.~Sj?strand, S.~Ask, J.~R. Christiansen, R.~Corke, N.~Desai, P.~Ilten et~al.,
  \emph{{An Introduction to PYTHIA 8.2}},
  \href{http://dx.doi.org/10.1016/j.cpc.2015.01.024}{\emph{Comput. Phys.
  Commun.} {\bf 191} (2015) 159--177},
  [\href{http://arxiv.org/abs/1410.3012}{{\tt 1410.3012}}].

\bibitem{Kim2015}
J.~S. Kim, D.~Schmeier, J.~Tattersall and K.~Rolbiecki, \emph{{A framework to
  create customised LHC analyses within CheckMATE}},
  \href{http://arxiv.org/abs/1503.01123}{{\tt 1503.01123}}.

\bibitem{Drees2013}
M.~Drees, H.~Dreiner, J.~S. Kim, D.~Schmeier and J.~Tattersall,
  \emph{{CheckMATE: Confronting your Favourite New Physics Model with LHC
  Data}},  \href{http://arxiv.org/abs/1312.2591}{{\tt 1312.2591}}.

\bibitem{Ovyn2009}
S.~Ovyn, X.~Rouby and V.~Lemaitre, \emph{{Delphes, a framework for fast
  simulation of a generic collider experiment}},
  \href{http://arxiv.org/abs/0903.2225}{{\tt 0903.2225}}.

\bibitem{DeFavereau2014}
J.~de~Favereau, C.~Delaere, P.~Demin, A.~Giammanco, V.~Lema{\^{\i}}tre,
  A.~Mertens et~al., \emph{{DELPHES 3: a modular framework for fast simulation
  of a generic collider experiment}},
  \href{http://dx.doi.org/10.1007/JHEP02(2014)057}{\emph{Journal of High Energy
  Physics} {\bf 2014} (feb, 2014) 57},
  [\href{http://arxiv.org/abs/1307.6346}{{\tt 1307.6346}}].

\bibitem{CMSCollaboration2014}
{CMS Collaboration}, \emph{{Measurement of the ratio B(t to Wb)/B(t to Wq) in
  pp collisions at sqrt(s) = 8 TeV}},
  \href{http://arxiv.org/abs/1404.2292}{{\tt 1404.2292}}.

\bibitem{ATLAS-CONF-2017-056}
{\scshape ATLAS Collaboration} collaboration, A.~Collaboration, \emph{{Direct
  top-quark decay width measurement in the $t\bar{t}$ lepton+jets channel at
  $\sqrt{s}=8~\text{TeV}$ with the ATLAS experiment}},  Tech. Rep.
  ATLAS-CONF-2017-056, CERN, Geneva, Jul, 2017.

\bibitem{Bhowmik:2017gzr-FCNCtop}
{\scshape ATLAS, CMS} collaboration, S.~Bhowmik, \emph{{Flavor Changing Neutral
  Current searches in the top quark sector}}, {\emph{PoS} {\bf CKM2016} (2017)
  126}.

\end{thebibliography}\endgroup

\end{document}